\documentclass{emulateapj}
\usepackage{colordvi}
\usepackage{epsfig}
\usepackage{graphicx}
\usepackage{amssymb}
\usepackage{lscape}
\newcommand{\mic}{~$\mu$m}

\newcommand{\HST}{{\it HST}}

\begin{document}

\title{UV-to-FIR analysis of {\it Spitzer}/IRAC sources in the Extended Groth Strip II:
Photometric redshifts, Stellar masses and Star formation rates}

\author{G. Barro\altaffilmark{1}, P.G. P\'erez-Gonz\'alez\altaffilmark{1,2}, J. Gallego\altaffilmark{1}, M. L. N. Ashby\altaffilmark{3},  M. Kajisawa\altaffilmark{4}, S. Miyazaki\altaffilmark{5}, V. Villar\altaffilmark{1}, T. Yamada\altaffilmark{4}, J. Zamorano\altaffilmark{1}}
\altaffiltext{1}{Departamento de Astrof\'{\i}sica, Facultad de CC. F\'{\i}sicas,
Universidad Complutense de Madrid, E-28040 Madrid, Spain}
\altaffiltext{2}{Associate Astronomer at Steward Observatory, The University of Arizona}
\altaffiltext{2}{Harvard-Smithsonian Center for Astrophysics, 60 Garden St., Cambridge, MA 02138}
\altaffiltext{3}{Astronomical Institute, Tohoku University, Aramaki, Aoba, Sendai 980–8578, Japan}
\altaffiltext{4}{National Astronomical Observatory of Japan, Mitaka, Tokyo 181-8588, Japan}
\slugcomment{Last edited: \today}
\date{Submitted: \today}
\label{firstpage}
\begin{abstract}

  Based on the ultraviolet to far-infrared photometry already compiled 
  and presented in a companion paper (Barro et al. 2011a, Paper I), 
  we present a detailed SED analysis of nearly 80,000 
  IRAC 3.6+4.5\,$\mu$m-selected galaxies in the Extended
  Groth Strip.  We estimate photometric redshifts, stellar masses, and
  star formation rates separately for each galaxy in this large sample.  
  The catalog includes 76,936 sources with [3.6]$\leq$23.75 (85\%
  completeness level of the IRAC survey) over 0.48\,deg$^{2}$.  The
  typical photometric redshift accuracy is $\Delta z/(1+z)$$=$0.034,
  with a catastrophic outlier fraction of just 2\%.  We quantify the
  systematics introduced by the use of different stellar population 
  synthesis libraries and IMFs in the calculation of stellar masses.
  We find systematic offsets ranging from 0.1 to 0.4\,dex,
  with a typical scatter of 0.3\,dex. We also provide UV- and IR-based
  SFRs for all sample galaxies, based on several sets of
  dust emission templates and SFR indicators.  We evaluate the
  systematic differences and goodness of the different SFR estimations
  using the deep FIDEL 70\mic\, data available in the EGS. Typical
  random uncertainties of the IR-bases SFRs are a factor of two, with
  non-negligible systematic effects at z$\gtrsim$1.5 observed when
  only MIPS 24\mic\, data is available. All data products 
(SEDs, postage stamps from imaging
  data, and different estimations of the photometric redshifts,
  stellar masses, and SFRs of each galaxy) 
described in this and the companion paper are publicly available, and
  they can be accessed through our the web-interface utility {\it
    Rainbow-navigator}.

\end{abstract}

\section{Introduction}\label{intro}

Multi-band catalogs are the fuel for studies aimed at exploring the
global evolution of galaxies over cosmic history. They have been
used to study the redshift evolution of the star formation rate (SFR)
density (e.g., \citealt{2006ApJ...651..142H},
\citealt{2008ApJS..175...48R}, \citealt{2009ApJ...705..936B}), and the
stellar mass assembly process (e.g., \citealt{2006ApJ...651..120B},
\citealt{2006A&A...459..745F}, \citealt{2008ApJ...675..234P},
\citealt{2009ApJ...701.1765M}).

The unprecedented sensitivity of modern surveys detect millions of
distant galaxies to faint flux levels that for all practical purposes
lie well beyond the capabilities of even the most recent multi-object
spectrographs at the largest telescopes.  As a consequence, their
intrinsic properties must be estimated through multi-band photometric
data using fitting techniques to stellar population templates, and/or
empirical relations.  Among the basic parameters needed to
characterize a galaxy, arguably the most important is the redshift,
which must be inferred from an analysis of its Spectral Energy
Distribution (SED).  Photometric redshift techniques are now
sufficiently accurate to derive statistically reliable conclusions for
high-redshift galaxy populations (e.g., \citealt{1998ApJ...509..103S},
\citealt{2003A&A...401...73W}, \citealt{2009ApJ...690.1236I}).

Many different codes have been developed to calculate photometric
redshifts based in the same principle: finding the galaxy spectral
template best fitting the observed photometry in several band-passes.
Some examples include HYPERZ \citep{2000A&A...363..476B}, BPZ
\citep{2000ApJ...536..571B} or LePHARE (Arnouts\&Ilbert; e.g.,
\citealt{2009ApJ...690.1236I}). The implementation is very sensitive
to the quality of the photometry and the capability of the observed
bands to probe key continuum features of the spectra (e.g., the Lyman
and Balmer breaks).  It also depends strongly on the availability of
templates that are statistically representative and successful in
characterizing the emission of galaxies. The impact of these factors
in the uncertainty of the estimations is not straightforward, and it
can lead to catastrophic errors beyond the simple propagation of the
statistical errors (\citealt{2008ApJ...689..709O},
\citealt{2008A&A...480..703H}).  In recent years, several techniques
have been developed to improve the reliability of the photometric
redshifts (e.g., Bayesian priors \citep{2000ApJ...536..571B},
template-optimization procedures \citep{2006A&A...457..841I} and
machine-learning neural networks \citep{2004PASP..116..345C}).  Recent
work including some of these advances have achieved remarkable
precision [e.g., $\Delta z/(1+z)<0.012$ in
  \citealt{2009ApJ...690.1236I}, and $\Delta z/(1+z)=0.06$ at z$>$1.5
  in \citealt{2009PASP..121....2V}].

Once a galaxy's redshift has been estimated, the most significant
physical properties that can then be derived from multi-wavelength
photometry are the stellar mass and the SFR.  However, estimates
derived from modeling of the observed SEDs involve significant random
and systematic uncertainties.  The estimate of the stellar mass by
fitting stellar population synthesis models is a widespread technique
(e.g., \citealt{2003ApJ...585L.117B}, \citealt{2007MNRAS.378.1550P},
\citealt{2008A&A...491..713W}) that requires making some assumptions
regarding the initial stellar mass function, the star formation
history or the extinction law.  Moreover, there exist significant
differences among stellar population libraries. These differences can
lead to discrepancies in the stellar mass estimation of a factor of a
few (\citealt{2006ApJ...652...85M}, \citealt{2007ASPC..374..303B}).

SFR estimates based on UV and/or IR luminosities are considered reasonably 
robust for large galaxy samples with multi-wavelength photometry, where 
other tracers, such as spectroscopy, are unavailable
(\citealt{2006ApJ...653.1004R}, \citealt{2007ApJS..173..267S},
\citealt{2007ApJ...670..156D}). A major problem with SFRs estimated
from UV data is the need for a extinction correction, which can be
highly uncertain and redshift-dependent
(\citealt{2007ApJ...670..279I},\citealt{2007MNRAS.380..986B},
\citealt{2009ApJ...700..161S}). On the other hand, IR-based SFRs
estimated by fitting the MIR-to-mm fluxes with dust emission templates
are model dependent (\citealt{2002ApJ...579L...1P},
\citealt{2005ApJ...633..857D}, \citealt{2006ApJ...637..727C}).
Furthermore, these tracers are based on the assumption that the bulk
of the IR emission traces warm dust heated by young star-forming
regions. Thus, if a fraction of the energy heating the dust originates
from an alternative source, such as deeply dust enshrouded AGNs or
diffuse radiation fields (\citealt{2007ApJ...670..156D},
\citealt{2009ApJ...700..161S}) the SFR will be overestimated.
Nevertheless, despite these second-order effects, the uncertainties in
the SFR are frequently driven by the absence of sufficient IR
photometry to constrain the models robustly.  In the last few years,
the studies of SFRs at high redshift have often been based on the
observed flux at MIPS 24\,$\mu$m only and, although IR monochromatic
luminosities are known to correlate well to total IR luminosity,
recent works based on more detailed IR coverage have demonstrated that
SFRs from MIPS 24\,$\mu$m data may present significant systematics
(\citealt{2007ApJ...668...45P}, \citealt{2007ApJ...670..156D},
\citealt{2008ApJ...675..262R}).

In this context, \citealt{2011arXiv1101.3308B} (hereafter Paper I)
presented a multi-band photometric and spectroscopic catalog
(including data from X-ray to radio wavelengths) in the Extended Groth
Strip (EGS), that can be used as a starting point for detailed
analysis of the galaxy population.  That paper describes the method
used to measure coherent multi-band photometry and presents the
general properties of the merged catalog, including an analysis of the
quality and reliability of the photometry.  Paper I also presents {\it
  Rainbow Navigator}, a publicly available web-interface that provides
access to all the multi-band data products.

In this paper, we focus on fitting the optical-to-NIR SEDs and IR
emission of all the sources presented in Paper I using
stellar population synthesis models (SPS) and dust emission templates.
We then use the SEDs and fits to estimate photometric redshifts, stellar
masses, and SFRs.  We also quantify the uncertainties attending
these estimations.  In particular, we assess the quality of the
photometric redshifts by comparing our results with spectroscopic
redshifts and with other photometric redshift compilations found in
the literature.  We explore the systematic uncertainties in the
stellar masses associated with the modeling assumptions, such as the
choice of SPS models or the initial mass function (IMF). Finally, we
study the systematic uncertainties in the IR-based SFRs estimated with
different IR templates and indicators (e.g., different total IR
luminosity-to-SFR calibrations).

The outline of this paper is as follows.  $\S$~\ref{multibandcat}
briefly reviews the available data and then summarizes the most
relevant steps of the photometric measurement and band-merging
procedure (presented in paper I), as well as the overall photometric
properties of the IRAC 3.6+4.5\,$\mu$m-selected
catalog. $\S$~\ref{analyzesed} describes the techniques developed to
perform the UV-to-IR SED fitting, and the methods used to estimate
redshifts.  $\S$~\ref{evalmass} describes the stellar masses
estimation technique, and quantifies the uncertainties introduced by
the modeling assumptions.  $\S$~\ref{evalSFR} describes the methods
used to fit the FIR emission to dust emission templates and the
estimation of IR luminosities and SFRs.  $\S$~\ref{datacatalogs}
presents tables containing all the data products presented in this
paper, as well as the public database created to facilitate the access
to these resources.

Throughout this paper we use AB magnitudes. We adopt the cosmology
$H_{0}=70$ km$^{-1}$s$^{-1}$Mpc$^{-1}$, $\Omega_{m}=0.3$ and
$\Omega_{\lambda}=0.7$. Our default choice of SED modeling parameters
are: the PEGASE \citep{1997A&A...326..950F} library, a
\citet{1955ApJ...121..161S} IMF (M~$\in$~[0.1-100]\,M$_{\odot}$), and
a \citet{2000ApJ...533..682C} extinction law.

\placetable{zeropoints}
\begin{deluxetable*}{lccccr}
\tabletypesize{\scriptsize}
\setlength{\tabcolsep}{0.04in} 
\tablewidth{0pt}
\tablecaption{\label{zeropoints}Photometric properties of the dataset}
\tablehead{\colhead{Filter} &  \colhead{$\lambda_{\mathrm{eff}}$}& \colhead{m$_{\mathrm{lim}}$[AB]}& \colhead{FWHM} & \colhead{Gal. ext} &\colhead{Offset} \\
\colhead{(1)}&\colhead{(2)}&\colhead{(3)}&\colhead{(4)}&\colhead{(5)}&\colhead{(6)}}
\startdata
GALEX-FUV                  &  153.9~nm & 25.6 &5.5\arcsec&      0.195&   0.04\\  
GALEX-NUV                  &  231.6~nm & 25.6 &5.5\arcsec&      0.101&   0.08\\  
MMT-$u$                   &  362.5~nm & 26.1 &1.0\arcsec&      0.049&  -0.09\\  
CFHTLS-$u^{*}$             &  381.1~nm & 25.7 &0.9\arcsec&      0.045&   -0.04\\ 
CFH-B                      &  439.0~nm & 25.7 &1.2\arcsec&      0.036&   0.04\\  
MMT-$g$                   &  481.4~nm & 26.7 &1.3\arcsec&      0.031&  -0.09\\  
CFHTLS-$g'$                &  486.3~nm & 26.5 &0.9\arcsec&      0.031&   0.03\\  
ACS-$V_{606}$              & 591.3~nm & 26.1 &0.2\arcsec&       0.022&   0.02\\  
CFHTLS-$r'$                &  625.8~nm & 26.3 &0.8\arcsec&      0.020&   0.03\\  
Subaru-R                   &  651.8~nm & 26.1 &0.7\arcsec&      0.019&   0.00\\  
CFH-R                      &  660.1~nm & 25.3 &1.0\arcsec&      0.019&  -0.03\\  
CFHTLS-$i'$                &  769.0~nm & 25.9 &0.8\arcsec&      0.015&   0.03\\  
MMT-$i$                   &  781.5~nm & 25.3 &1.0\arcsec&      0.015&   0.00\\  
ACS-$i_{814}$              &  813.2~nm & 26.1 &0.2\arcsec&      0.014&   0.00\\  
CFH-I                      &  833.0~nm & 24.9 &1.1\arcsec&      0.013&   0.02\\  
CFHTLS-$z'$                &  887.1~nm & 24.7 &0.8\arcsec&      0.012&  -0.02\\  
MMT-$z$                   &  907.0~nm & 25.3 &1.2\arcsec&      0.011&  -0.11\\  
NICMOS-$J_{110}$           &  1.10~$\mu$m  & 23.5 &0.7\arcsec&  0.008&   0.00\\  
$\Omega2k-J$               &  1.21~$\mu$m  & 22.9 &1.0\arcsec&  0.007&  -0.16\\  
WIRC-$J^{\dagger}$          &  1.24~$\mu$m  & 21.9 &1.0\arcsec  &  0.007&  0.01\\   
NICMOS-$H_{160}$           &  1.59~$\mu$m  & 24.2 &0.8\arcsec&  0.005&   0.00\\  
$\Omega'-K$                &  2.11~$\mu$m  & 20.7 &1.5\arcsec&  0.003&   -0.10\\ 
Subaru-MOIRCS-$Ks$         &  2.15~$\mu$m  & 23.7 &0.6\arcsec&  0.003&   -0.04\\ 
WIRC-$K^{\dagger}$  &  2.16~$\mu$m  & 22.9 &1.0\arcsec  &  0.003&   0.00  \\
IRAC-36                 &  3.6~$\mu$m &  23.7&2.1\arcsec  &  0.001&   0.00  \\
IRAC-45                 &  4.5~$\mu$m &  23.7&2.1\arcsec  &  0.001&   0.00  \\
IRAC-58                 &  5.8~$\mu$m &  22.1&2.2\arcsec  &  0.001&   0.12  \\
IRAC-80                 &  8.0~$\mu$m &  22.1&2.2\arcsec  &  0.000&   0.12  
\enddata
\tablecomments{\\
  $\dagger$ The photometry was not  measured, but taken from a published catalog.\\
  Col(1) Name of the observing band and instrument.\\
  Col(2) Effective wavelength of the filter calculated by convolving the Vega spectrum \citep{1994AJ....108.1931C} with the transmission curve of the filter+detector.\\
  Col(3) Limiting AB magnitude of the image estimated as the magnitude of a SNR$=$5 detection (see \S~\ref{photometry} for details on the flux measurement).\\
  Col(4) Median FWHM of the PSF in arcseconds measured in a large number of stars (see \S~5.4 of paper I for details on the stellarity criteria).\\
  Col(5) Galactic extinction estimated from the \citet{1998ApJ...500..525S} maps and assuming and average value of E(B-V)=0.004.\\
  Col(6) Zero-point corrections applied to the photometric bands, computed by comparing observed and synthetic magnitudes for spectroscopic galaxies (see \S~\ref{zeropoint}).}
\end{deluxetable*}

\begin{figure}
\centering
\includegraphics[width=8.5cm,angle=0.]{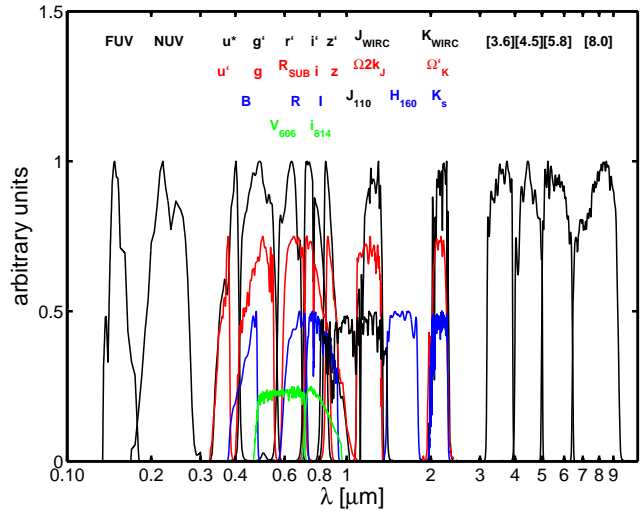}
\caption{\label{filters} Filter transmission for the photometric bands
  included in the dataset. The curves include the atmospheric
  transmission (for ground based observations), quantum efficiency,
  and the transmission of the optical elements. The curves are
  normalized at the maximum value of the transmission and scaled
  arbitrarily for visualization. The color code for each filter
  corresponds to the labels shown above. The optical bands depicted
  from top to bottom are those of CFHTLS, MMT and CFHT12k,
  respectively. The NIR bands, also shown from top-to-bottom are those
  of WIRC, CAHA and {\it HST}/NICMOS. The GALEX (FUV, NUV) and IRAC
  ([3.6],[4.5],[5.8],[8.0]) filters are also listed in the top row.}
\end{figure}

\section{Multi-wavelength catalog}\label{multibandcat}

The present work is based on the multi-wavelength catalog of IRAC
3.6+4.5\,$\mu$m-selected galaxies in the Extended Groth Strip (EGS;
$\alpha=14^{h}14^{m}$, $\delta=+53^{\circ}30'$) presented in Paper I.
The catalog contains all the publicly available data provided by the
All-Wavelength Extended Groth Strip International Survey (AEGIS)
collaboration and some proprietary data including the following bands:
GALEX FUV and NUV, CFHTLS~$u^{*}$$g'$$r'$$i'$$z'$, MMT-$u'giz$,
CFHT12k~BRI, ACS~$V_{606}$$i_{814}$, Subaru~R,
NICMOS~$J_{110}$$H_{160}$, MOIRCS~$K_{s}$, CAHA-$JK_{s}$, WIRC~$J$$K$,
the four IRAC bands at 3.6, 4.5, 5.8, and 8.0\,$\mu$m, and lastly MIPS
24 and 70\,$\mu$m.  We cross-correlated our IRAC-selected catalog with
the X-ray (Chandra) and radio (VLA/20cm) catalogs of
\citet{2009ApJS..180..102L} and \citet{2007ApJ...660L..77I}, and with
all the spectroscopic redshifts from DEEP2 DR3 and a small sample of
238 spectroscopically confirmed Lyman break galaxies (LBG) from
\citet{2003ApJ...592..728S}.  The reader is referred to Paper I and
\citealt{2007ApJ...660L...1D}, and references therein, for a detailed
description of all these datasets. Figure~\ref{filters} illustrates
the different filter transmission profiles for each band, and
Table~\ref{zeropoints} presents the effective filter wavelengths, the
survey depths and image quality achieved in each band, and the (small)
zero-point re-calibrations (Section~\ref{zeropoint}).

The photometric coverage of the EGS is largely inhomogeneous, with
each band covering a different portion of the IRAC mosaic
(\citealt{2007ApJ...660L...1D}).  Fortunately, there is a natural way
to divide the field into two smaller sub-regions.  The main region,
defined by the overlapping area of the CFHTLS and IRAC frames
(0.35\,deg$^{2}$), presents the densest coverage ($\sim$19 bands,
including GALEX, HST and MOIRCS). This region is essentially a field
with the side edges following the contours of the IRAC image, i.e.,
inclined by 50$^{\circ}$ east of north, and upper and lower boundaries
limited by 52.16$^{\circ}$$<\delta<$53.20$^{\circ}$. The bottom-right
side is also restricted to $\alpha>$214.04$^{\circ}$ due to the
intersection with the CFHTLS mosaic (a square field oriented North up,
East left).

The 0.13\,deg$^{2}$ outside of the main region (hereafter referred to
as flanking regions) also have solid optical-to-NIR coverage. However,
the overall data quality is slightly lower than in the main region.
The median coverage includes only 11 bands, and for the most part
lacks the deepest, highest-resolution imaging.  As a result, the
quality of the SED coverage in the flanking regions is significantly
lower than in the main region.  For these reasons we focus in this
contribution on the main region.

\subsection{Multi-band identification and photometry}
\label{photometry}

The procedure followed to build consistent UV-to-FIR SEDs from the
multiple datasets is described detail in Paper I (see also
\citealt{2005ApJ...630...82P} and \citealt{2008ApJ...675..234P},
hereafter PG05 and PG08).  This Section summarizes the most relevant
elements of the method, so that the impact of the photometric uncertainties
on the parameters estimated from the SED modeling can be
assessed (\S~\ref{evalmass} and \S~\ref{evalSFR}).


First, multi-band identification is carried out by cross-correlating
the 3.6+4.5\,$\mu$m selection with all other optical/NIR catalogs
(pre-computed with SExtractor; Bertin \& Arnouts 1996) using a
2\arcsec\ search radius.  The MIPS, Radio and X-ray catalogs required
a different approach.  For the MIPS and radio catalogs we used a
2.5\arcsec\ and 3\arcsec\ matching radius, respectively.  For the
X-ray catalog we used a 1 or 2\arcsec radius depending on whether the
X-ray sources were pre-identified in any other band (Laird et
al. 2009).  When two or more optical/NIR counterparts separated by
$>$1\arcsec (approximately half the FWHM in IRAC-3.6) are identified
within the search radius, we apply a de-blending procedure to
incorporate the multiple sources in the catalog (e.g., irac070100
would become irac070100\_1 and irac070100\_2).  Roughly 10\% of the
IRAC sources present 2 or more counterparts in the ground-based
images.

Once the sources are identified, the photometry was computed
separately in all bands, to properly account for the significant
differences in spatial resolution.
The fluxes were then combined to derive the merged SED.
The procedure is carried out using our custom software {\it Rainbow}
based on the photometric apertures obtained from a previous
SExtractor run.

For the optical and NIR bands, total fluxes were estimated using Kron
(1980) elliptical apertures. The properties of the aperture are the
same in all bands (although different between objects) and are defined
from a {\it reference} image, which is chosen by sorting the bands
according to depth and picking the first band with a counterpart
positive detection. Thus, this image is usually among the deepest, and
presents a spatial resolution representative of the entire dataset
(typically SUBARU-R or CFHTL-$i'$). Nevertheless, as a precaution, we
established a minimum aperture size equal to the coarsest seeing in
all bands (1.5\arcsec).  Although the choice of reference band depends
on the cross-identification, the flux is measured in all bands
independently of the counterpart detection.  If a source is detected
by IRAC only (i.e., there is no optical/NIR reference image), we use a
fixed circular aperture of minimum size. If the source is detected in
just a few optical/NIR bands (e.g., it is detected in $r$ but not in
$z$) we still use the reference aperture in the un-detected bands.  In
this way we recover fluxes for very faint sources not detected by
SExtractor.  If the forced measurements do not return a positive flux,
the background flux from the sky {\it rms} within the aperture
instead.  These non-detections were not used for the subsequent SED
fitting procedure.

The IRAC photometry was computed using circular apertures of
2\arcsec\ radius and applying an aperture corrections estimated from
empirical PSF growth curves. The measurement is carried our
simultaneously in the four IRAC channels, using the 3.6+4.5\,$\mu$m
positions as priors for the 5.8 and 8.0\,$\mu$m bands, which are much
less sensitive. In the case of blended IRAC sources (i.e., those with
multiple optical/NIR counterparts), we recomputed the photometry
applying a deconvolution method similar to that used in
\citet{2006A&A...449..951G} or \citet{2008ApJ...682..985W}, which
essentially relies in using smaller 0.9\arcsec\ radius apertures with
larger aperture corrections.  Paper I describes the accuracy of the
deblending technique. For the GALEX (FUV, NUV) bands we drawn the
photometry from the source catalog of the public data release GR3.
This is computed with aperture photometry based on SExtractor
\citep{2007ApJS..173..682M}. For the IRAC sources missed in this
catalog (only $\sim$8\% and 25\% of the IRAC catalog is detected in
the FUV and NUV bands, respectively; see Table 4 of paper I) we used
the forced measurement method described above. The photometry in the
MIPS (24\,$\mu$m, 70~$\mu$m) bands was carried out using PSF fitting
with IRAF-DAOPHOT and aperture corrections (see PG05 and PG08 for more
details).


The photometric uncertainties were computed simultaneously with the
flux measurement.  Although the {\it Rainbow} measurements are
SExtractor-based, the SExtractor photometric errors were not used,
because these are often underestimated due to correlated signal in
adjacent pixels (\citealt{2003AJ....125.1107L},
\citealt{2006ApJS..162....1G}).  Instead, we used three different
approach that range from a SExtractor-like method to a procedure
similar to that described in \citet[][i.e., measuring the sky {\it
    rms} in empty photometric apertures at multiple
  positions]{2003AJ....125.1107L}.  The photometric uncertainty was
set to the largest value thereby derived.

The final multi-wavelength catalog contains 76,493 and 112,428 sources
with [3.6]$<$23.75\,mag and [3.6]$<$24.75\,mag, respectively (these
magnitude cuts correspond to the 85\% and 75\% completeness levels of
the IRAC mosaics).  Approximately 68$\%$ of the sources are located in
the main region (52,453; [3.6]$<23.75$).  Spectroscopic redshifts have
been assigned to 10$\%$ of the sample (only 120 are at z$>$1.5).  A
total of 2913 stars have been identified based on several optical/NIR
color criteria (see Section 5.4 of paper I). A source was identified
as a star only if 3 or more criteria were satisfied. The stellarity
value (as the total number of criteria satisfied) is given in
Table~\ref{dataredshift} (See \S~\ref{datacatalogs}). The fractions of
IRAC sources detected at 24~$\mu$m and 70~$\mu$m are 20$\%$ and 2$\%$,
respectively.  Finally, a total of 990 and 590 sources are detected in
the X-ray and radio catalogs of \citet{2009ApJS..180..102L} and
\citet{2007ApJ...660L..77I}, respectively.

In the following, we analyze the SEDs and physical properties of the
IRAC sources with [3.6]$<$23.75\,mag (typically SNR$\gtrsim$10).
Nevertheless, the catalog contains sources up to [3.6]$<$24.75
(3$\sigma$ limiting magnitude).  The complete catalog is available in
the electronic edition of the journal or through our web interface
{\it Rainbow navigator} (see Paper I for more details).

\subsection{Galactic Extinction}

The EGS field lies at high galactic latitude benefiting from low
extinction and low Galactic/zodiacal infrared emission.  We derive an
average E(B-V)=0.004 based on the maps of \citet{1998ApJ...500..525S}
based on several positions evenly spaced along the strip, centered at
$\alpha=$241.80$^{\circ}$, $\delta=$52.80$^{\circ}$.  In our analysis,
a differential galactic extinction for each band is computed assuming
a \cite{1989ApJ...345..245C} curve with R=3.1. These corrections,
summarized in Table 1, are not included in the photometric catalog
(presented in paper I) but these are applied before applying the SED
fitting procedure.

\begin{figure}
\centering
\includegraphics[width=9cm,angle=-90.]{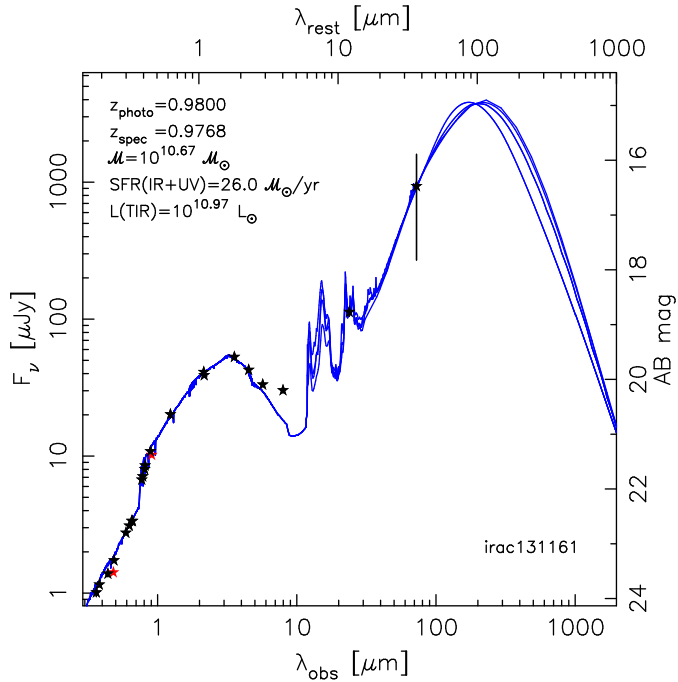}
\caption{\label{example_sed} Example of the full Spectral Energy
  Distribution (SED) of a galaxy in our sample (black dots), and fit
  (blue lines) of the observed UV-to-MIR photometry to a set of
  empirical templates computed from PEGASE 2.0 models
  (\citealt{1997A&A...326..950F}) assuming a Salpeter IMF
  (M~$\in$~[0.1-100]M$_{o}$), and \citet{2000ApJ...533..682C}
  extinction law (see \S~\ref{templates}), and the FIR photometry
  (MIPS 24 and 70~$\mu$m) to dust emission models of
  \cite{2001ApJ...556..562C}, \cite{2002ApJ...576..159D} and
  \cite{2009ApJ...692..556R} (see Section~\ref{irfitting}). The
  multiple lines in the FIR region correspond to best-fitting template
  from each of the dust emission models, and the average value of the
  three. In the upper left corner, we indicate the photometric
  redshift, and the stellar mass, IR-based SFR and total IR luminosity
  estimated from fitting procedure.}
\end{figure}

\section{SED analysis: Photometric redshifts}
\label{analyzesed}

\subsection{{\it Rainbow} code}
\label{fitting}

We computed photometric redshifts for all IRAC sources from the
multi-color catalog presented in Paper I using our own dedicated
template fitting code ({\it Rainbow} software hereafter; see PG05 and
PG08).  The program creates a grid of redshifted galaxy templates in
steps of $\delta$z$=$0.01 and then applies a $\chi^{2}$ minimization
algorithm to find the template best-fitting the multi-band
photometry. Upper limit detections and fluxes with uncertainties
larger than 0.5\,mag are not included in the fit. The $\chi^{2}$
definition takes into account the flux uncertainties of each band,
being defined as:

\begin{equation}
  \chi^{2}=\sum_{i=0}^{N(\mathrm{band})}\Big[\frac{F_{\mathrm{obs},i}-A\cdot F_{\mathrm{temp},i}}{\sigma_{i}}\Big]^{2}
\end{equation}

\noindent where $F_{obs,i}$ is the observed flux in the {\it i} filter
and $\sigma_{i}$ is its uncertainty, $F_{temp,i}$ is the flux of the
redshifted template in the {\it i} filter (obtained by convolving the
template with the filter transmission curve).  A scaling factor is
applied to the input template to fit the observed photometry. This
normalization parameter A is used to compute quantities such as the
stellar masses, absolute magnitudes or SFRs (see Sections~\ref{masses}
and \ref{sfrs}).

Prior to the $\chi^{2}$ minimization procedure, the {\it Rainbow} code
gets rid of deviant and redundant photometric data points. The fluxes
presenting a very steep gradient with respect to the surrounding bands
are flagged and removed before attempting the final fit.

By analyzing the $\chi^{2}$(z) distribution of the best fit in the
model grid, we built the redshift probability distribution function
(zPDF), from which we computed the most probable redshift and
1$\sigma$ errors, $z_{\mathrm{best}}$ and $\sigma_z$. The single value
that minimizes $\chi^{2}$(z) is $z_{\mathrm{peak}}$. We found that
$z_{\mathrm{best}}$ provided the most accurate results presenting less
outliers and a smaller scatter when compared with spectroscopic
redshifts. The uncertainties in the photometric redshifts are used to
compute the uncertainties in the stellar parameters derived from the
best-fitting template.

The {\it Rainbow} code also analyzes the dust emission on sources with
at least one flux measurement beyond rest-frame 8\,$\mu$m, i.e., the
MIPS 24 and 70~$\mu$m bands (see
Section~\ref{sfrs}). Figure~\ref{example_sed} shows the combined
optical and IR SED along with the estimated physical parameters for a
galaxy at z$\sim$1 as an example of the optical and IR fitting
techniques described here and in \S\ref{sfrs}. The best fit optical
template to the data was used to estimate the photometric redshift,
stellar mass (see Sections~\ref{analyzesed} and \ref{evalmass}) and
also the rest-frame UV flux. Moreover, IR luminosities and star
formation rates were obtained from the best fit IR template to the
data at rest-frame $\lambda>$5$\mu$m (see Section~\ref{irfitting}).

\subsection{Stellar population templates}
\label{templates}

\begin{figure*}
\centering
\includegraphics[width=8.2cm,angle=0.]{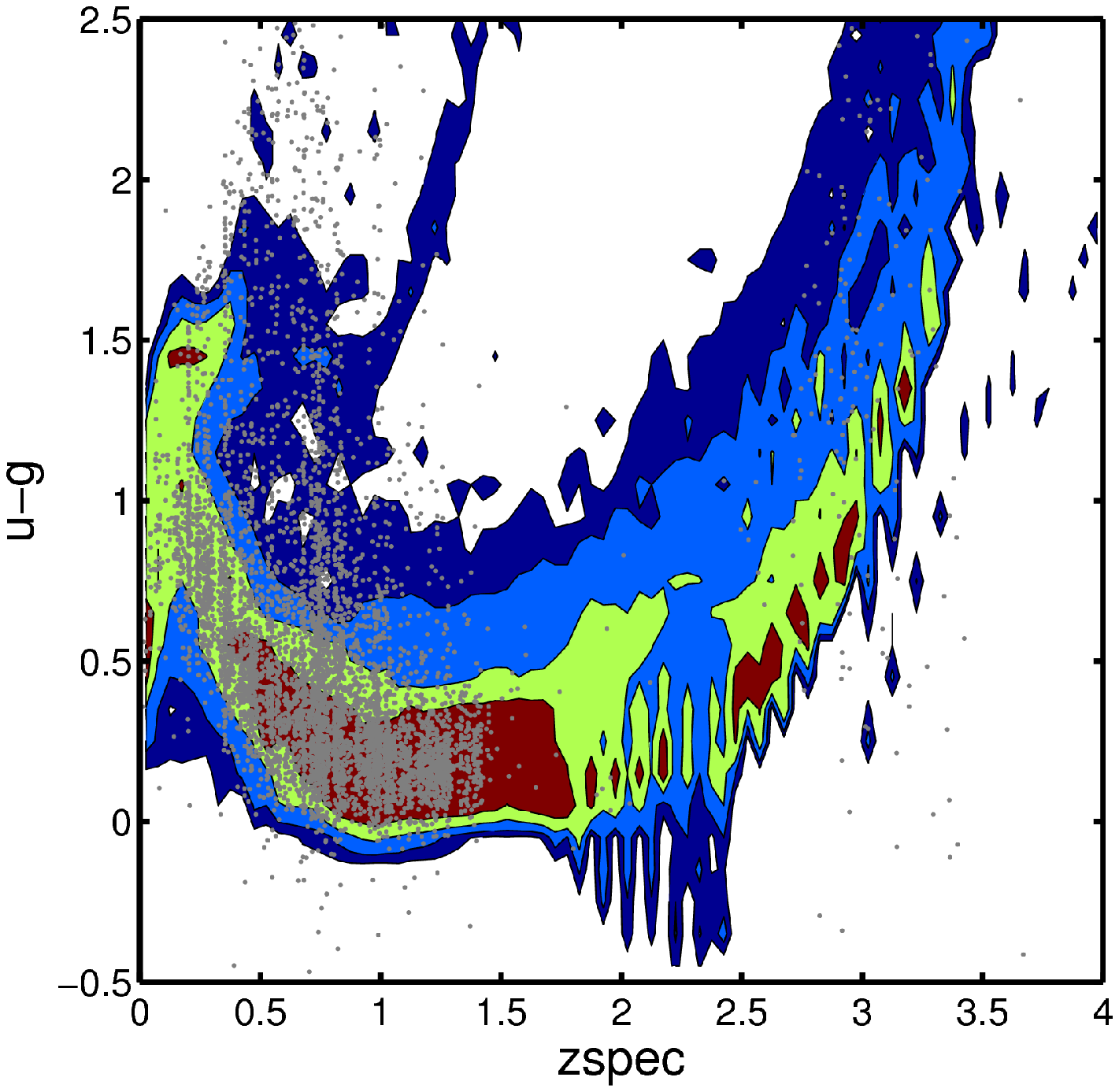}
\hspace{0.5cm}
\includegraphics[width=8.2cm,angle=0.]{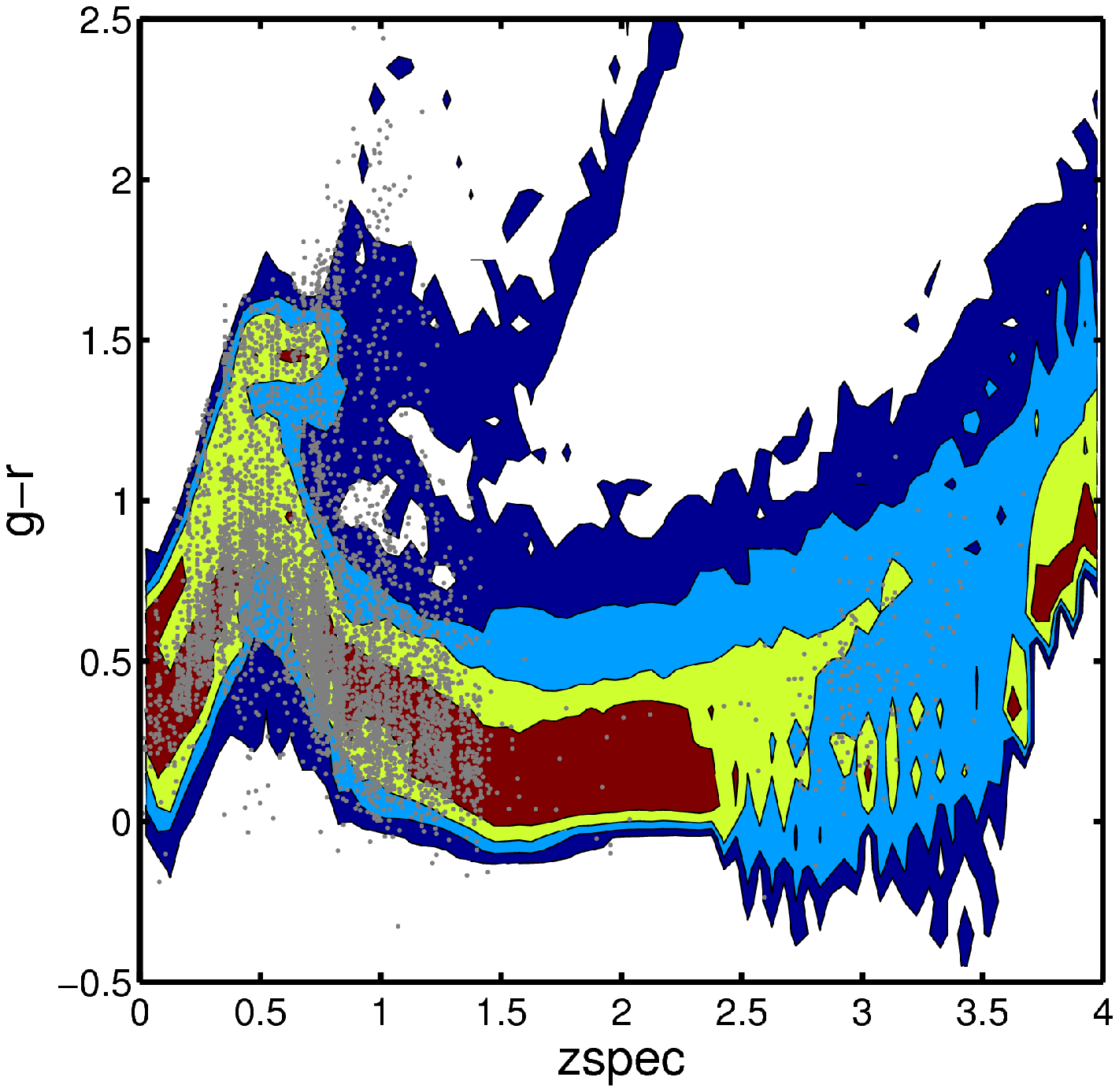}\\
\includegraphics[width=8.2cm,angle=0.]{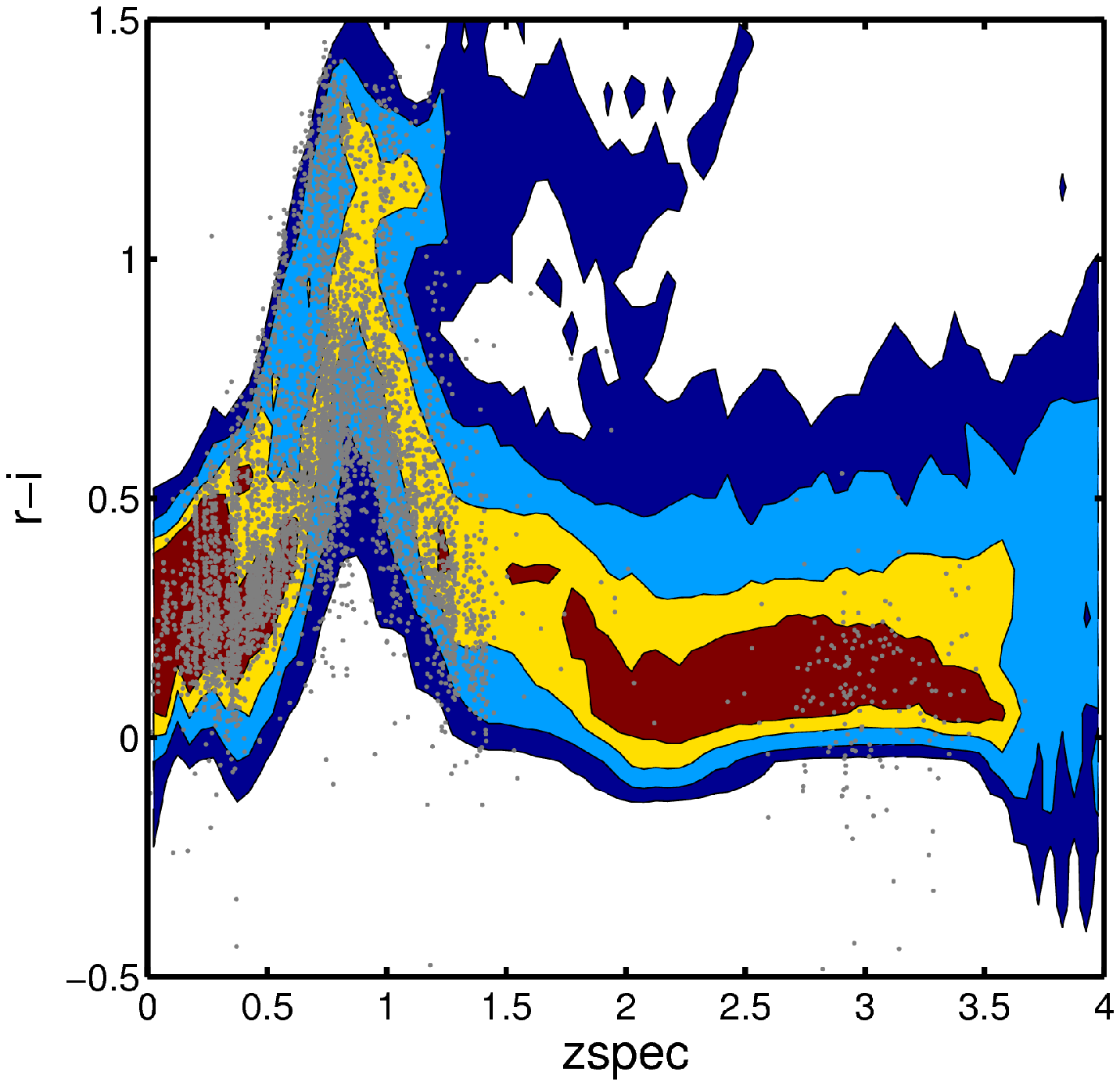}
\hspace{0.5cm}
\includegraphics[width=8.2cm,angle=0.]{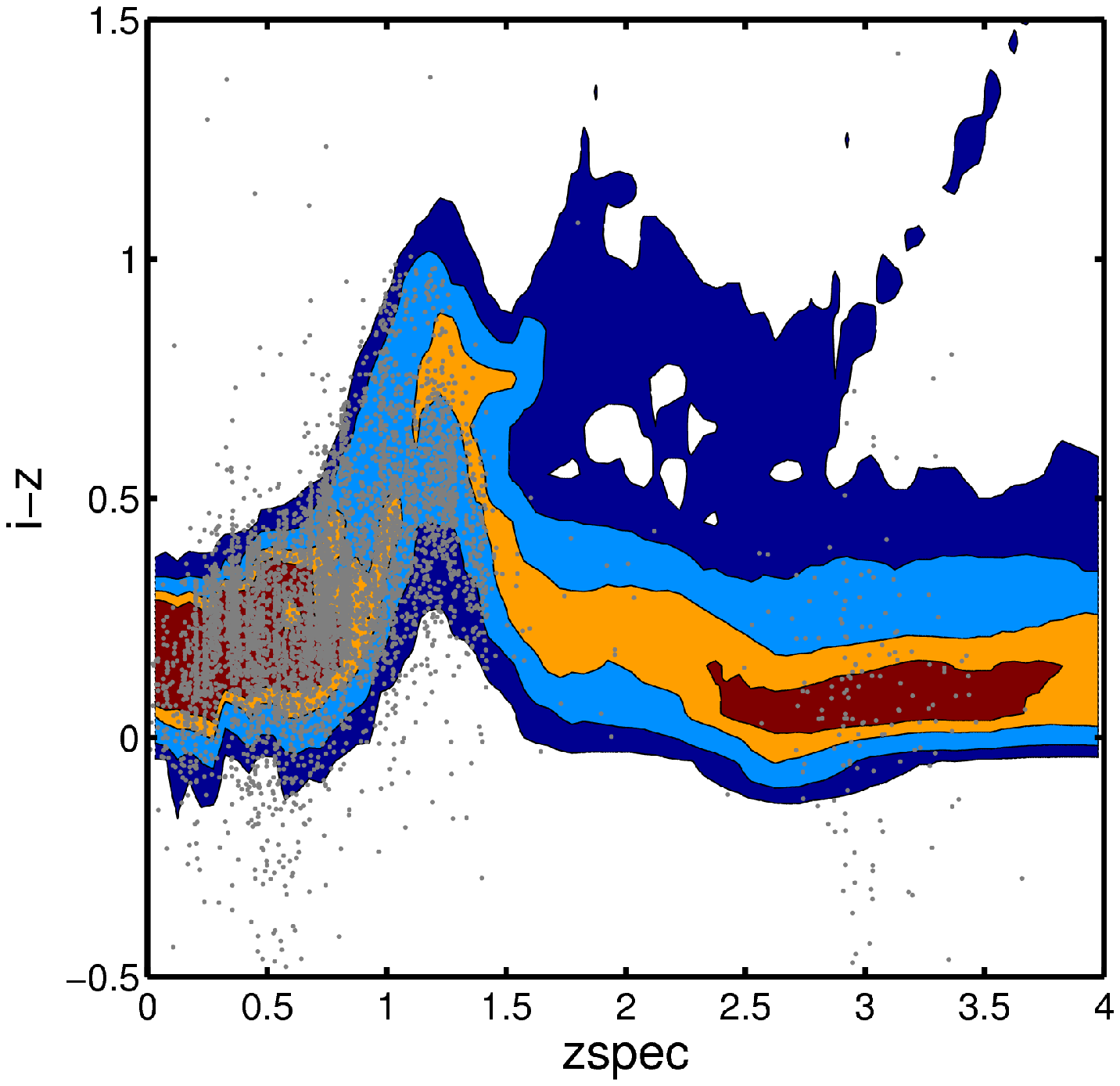}\\
\includegraphics[width=8.2cm,angle=0.]{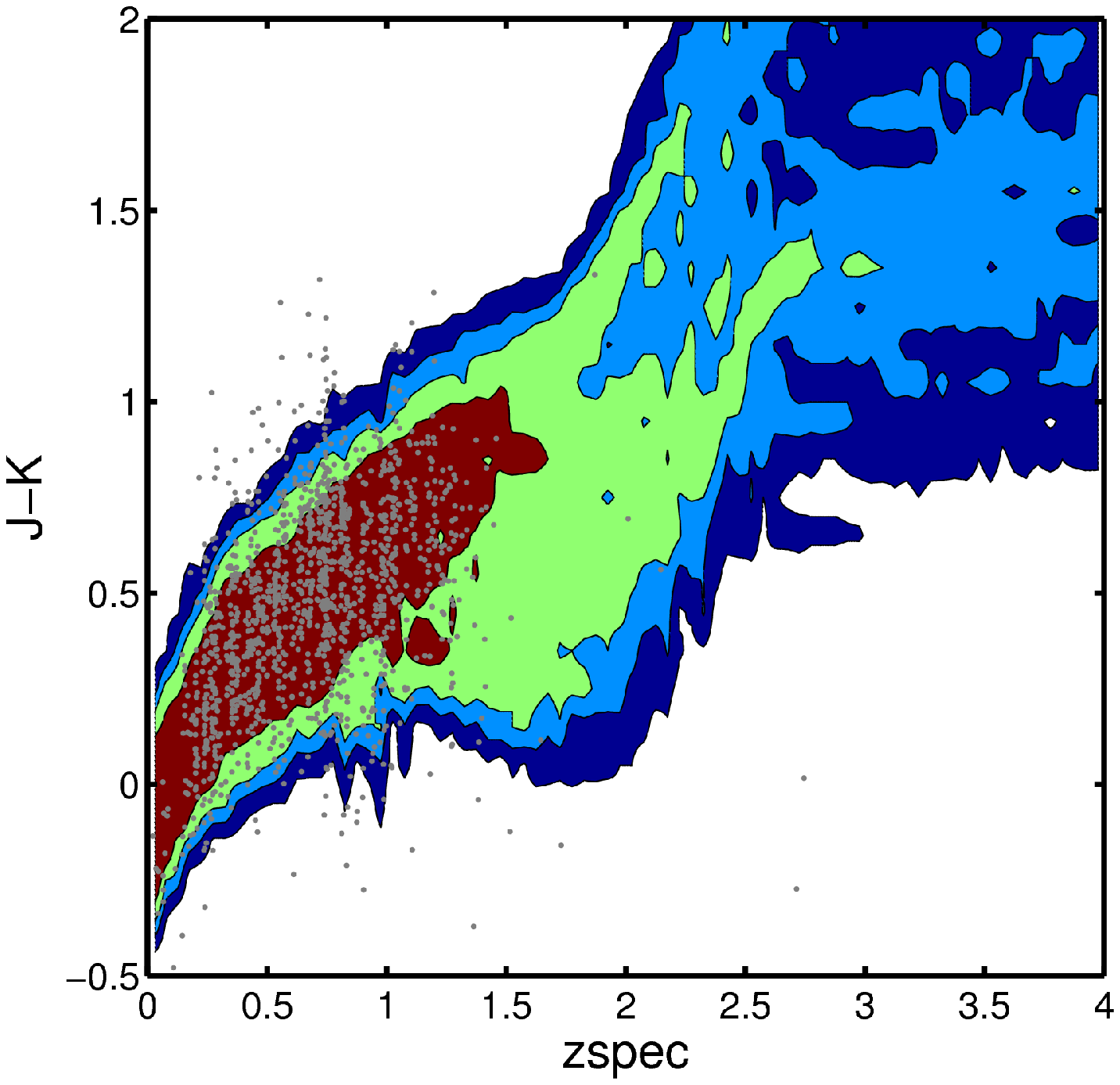}
\hspace{0.5cm}
\includegraphics[width=8.2cm,angle=0.]{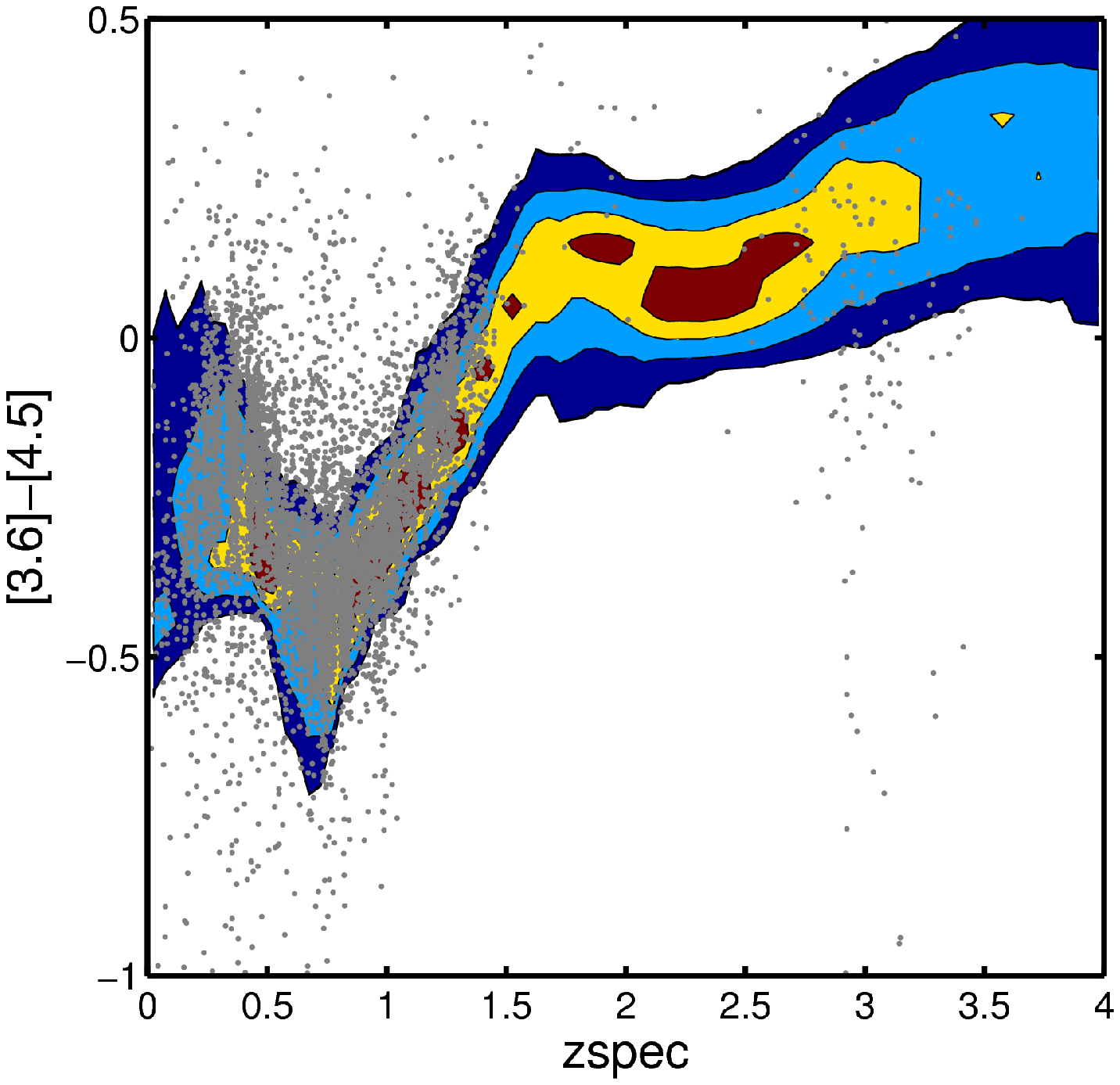}\\
\caption{\label{templates} Comparison of different optical and NIR
  observed colors as a function of the spectroscopic redshift (grey
  dots) versus the predicted colors for our 1876 galaxy templates
  (density map). Each density contour contains (from inside out) 25\%,
  50\%, 75\% and 90\% of the values.}
\end{figure*}

The stellar templates used by the minimization code are extracted from
a library of synthetic templates built by fitting stellar population
synthesis and dust emission models to a representative sample of
galaxies at different redshifts. This reference sample is drawn from
the GOODS-N and GOODS-S IRAC surveys and have highly reliable
spectroscopically confirmed redshifts (0$<$z$<$3) and at least 10
measurements of the SEDs from the UV to the MIR. A detailed
description on the modeling of these templates is given in PG08. Here
we briefly summarized their most relevant characteristics.

The stellar emission of the reference template set was characterized
using the PEGASE 2.0 models (\citealt{1997A&A...326..950F}) assuming a
Salpeter IMF (M~$\in$~[0.1-100]M$_{o}$), and
\citet{2000ApJ...533..682C} extinction law.  We also considered the
contribution from emission lines and the nebular continuum emitted by
ionized gas. The models were obtained assuming a single population
(1-POP models), characterized by an exponential star formation law.
As a result, each template is characterized by 4 parameters in the
1-POP case, namely the time scale $\tau$, age $t$, metallicity $Z$ and
dust attenuation $A$(V). The MIR/FIR region of some templates includes
a contribution from a hot dust component that was computed from dust
emission models using a similar procedure to that described in
\S~\ref{irfitting} of this paper.

Defining a representative spectral library is a critical issue for
photometric redshift codes, specially when NIR selected samples are
studied \citep{2008ApJ...677..219K}. The reference sample should span
a wide range of redshifts and galaxy colors that probe the parameter
space of the magnitude limited sample in sufficient detail.  This is
why we included in the template set a few z$>$1.5 galaxies which could
not be fitted accurately with low-z templates. Furthermore, we
complemented our synthetic templates with QSO and AGN empirical
templates drawn from \cite{2007ApJ...663...81P} that account for the
galaxy population whose UV-to-NIR emission is not dominated by stars
but by an AGN.

The template library contains a total of 1876 semi-empirical templates
(see PG08 for more details and examples of the SEDs) spanning a wide
range of colors and physical parameters. Figure~\ref{templates} shows
that the loci of the observed and template colors present an overall
good agreement for the majority of the spectroscopic galaxies in a
wide range of optical and NIR colors. The combination of colors based
on the CFHTLS filters (panels 1 to 4) are consistent with Figure 2 of
\cite{2006A&A...457..841I}, that presents the same colors for a
sub-sample of i-band selected galaxies in the CFHTLS-D1 field.
  
On the other hand, we find small discrepancies between templates and
observations in the [3.6]-[4.5] IRAC color at low redshift
(Figure~\ref{templates}, panel 6). This is not surprising considering
that these bands are probing the rest-frame NIR (see
e.g.,\citealt{2004ApJS..154...44H}; \citealt{2006ApJ...651..791B} for
similar examples), a wavelength region where the predictions from
stellar population synthesis models tend to be more uncertain
\citep{2005MNRAS.362..799M}. Furthermore, these differences tend to
increase at $\lambda\gtrsim$3\,$\mu$m rest-frame, where galaxies can
exhibit a significant contribution from hot dust or PAH emission
features that are not contemplated in the optical templates and
therefore require more complicated modeling procedures
(\citealt{2008ApJ...681..258M}, \citealt{2009ApJ...706.1020M}).

\subsection{Zero-point corrections and Template error function}
\label{zeropoint}

\begin{figure*}
\centering
\includegraphics[width=8.5cm,angle=0.]{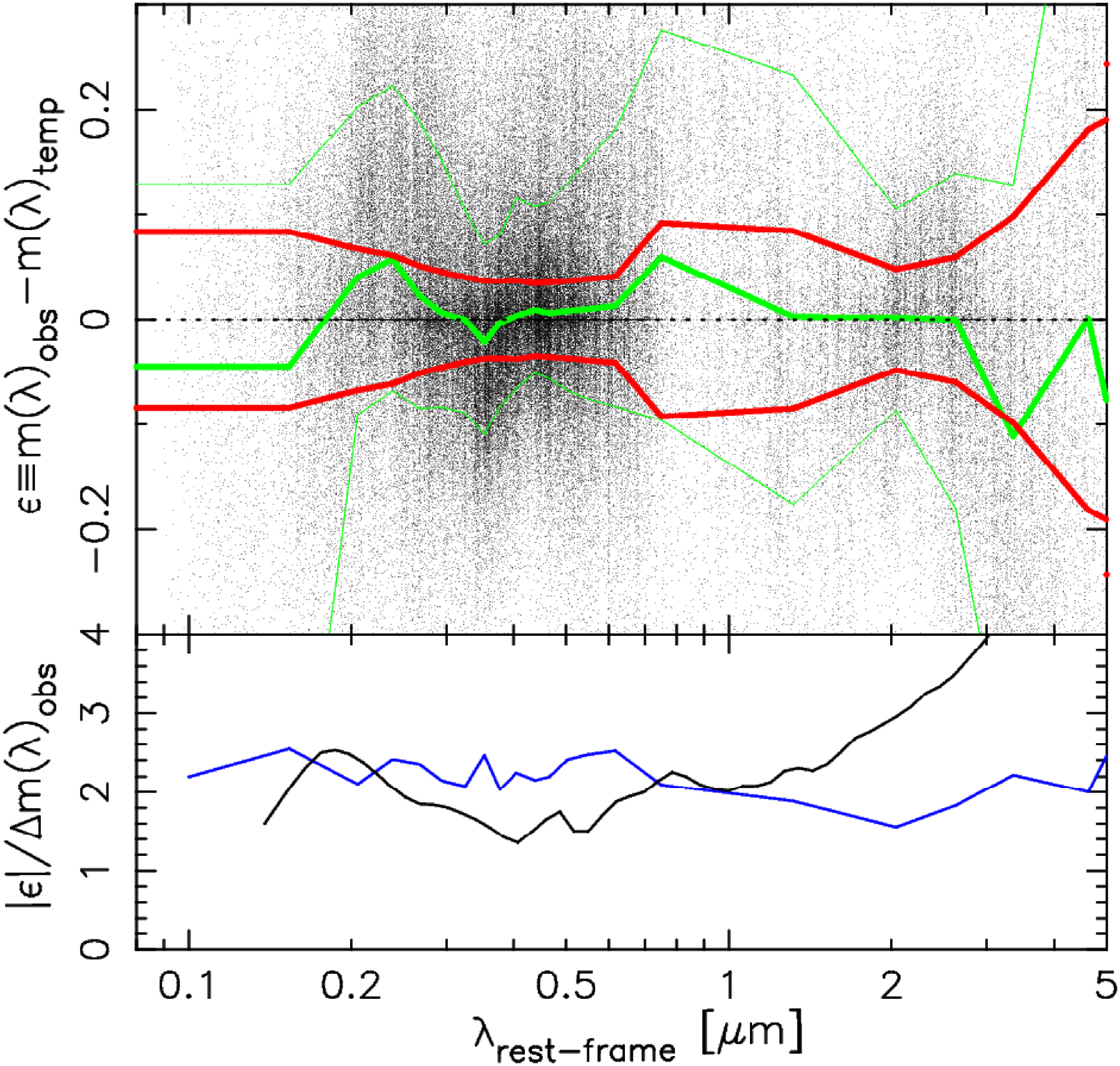}
\hspace{0.5cm}
\includegraphics[width=8.5cm,angle=0.]{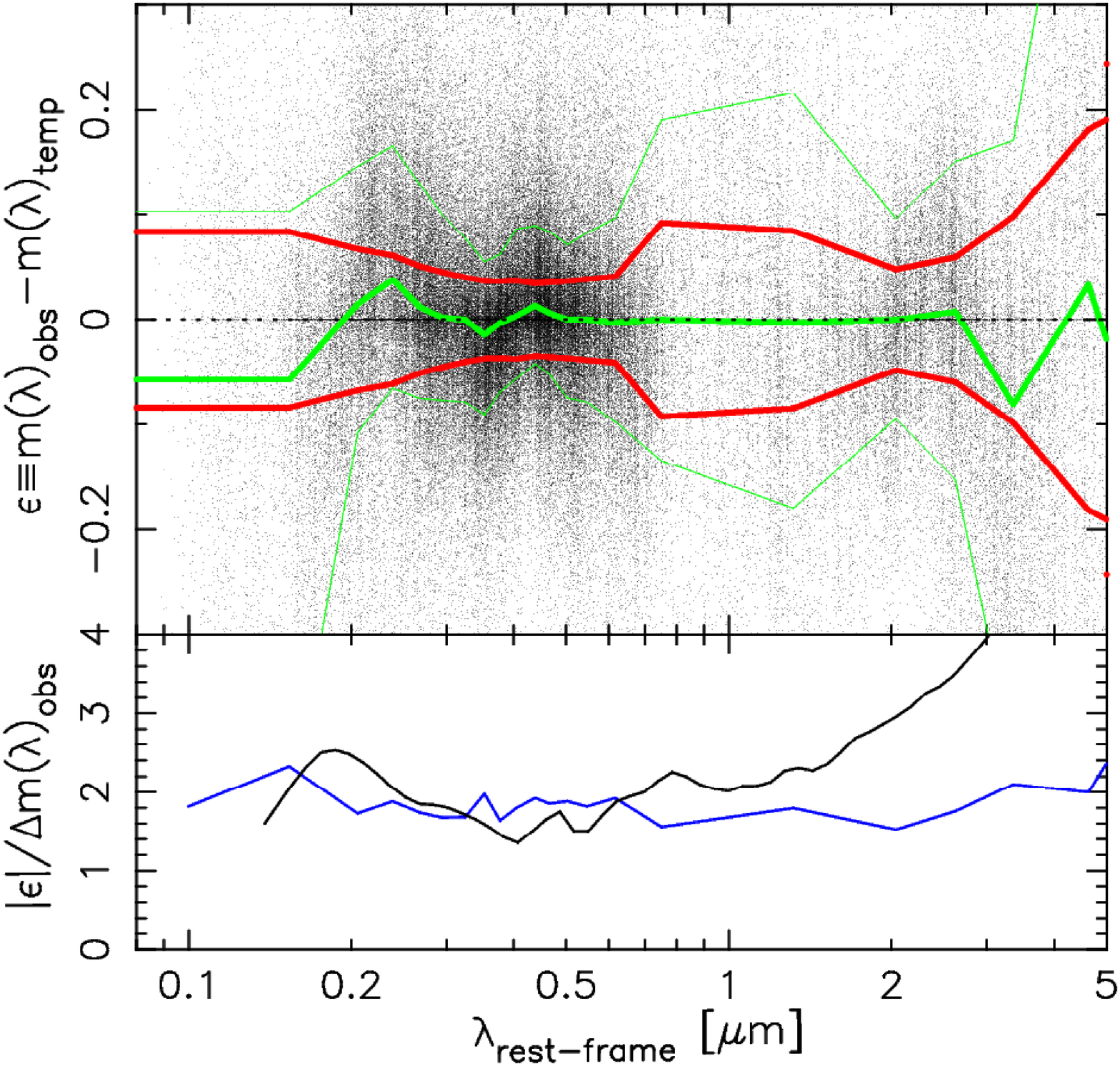}
\caption{\label{template_error} {\it Top}: Residuals of the comparison
  between observed and synthetic magnitudes for a sub-sample of
  galaxies with spectroscopic redshift and SNR$>$5 photometry in the
  main region. The residuals are shifted into rest-frame wavelengths
  based on the effective wavelength of the filters and the
  redshift. The Figure on the {\it left} shows the raw residuals
  before applying the zero-point corrections nor the template error
  function to the fitting procedure. The Figure on the {\it right}
  shows the final result of the iterative process to compute the
  zero-point corrections and the template error function.  The thick
  green line depicts the median value of the residuals per redshift
  bin. The upper and lower red lines indicate the median value of the
  photometric error at each redshift. The upper and lower thin green
  lines em-compasses 68\% (1$\sigma$) of the residual distribution
  around the median value. {\it Bottom}: The blue line depicts the
  median absolute value of the residuals in the top panel divided by
  the photometric error and by 0.67 to scale the median (50\%) to a
  1$\sigma$(68\%) confidence interval. The black line shows the
  template error function of \cite{2008ApJ...686.1503B} divided by the
  median photometric error (adapted from Figure~3 of their paper).}
\end{figure*}

An improvement introduced in the current work on EGS over the previous
{\it Rainbow} photometric redshifts in GOODS-N, GOODS-S and Lockman
Hole (PG05, PG08) is the fine-tuning of the photometric zero-points
and the use of a template error function.  Both procedures are based
on the comparison of the observed fluxes to synthetic photometry
derived from the convolution of the filter transmission curves with
the best fitting templates for the galaxies with reliable
spectroscopic redshifts. As demonstrated by
\cite{2006ApJ...651..791B}, \cite{2006A&A...457..841I} and
\cite{2009ApJ...690.1236I}, the comparison between the observed
apparent magnitudes and synthetic fluxes often shows small offsets
that can lead to systematic errors in the calculation of the
photometric redshifts. These offsets can be the result of small
systematic errors in the absolute calibration, uncertainties in the
filter transmission curves, or they can be the result of intrinsic
limitations of the templates in reproducing the observed SEDs
\citep{2008ApJ...686.1503B}.

To tackle these issues and improve the photometric redshift
estimation, we fit the SEDs of the galaxies with secure spectroscopic
redshift to our template set fixing the photometric redshift to the
spectroscopic value.  Then, we compute the difference between the
observed fluxes and the template fluxes for each band, and we consider
this residual value as a function of the rest-frame wavelength. The
left panel on Figure~\ref{template_error} shows the result of applying
this process to the sub-sample of galaxies with spectroscopic redshift
and photometric fluxes with SNR$>$5 in the main region.  The median of
the residuals (thick green line) shows an overall good agreement
between templates and observations, with an {\it rms} (thin green
lines) of $\sim$2 times the median value of the photometric
uncertainty across all the wavelength range (red lines).  However,
significant deviations appear in the rest-frame wavelengths around
200~nm, the 500-1000~nm region and the mid-IR ($\lambda>3~\mu$m).

To diminish the effect of these discrepancies, we considered two 
corrections: 1) we applied small calibration offsets in each band
based on the residuals of the comparison with synthetic magnitudes
(note that these corrections refer to observed wavelengths); and 2) we
used a template error function such as that introduced in
\citet{2008ApJ...686.1503B}. 

\begin{figure*}
\includegraphics[width=9cm,angle=0.]{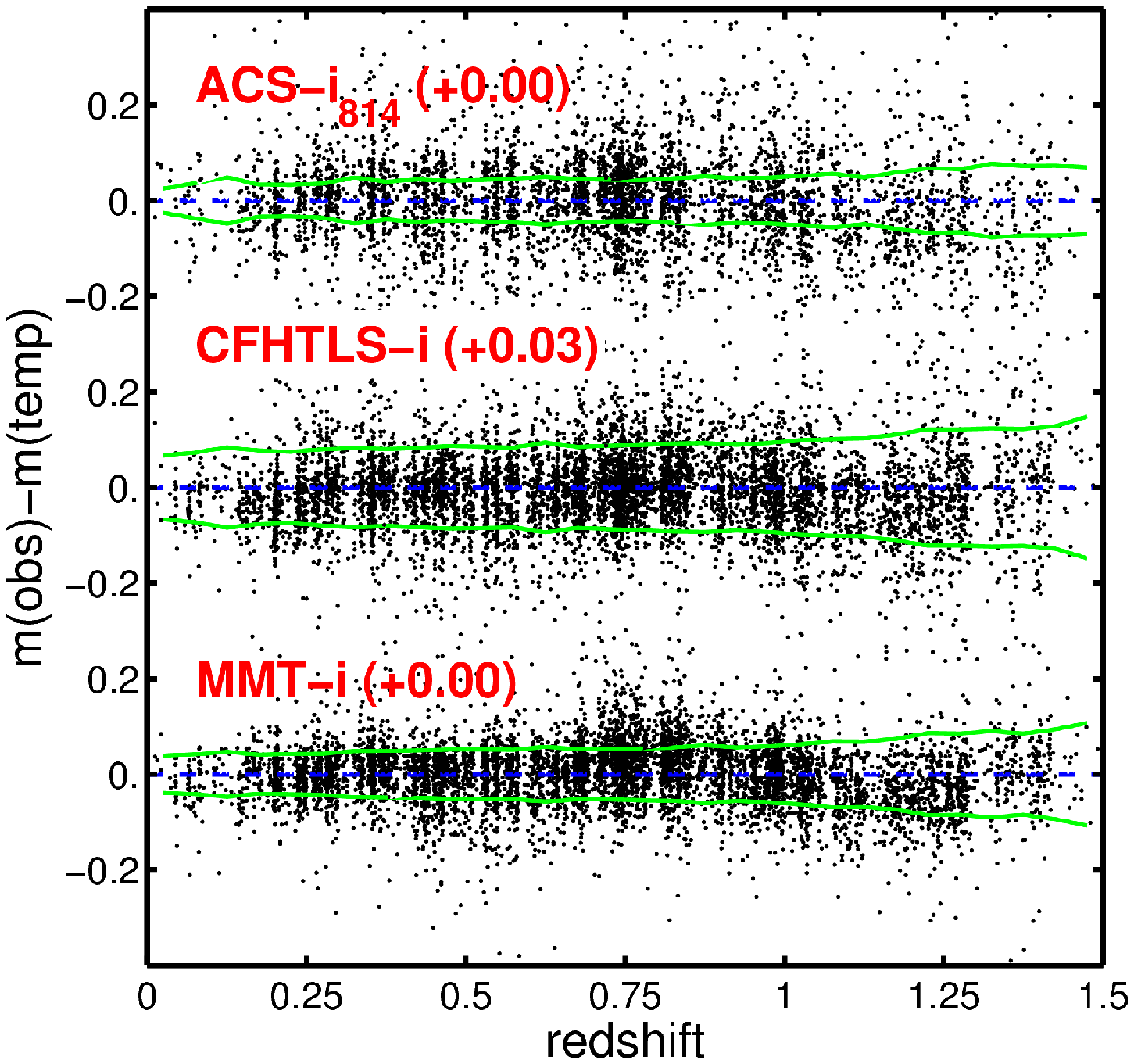}
\includegraphics[width=9cm,angle=0.]{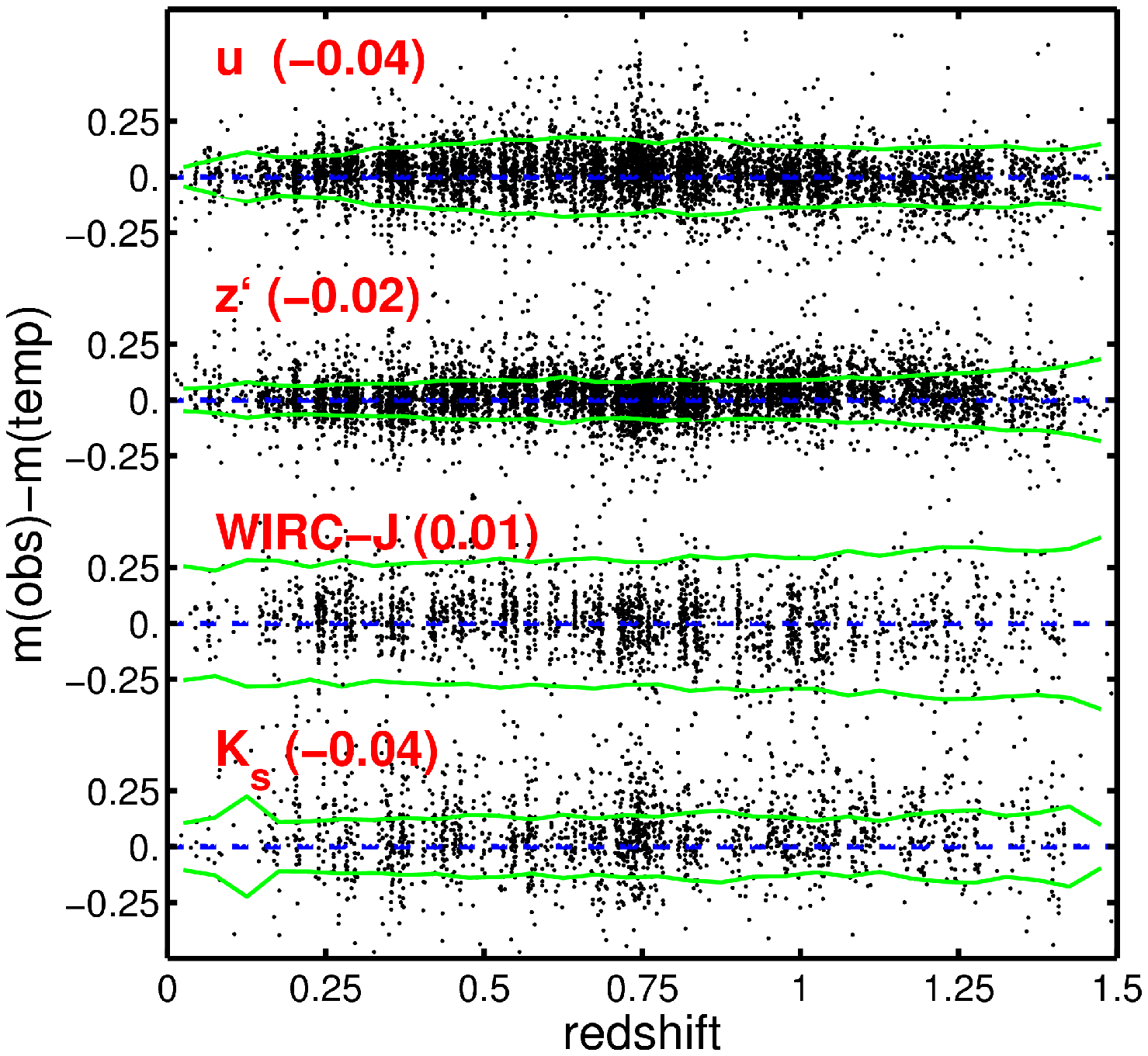}
\caption{\label{zeropt} {\it Left} - Differences between observed and
  synthetic magnitudes as a function of redshift in the ACS-$i_{814}$,
  CFHTLS-$i'$ and MMT-$i$ bands. The values quoted in the parenthesis
  indicate the zero-point correction applied to these bands. The
  dashed blue line depicts the median difference between observed and
  synthetic photometry after the zero-point correction has been
  applied. The green lines show the median photometric uncertainty in
  each band as a function of redshift multiplied by a factor 2. At
  z$\sim$1 the residuals are dominated by a systematic offset in the
  templates instead of by deviations in the photometric
  calibration. . {\it Right} - Same as the left panel for the
  MMT-$u^{*}$,MMT-$z'$,WIRC-$J$ and MOIRCS-$K_{s}$ bands.}
\end{figure*}

Figure~\ref{zeropt} shows the comparison between observed and
synthetic magnitudes for three different i-bands (ACS, CFHTLS and MMT;
left panel) and the $u*$,$z'$,$J$ and $K$ bands (from MMT, WIRC and
MOIRCS; right panel) as a function of redshift. The values in the
parenthesis quote the median correction applied to each band to
minimize the differences with respect to the synthetic fluxes. Note
that the three i-bands present a similar trend at z$\gtrsim$1, where
the observations are slightly brighter than the predictions from the
templates. This suggest that the feature is related to the templates
and not to the absolute calibration of the bands. At z$\gtrsim$1 the
i-band ($\lambda_{\mathrm{eff}}\sim$800~nm) probes rest-frame
wavelengths around $\sim$300-400~nm, where the overall quality of the
fit to templates is reduced.

The overall shape of the residual distribution, shown in the
left-panel of Figure~\ref{template_error}, is very effective for
identifying systematic deviations in the templates. This is because
small zero-point errors in any of the individual bands are smoothed
over the rest-frame wavelength range due to the mixed contribution
from multiple bands. Therefore, based on the overall scatter in the
residual with respect to the median photometric errors, we can compute
a template error function that parametrizes the overall uncertainties
in the templates as a function of wavelength. As demonstrated by
\cite{2008ApJ...686.1503B}, this function can be efficiently used as a
weight term in the $\chi^{2}$ function of the SED fitting procedure to
minimize the impact of the template uncertainties in some wavelength
ranges. The bottom of the right panel of Figure~\ref{template_error}
we show the median value of the absolute difference between observed
and template fluxes divided by the photometric error and multiplied by
0.67 to scale the median (50\%) to a 1~$\sigma$(68\%) confidence
interval, as done in \citealt{2008ApJ...686.1503B}. Compared to the
results of this work our combination of templates and filters present
a slightly better agreement in the rest-frame UV and NIR (between
1-2~$\mu$m), probably as a result of our larger template set, which
present more diversity in their spectral shapes.

In principle, the zero-point corrections and the effects of the
template error function produce similar effects. Moreover, the
re-calibration of adjacent (sometimes very similar) bands tend to
modify the residual of both fits. Therefore, in order to obtain
consistent results, both the template error function and the
zero-point corrections are computed iteratively repeating the fitting
process until we obtain variations smaller than 1-2\% (typically after
a couple of iterations). The zero-point offsets are summarized in
Table~\ref{zeropoints} . Virtually all of the corrections are smaller
than 0.1~mag, and some of them are exactly zero. The final results of
the procedure are shown in the right panel of
Figure~\ref{template_error}. The application of the zero-point offsets
results in the flattening of the median difference between observed
and template magnitudes for the whole wavelength range in our SEDs
except in two regions, one around 200~nm and the other at
$\sim$3~$\mu$m.

The poor agreement at 3~$\mu$m is most probably associated with
limitations in the NIR-MIR range of stellar population templates and
the contribution from PAH emission, which is also not properly taken
into account in the SPSs models nor the dust emission templates.  In
addition, there is a small peak/bump at $\sim$350-450~nm, which is
very similar to the feature reported by \cite{2007MNRAS.381..543W} and
\cite{2008A&A...491..713W}. In these papers, they explain this effect
with an excess in the strength of the Balmer break in the models by
\citet{2003MNRAS.344.1000B}, relative to the observed values. The peak
at 200~nm can be partly related with the use of a
\citet{2000ApJ...533..682C} extinction curve in the modeling of the
the galaxy templates. This parametrization lacks the silicate
absorption at 2175~\AA~, which appears in other extinction curves,
such as that of \citet{1989ApJ...345..245C}, which has been claimed to
be produced by PAHs. The presence of this absorption bump has been
reported on some studies of high-z galaxies
(\citealt{2007A&A...472..455N}; \citealt{2009A&A...503..765N}).  

As an additional check of the accuracy of the method we compare our
photometry against the fluxes of a control sample of $\sim$300 bright
unsaturated stars in common with the SDSS. In particular, we restrict
the comparison to relatively blue sources ($u'$-g$=$1.2, in the MMT
bands) in order to avoid large color corrections in the filter
transformations.  These color terms were computed by convolving the
filter transmissions with the spectra of F, G and K class stars
\citep{1992IAUS..149..225K}, which makes up for most of our sample of
stars. The transformation with respect to the MMT bands, which present
a filter system similar to that of SDSS, are
\begin{eqnarray}\nonumber
u_{\mathrm{MMT}}&=&u_{\mathrm{SDSS}}-0.095\cdot[u-g]_{\mathrm{SDSS}}+0.070\\ \nonumber
g_{\mathrm{MMT}}&=&g_{\mathrm{SDSS}}-0.063\cdot[g-i]_{\mathrm{SDSS}}\\ \nonumber
i_{\mathrm{MMT}}&=&i_{\mathrm{SDSS}}-0.203\cdot[i-z]_{\mathrm{SDSS}}-0.002\\ \nonumber
z_{\mathrm{MMT}}&=&z_{\mathrm{SDSS}}-0.087\cdot[i-z]_{\mathrm{SDSS}}-0.002
\end{eqnarray}
After applying these corrections, we find zero-point offsets of
$\Delta$u'=-0.05,$\Delta$g=-0.10,$\Delta$i=-0.01 and $\Delta$z=-0.09
with respect to the SDSS. The values are roughly consistent with our
previous results based on galaxy templates. Only the $u'$ and $z$
bands present slightly lower values of the correction. These could be
an additional effect of the template uncertainties (at 250 and 450~nm
rest-frame), specially for the u' band .Also, it is worth noting that
the zero-point offsets are estimated simultaneously and iteratively
for all bands whereas the comparison to SDSS is done separately for
each band.

\subsection{Photometric redshift accuracy} 
\label{evalpzeta}

In this section we analyze the overall accuracy of the photometric
redshifts ($z_{\mathrm{phot}}$) by comparing them against
spectroscopic redshifts ($z_{\mathrm{spec}}$). In particular, we study
the quality of our results as a function of the spectroscopic redshift
and the observed magnitude in optical and NIR bands, and we provide
specific results for different groups of galaxies such us X-ray, MIPS
or Radio sources.  For the 76,936 galaxies ([3.6]$<23.75$) in the
sample we identify 7,636 ($\sim$10\%) spectroscopically confirmed
sources from the DEEP2 catalog (mostly at z$<$1) and from a small
sample of LBGs (z$\sim$3) presented in \citet{2003ApJ...592..728S}.

\subsubsection{$z_{\mathrm{phot}}$ versus $z_{\mathrm{spec}}$: DEEP2 sample}

\begin{figure*}
\centering
\includegraphics[width=8.7cm,angle=0.]{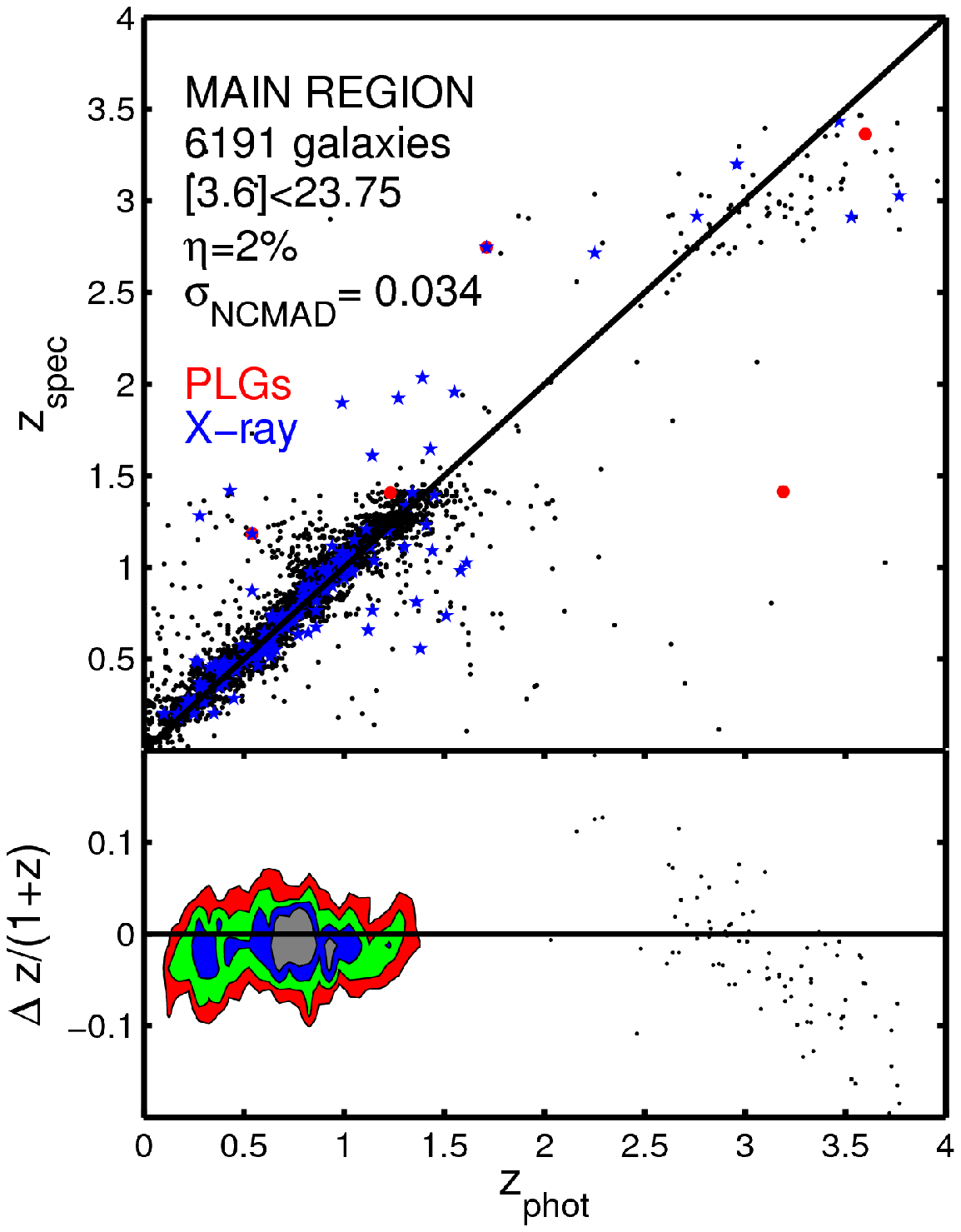}
\hspace{0.2cm}
\includegraphics[width=8.7cm,angle=0.]{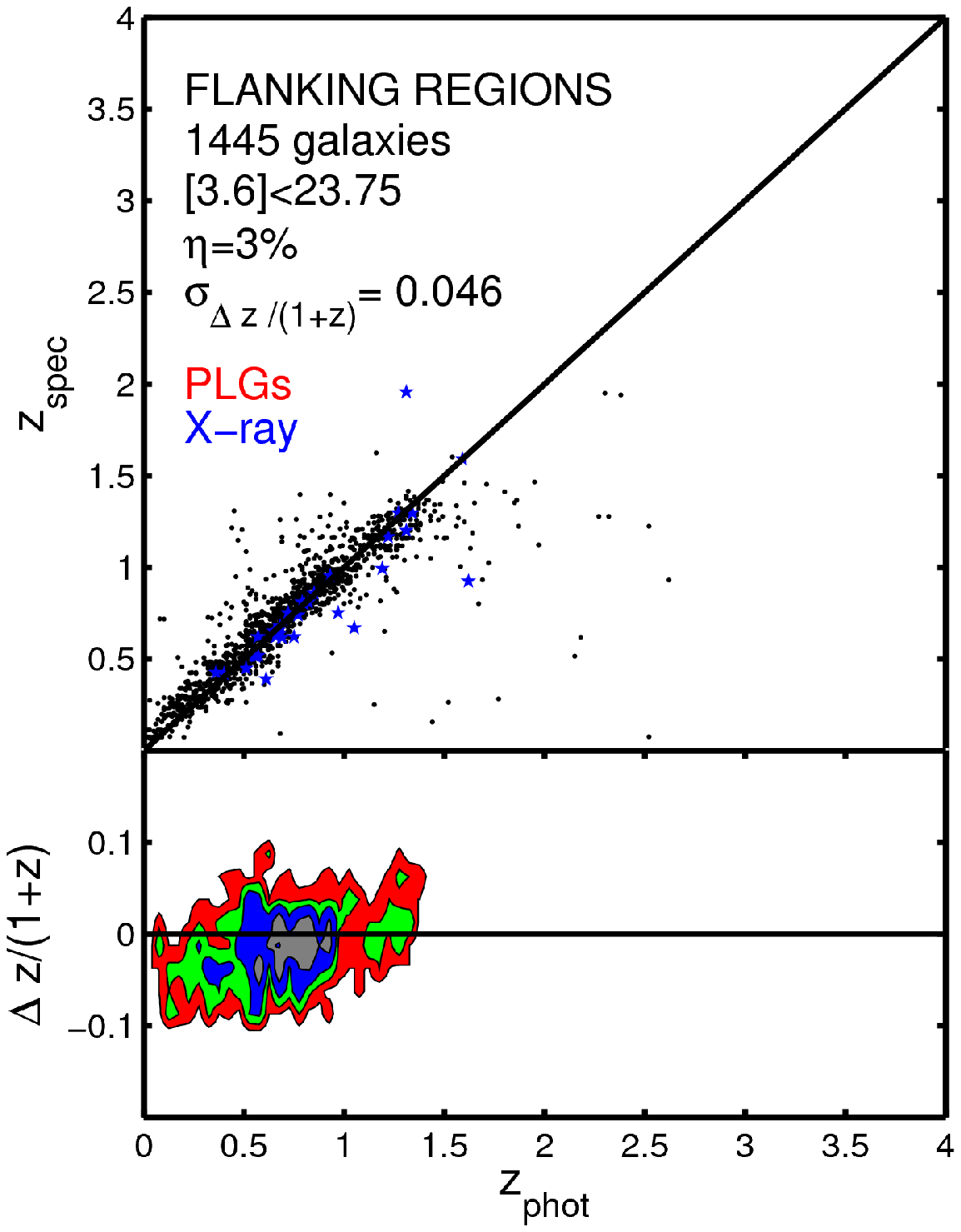}
\caption{\label{zaccuracy_c23}{\it Top} panels: Photometric redshifts
  versus spectroscopic redshifts for [3.6]$<$23.75\,mag sources in the
  main region (left) and flanking regions (right). Blue stars indicate
  sources detected in the X-ray catalog of
  \citet{2009ApJS..180..102L}. Red dots correspond to galaxies with a
  power-law spectrum in the IRAC bands. This feature is frequently
  used to identify obscured AGNs (\citealt{2004ApJS..154..155A},
  \citealt{2007ApJ...660..167D}), that usually under-perform in the
  photometric redshift procedure.{\it Bottom} plots: Density plots of
  the scatter in $\Delta z/(1+z)$ as a function of redshift for the
  main region (left) and flanking regions (right). Each contour
  contains (from the inside out) 25\%, 50\%, 75\% and 90\% of the
  spectroscopic sample, respectively.}
\end{figure*}

Figure \ref{zaccuracy_c23} shows the comparison between
$z_{\mathrm{phot}}$ and $z_{\mathrm{spec}}$ for 6,191 and 1,445
sources with reliable spectroscopic redshift in the main and flanking
regions, respectively.

Following \cite{2006A&A...453..809I}, we quantified the redshift
accuracy using the normalized median absolute deviation
($\sigma_{\mathrm{NMAD}}$) of $\Delta
z=z_{\mathrm{phot}}-z_{\mathrm{spec}}$

\begin{equation}
\sigma_{\mathrm{NMAD}}=1.48\times\mathrm{median}\left(\left|\frac{\Delta z-\textrm{median}(\Delta z)}{1+z_{\mathrm{spec}}}\right|\right)
\end{equation}

This quantity is equal to the {\it rms} for a Gaussian distribution
and it is less sensitive to the outliers than the usual $rms$ divided
by (1+z) \citep{2006A&A...453..809I}. We define $\eta$ as the fraction
of catastrophic outliers (those sources having $|\Delta
z|/(1+z)>$0.20).

Table~\ref{photozquality1} and ~\ref{photozquality2} summarize the
quality of $z_{\mathrm{phot}}$ as a function of redshift in the main
and flanking regions. The overall scatter and median systematic
deviation are $\sigma_{\mathrm{NMAD}}$=0.034 and 0.046, and $\Delta
z$/(1+z)=0.010 and 0.013 for each region, respectively. As expected,
the {\it rms} in the flanking regions, where the overall photometric
quality is slightly lower, is higher ($\sim$20\%) than in the main
region. Nonetheless, the outlier fraction is only 1\% worse.

The bottom panels of Figure~\ref{zaccuracy_c23} show the density plot
of $\Delta z/(1+z)$ as a function of redshift. The subset of LBGs at
$z_{\mathrm{spec}}>$2.5 are shown as dots. These sources are
explicitly discussed in the following section.  The scatter
distribution indicates that the accuracy of $z_{\mathrm{phot}}$ does
not depend strongly on the redshift up to the limit of the DEEP2
sample. The systematics in both regions are fairly similar presenting
a minimum scatter at 0.5$<$z$<$1, around the peak of the
$z_{\mathrm{phot}}$ distribution (see \S~\ref{zdistrib}), and
increasing by a factor of $\sim$1.3 at lower and higher redshifts
($z_{\mathrm{spec}}\lesssim$1.5). We find that the slightly worse
performance at z$<$0.5 is associated to the use of 4 IRAC bands in the
fitting of the SEDs. Although the template error out-weights the
contribution of these bands (mostly at $\lambda>$3~$\mu$m rest-frame;
see \S~\ref{zeropoint}) their contribution cause a broadening of the
zPDF that tends to increase the scatter. Nonetheless, this effect does
not increase the outlier fraction at z$<$0.5, which is comparatively
lower than at 1$<$z$<$1.5, for similar values of
$\sigma_{\mathrm{NMAD}}$.

\begin{figure*}
\centering
\includegraphics[width=8.cm,angle=0.]{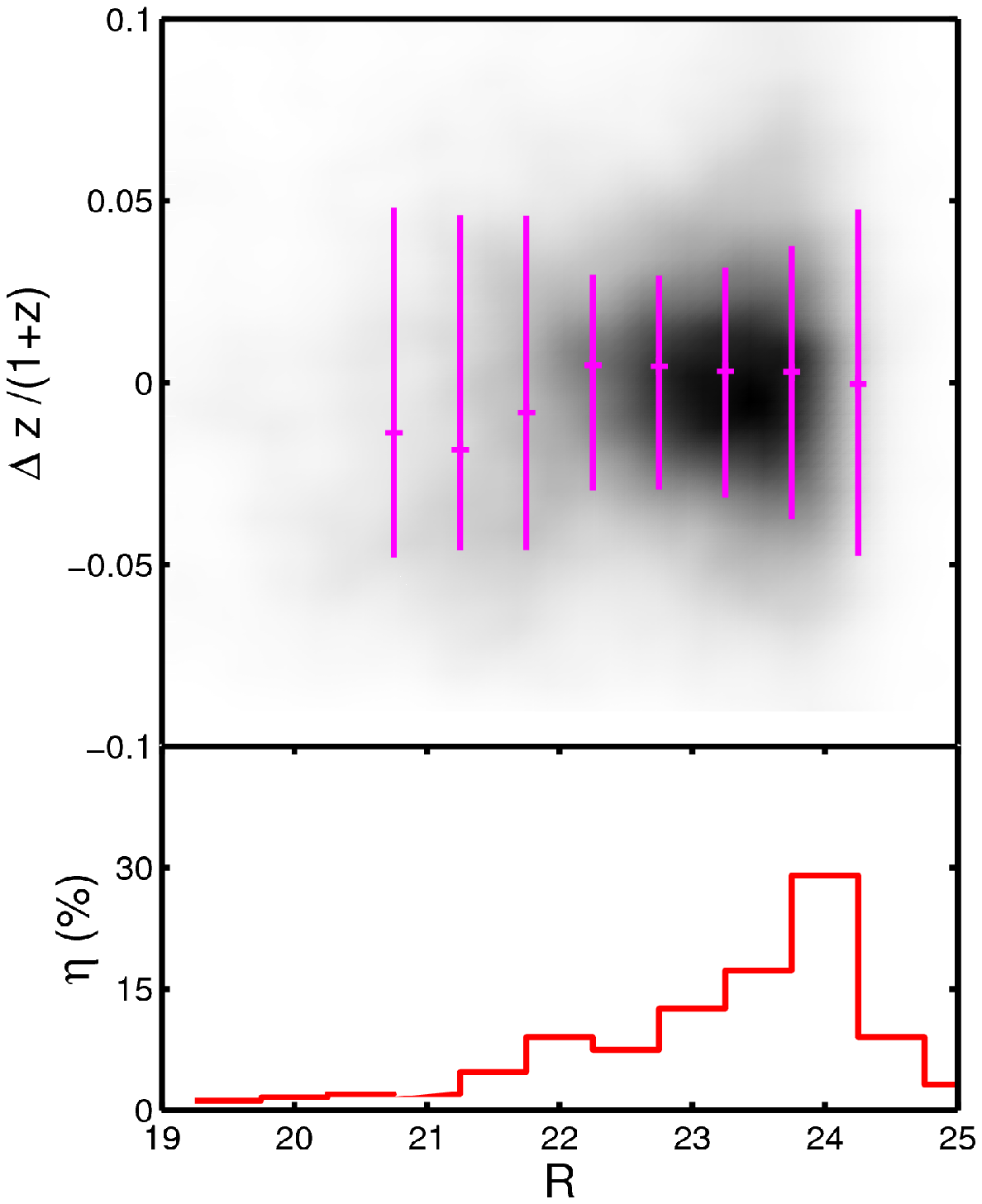}
\hspace{0.5cm}
\includegraphics[width=8.cm,angle=0.]{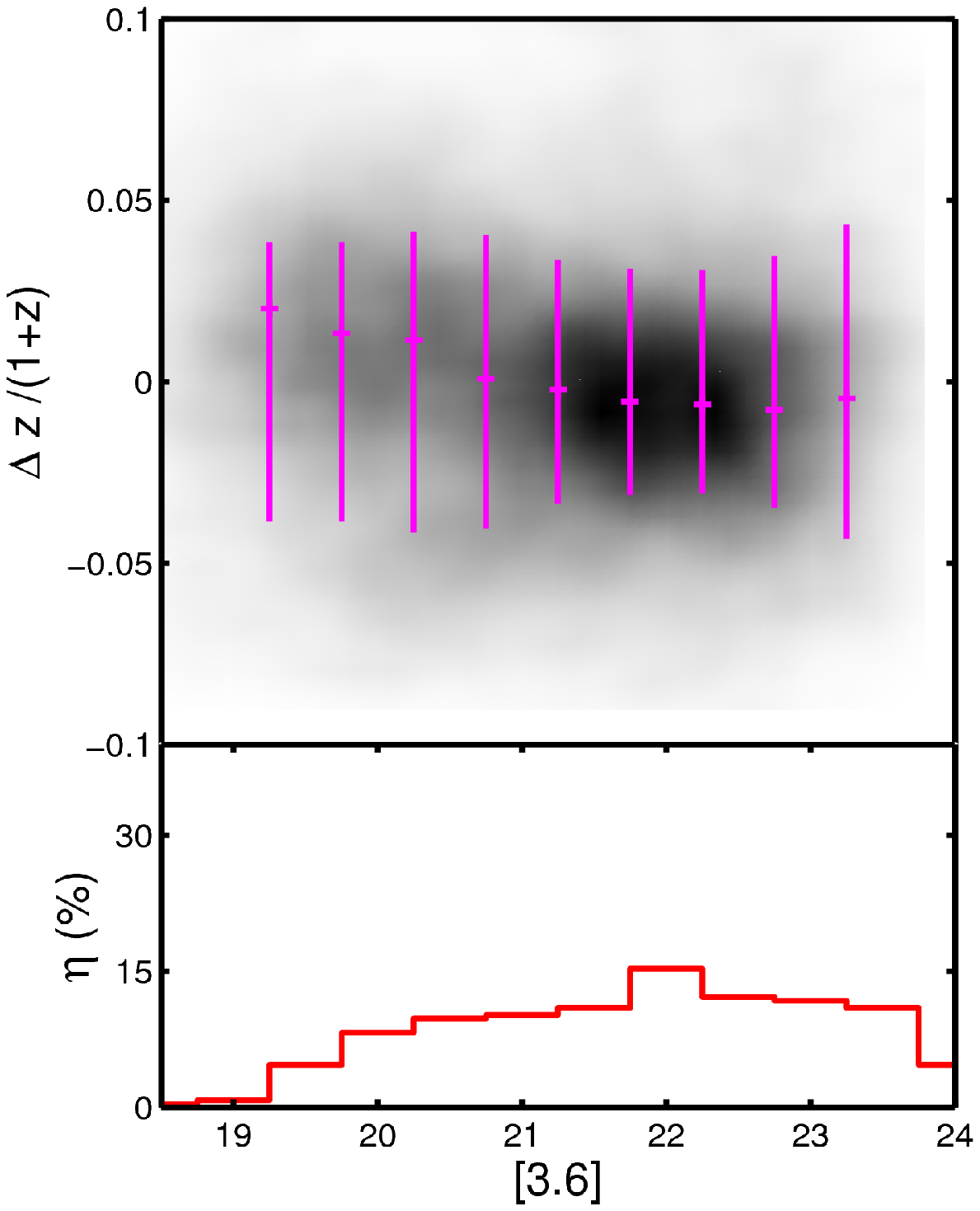}
\caption{\label{ncmad_magnitude} Density plot of the scatter in
  $\Delta z/(1+z)$ as a function of magnitude in the R (left) and
  [3.6] (right) bands. The magenta bars depict the median value of
  $\Delta z/(1+z)$ and $\sigma_{\mathrm{NCMAD}}$ (with respect zero)
  for each magnitude bin. The lower panel of each plot shows the
  fraction of photometric redshift outliers ($\eta$) as a function of
  magnitude.}
\end{figure*}

We also analyze the quality of $z_{\mathrm{phot}}$ as a function of the
optical and NIR magnitudes. As the efficiency of $z_{\mathrm{phot}}$
mostly relies on the detection of strong continuum features, the
estimates are highly sensitive to overall consistency of the
multi-band coverage.  Figure~\ref{ncmad_magnitude} shows the scatter
in $\Delta z$/(1+z) as a function the observed magnitudes in the $R$
and [3.6] bands for sources in the main region.  The results in the
flanking fields are similar, but with a larger scatter. We choose
these bands to be representative of the brightness of the sources in
the optical and NIR, and ultimately of the overall band coverage. Note
that, although this is NIR selected sample, most of the photometric
coverage consist on optical bands. Thus, galaxies with faint optical
magnitudes tend to present worse photometric redshifts.  The magenta
bars depict the median deviation and $\sigma_{\mathrm{NCMAD}}$ per
magnitude bin. We have corrected both plots by a median offset of
$\Delta z$/(1+z)$=$0.01.  In the R-band, the scatter increases
monotonically with the optical magnitude from
$\sigma_{\mathrm{NCMAD}}=0.03-0.06$ for R$=$22-25, and $>$50\% of the
outliers are located at R$>$23.5. The scatter is also wider at R$<$22.
However, since most of these bright galaxies lie at low-z, this trend
is essentially the same one mentioned above for sources at
$z<$0.5. Interestingly, there is weaker dependence in the scatter (and
the outlier fraction) with the [3.6] magnitude than with the R-band
magnitude.  This is because, the overall quality of the optical
photometry is more relevant for constraining the shape of the SED, and
there is typically a wide range of optical brightnesses for any given
[3.6] magnitude (see, e.g., Figure~6 of Paper I).

\subsubsection{$z_{\mathrm{phot}}$ versus $z_{\mathrm{spec}}$: LBGs sample}
\label{highz}

Given that the DEEP2 spectroscopic catalog consist mostly of low
redshift galaxies (68\% is located at z$\lesssim$0.9), we have
included in our sample spectroscopic redshifts drawn from the LBG
catalog of \citet{2003ApJ...592..728S} to specifically study the
accuracy of $z_{\mathrm{phot}}$ beyond the classic spectroscopic
limit. This catalog contains 334 LBGs galaxies, 193 of them with
confirmed spectroscopic redshift. To check the quality of our
$z_{\mathrm{phot}}$ at z$>$2.5, we first compare our results to their
$z_{\mathrm{spec}}$, and then we check that the our photometric
redshift distribution for the whole LBG sample is consistent with the
average redshift of this population.

We identify IRAC counterparts for 91(147) of the spectroscopic LBGs
with [3.6]$<$23.75\,mag (24.75). The rest were missed mainly because
they lie out of the observed area in the IRAC survey; only 10 galaxies
were lost due to their faintness in the IRAC bands. Note that,
although these LBGs are relatively bright in the optical
($R$$<$25.5\,mag), most of them are intrinsically faint in the IRAC
bands, $\sim$50$\%$ and 20$\%$ are fainter than [3.6]$=$23.75\,mag and
24.75\,mag, respectively. In general, LBGs are known to span a wide
range of IRAC magnitudes (\citealt{2005ApJ...634..137H},
\citealt{2006ApJ...648...81R}), and they exhibit a clear dichotomy in
the R-[3.6] color, with red (R$-$[3.6]$>$1.5) sources showing brighter
IRAC magnitudes than blue sources \citep{2008MNRAS.386...11M}. We find
that the median magnitudes and colors for the LBGs in our sample are
[3.6]$=$22.74, R$-$[3.6]$=$2.06 and [3.6]$=$23.80, R$-$[3.6]$=$0.88
for red and blue galaxies, respectively, in good agreement with the
values of \citet{2008MNRAS.386...11M} for a large sample of LBGs also
drawn from the LBG catalog of \citet{2003ApJ...592..728S}.

The quality of $z_{\mathrm{phot}}$ for the spectroscopic LBGs is
summarized in Table~\ref{photozquality1}. For the galaxies with
[3.6]$<$23.75, both the scatter and the outlier fraction
($\sigma_{\mathrm{NCMAD}}=0.063$, $\eta=10\%$) are slightly worse than
the median of the sample, as expected by their intrinsic faintness in
several optical and NIR bands. Nonetheless, the statistics are similar
to the results of other authors at high redshift (e.g.,
\citealt{2008ApJ...682..985W}) indicating that our $z_{\mathrm{phot}}$
still provide reasonably consistent values beyond z$>$1.5. If we also
consider the faintest sources ([3.6]$<$24.75), the statistics do not
degrade much ($\sigma_{\mathrm{NCMAD}}=0.069$), even though we are
including 60\% more sources. We have visually inspected the outliers
and at least 4 of them present flux contamination from close-by
sources and another 3 are strong AGNs detected in the X-rays. The rest
of them present a high-z solution in the zPDF, but the flux at [5.8]
and [8.0] is too faint to reliably identify the rapid decline of the
stellar component at $\lambda>$1.6$\mu$m, which results on favoring
the low-redshift solution.

\begin{figure}
\includegraphics[width=8.5cm,angle=0.]{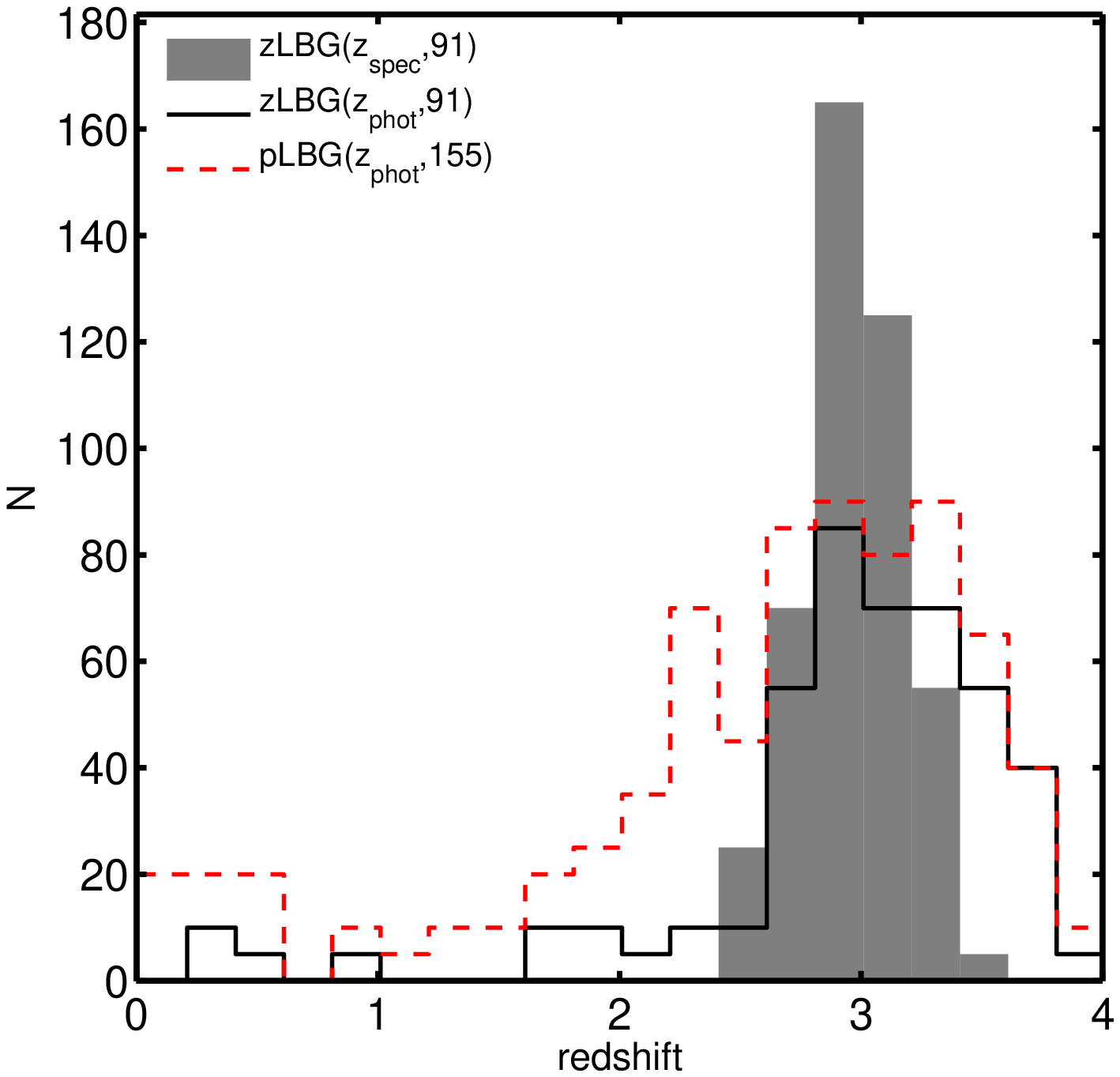}
\caption{\label{lbgplot} Photometric redshift distribution of the
  photometric (155, red dashed line) and spectroscopically confirmed
  (91, black line) LBGs (pLBG,zLBG) with [3.6]$<$23.75 in common with
  the sample of \citet{2003ApJ...592..728S}. The spectroscopic
  redshift distribution for the zLBG is shown as a filled grey
  histogram for comparison.}
\end{figure}

We also compared the photometric redshift distribution of the 155
galaxies with [3.6]$<$23.75 identified in the whole LBG
catalog. Figure~\ref{lbgplot} shows the redshift distribution of the
photometric and spectroscopic LBGs in our sample. The median value and
quartiles for the photometric LBGs with [3.6]$<$23.75 is
$\tilde{z}_{\mathrm{phot}}$$=$2.8$\pm^{0.4}_{0.6}$ consistent with the
median redshift of the spectroscopic sample ($\tilde{z}$$=$2.95) and
with the typical width of the redshift distribution for the LBG
criteria ($\tilde{z}$$=$3.0$\pm$0.3; \citealt{2004ApJ...604..534S},
\citealt{2005ApJ...633..748R}).  About 14$\%$ of these sources lie at
redshift z$<$1.5, similarly to the outlier fraction of the
spectroscopic sample.


\subsubsection{$z_{\mathrm{phot}}$ versus $z_{\mathrm{spec}}$: X-ray, Power-law, MIPS and Radio galaxies}

\placetable{photozquality}
\begin{deluxetable*}{lccccccccccc}
\centering
\tabletypesize{\footnotesize}
\setlength{\tabcolsep}{0.02in} 
\tablewidth{0pt}
\tablecaption{\label{photozquality1}Photometric redshift accuracy in the Main region [3.6]$<$23.75}
\tablehead{
& &  \multicolumn{4}{c}{{\it Rainbow}} & & \multicolumn{5}{c}{EAZY} \\
\cline{3-6}\cline{8-12}\\
\colhead{Redshift} &   \colhead{No.} & \colhead{$\Delta$z/(1+z)} & \colhead{$\sigma_{\mathrm{NCMAD}}$}  &  \colhead{$\eta$} &  \colhead{$\Delta z_{\mathrm{phot}}$} & &
\colhead{$\Delta$z/(1+z)} & \colhead{$\sigma_{\mathrm{NCMAD}}$}  &  \colhead{$\eta$} &  \colhead{$\Delta z_{\mathrm{phot}}$} & \colhead{$Q_{z}$}\\
(1)&(2)&(3)&(4)&(5)&(6)&&(7)&(8)&(9)&(10)&(11)
}
\startdata
All           & 6191&  0.010 & 0.034 &  2\% & 0.068 &   &0.019& 0.029&  3\%& 0.059& 94\%\\ 
\hline\\
0.$<$z$<$0.5  & 1637&  0.015 & 0.040 &  2\% & 0.070 &   & 0.019& 0.031&  1\%& 0.049& 97\%\\ 
0.5$<$z$<$1.0 & 3171&  0.007 & 0.028 &  2\% & 0.061 &   & 0.018& 0.025&  2\%& 0.055& 95\%\\ 
1.0$<$z$<$2.5 & 1292&  0.017 & 0.035 &  5\% & 0.083 &   & 0.021& 0.032&  4\%& 0.078& 89\%\\ 
z$>$2.5 (LBGs)&   91& -0.023 & 0.063 & 10\% & 0.110 &   &-0.014& 0.043& 15\%& 0.107& 42\%\\ 
z$>$2.5 (LBGs [3.6]$<$24.75) & 147& -0.027& 0.069& 12\%& 0.105&  &0.012&  0.060& 23\%& 0.119& 33\%\\ 
\hline\\
X-ray         & 142&   0.003 & 0.038 & 10\% & 0.070 &  & 0.008& 0.034& 10\%& 0.057& 82\%\\ 
PLGs          &   8&   0.031 & 0.142 & 50\% & 0.092 &  & 0.018& 0.094& 25\%& 0.108& 50\%\\ 
MIPS-24$\mu$m & 1955&  0.010 & 0.033 & 3\%  & 0.068 &  & 0.023& 0.026&  3\%& 0.055& 94\%\\ 
MIPS-70$\mu$m &  262&  0.015 & 0.045 & 1\%  & 0.071 &  & 0.031& 0.028&  2\%& 0.050& 95\%\\ 
Radio         &   85& -0.001 & 0.052 & 5\%  & 0.066 &  & 0.017& 0.032&  2\%& 0.048& 92\%
\enddata
\tablecomments{ 
 Photometric redshift quality with the estimates with {\it Rainbow} and EAZY (see \S~\ref{eazy}).\\
(1) Spectroscopic redshift range.\\
(2) Number of sources in the redshift bin.\\
(3,7) Median systematic deviation in $\Delta$z/(1+z); $\Delta z=z_{\mathrm{phot}}-z_{\mathrm{spec}}$.\\
(4,8) Normalized median absolute deviation.\\
(5,9) Percentage of catastrophic outliers ($|\Delta z|/(1+z)>$0.20).\\
(6,10) 68\% confidence interval in the zPDF around the most probable $z_{\mathrm{phot}}$.\\
(11) Percentage of the sources with Q$_{z}$$>$1 in EAZY (high quality flag according to Brammer et al. 2008). }
\end{deluxetable*}

\placetable{photozquality}
\begin{deluxetable*}{lccccccccccc}
\centering
\tabletypesize{\footnotesize}
\setlength{\tabcolsep}{0.02in} 
\tablewidth{0pt}
\tablecaption{\label{photozquality2}Photometric redshift quality in the Flanking regions [3.6]$<$23.75}
\tablehead{
& &  \multicolumn{4}{c}{{\it Rainbow}} & & \multicolumn{5}{c}{EAZY} \\
\cline{3-6}\cline{8-12}\\
\colhead{Redshift} &   \colhead{No.} & \colhead{$\Delta$z/(1+z)} & \colhead{$\sigma_{\mathrm{NCMAD}}$}  &  \colhead{$\eta$} &  \colhead{$\Delta z_{\mathrm{phot}}$} & &
\colhead{$\Delta$z/(1+z)} & \colhead{$\sigma_{\mathrm{NCMAD}}$}  &  \colhead{$\eta$} &  \colhead{$\Delta z_{\mathrm{phot}}$} & \colhead{$Q_{z}$}\\
(1)&(2)&(3)&(4)&(5)&(6)&&(7)&(8)&(9)&(10)&(11)
}
\startdata
All           & 1445&  0.013 & 0.046 &  3\% & 0.065 & &  0.027& 0.050&  4\%& 0.073& 87\%\\ 
\hline\\
0.$<$z$<$0.5  &  373&  0.021 & 0.058 &  2\% & 0.065 &  &  0.037& 0.058&  5\%& 0.052& 83\%\\ 
0.5$<$z$<$1.0 &  785&  0.015 & 0.040 &  3\% & 0.059 &  &  0.014& 0.069&  8\%& 0.077& 75\%\\ 
1.0$<$z$<$2.5 &  274& -0.001 & 0.058 &  8\% & 0.082 &  &  0.009& 0.043&  8\%& 0.079& 70\%\\ 
\hline\\
X-ray         &   33&  -0.019 & 0.031 & 9\% & 0.057&  &  0.000& 0.035&  9\%& 0.061& 88\%\\ 
PLGs          &    0&       - &    -  & -   &     -&  &      -&     -&    -&     -&  -\\ 
MIPS-24$\mu$m &  416&   0.011 & 0.046 & 4\% & 0.061&  &  0.028&  0.048& 5\%& 0.068& 88\%\\ 
MIPS-70$\mu$m &   66&   0.008 & 0.050 & 2\% & 0.063&  &  0.026&  0.052& 3\%& 0.055& 89\%\\ 
Radio         &   13&   0.000 & 0.055 & 8\% & 0.071&  &  0.035&  0.015& 0\%& 0.060& 85\%
\enddata
\tablecomments{Same as Table~\ref{photozquality1}}
\end{deluxetable*}

We analyze in detail the quality of the $z_{\mathrm{phot}}$ for
samples of galaxies that are known to present particularly different
SEDs from the majority of the templates (e.g., X-ray or AGNs), which
could cause a degradation of the redshift estimate. These sources are
shown with different makers and colors in Figure~\ref{zaccuracy_c23}
and their $z_{\mathrm{phot}}$ statistics are summarized in
Table~\ref{photozquality1} and ~\ref{photozquality2}.

The blue stars show galaxies identified in the X-Ray {\it
  Chandra}/ACIS catalog in EGS \citep{2009ApJS..180..102L}, probably
indicating the presence of an AGN. The SED of these sources is likely
affected by the AGN emission, which in principle should decrease the
efficiency of the template fitting procedure. In spite of showing a
larger outlier fraction (particularly at z$>$1.5), the
$z_{\mathrm{phot}}$ for X-ray sources are quite accurate, with a
scatter similar to that of the full sample.

The red dots in Figure~\ref{zaccuracy_c23} depict galaxies satisfying
the power-law criteria (PLG) commonly used to identify obscured AGNs
(\citealt{2004ApJS..154..155A}, \citealt{2007ApJ...660..167D}), a good
fraction of them being undetected in the X-rays.  We find a surface
density of 0.26\,arcmin$^{-2}$ for PLGs, in good agreement with the
0.22\,arcmin$^{-2}$ given in \citet[][we apply a similar criteria
  restricted to P$_{\chi}>$0.1 and a slope
  $\alpha<$-0.5]{2007ApJ...660..167D}. However, less than 2\% of these
sources present a spectroscopic redshift. Comparatively, PLGs present
a lower accuracy and higher outlier fraction than the X-ray sources as
a consequence of having their SED more contaminated by the AGN
emission.

Similarly to the PLG, we find that sources with red colors in the IRAC
bands (f$_{[3.6]}$$<$f$_{[4.5]}$$<$f$_{[5.8]}$$<$f$_{[8.0]}$), but not
strictly satisfying the PLG criteria, make up for up to 15\% of the
total outliers. This is not surprising given that, for the typical
galaxy at z$\leq$2, the presence of the stellar bump (at 1.6\,$\mu$m)
causes the flux in the last two IRAC bands to be lower than in the
previous two. Thus, for these sources, the code will try to assign
incorrect high redshift values of z$_{\mathrm{phot}}$.

Table~\ref{photozquality1} and ~\ref{photozquality2} also quote
numbers and $z_{\mathrm{phot}}$ statistics for the spectroscopic
sub-samples of galaxies detected in MIPS 24~$\mu$m
(f(24)$>$60~$\mu$Jy) , MIPS 70~$\mu$m (f(70)$>$3.5~mJy) and in the
catalog of Radio sources of \citet{2007ApJ...660L..77I}. The latter
present a slightly worse accuracy than the median of the sample,
whereas the MIPS detected galaxies present essentially the same
quality as the rest of the spectroscopic galaxies. This indicates that
for most of them the IR emission does not contribute significantly to
the NIR-MIR region fitted with the optical templates.

\subsubsection{Error analysis}\label{goodness}

The 1$\sigma$ uncertainty of the photometric redshifts, $\Delta
z_{\mathrm{phot}}$, is computed from the zPDF as the semi-width of the
redshift range corresponding to a 68$\%$ confidence interval around
the probability peak.  This value allows to provide an estimate of the
accuracy for sources without a spectroscopic redshift, which are
$>$90\% of the sample.

\begin{figure}
\includegraphics[width=8.5cm,angle=0.]{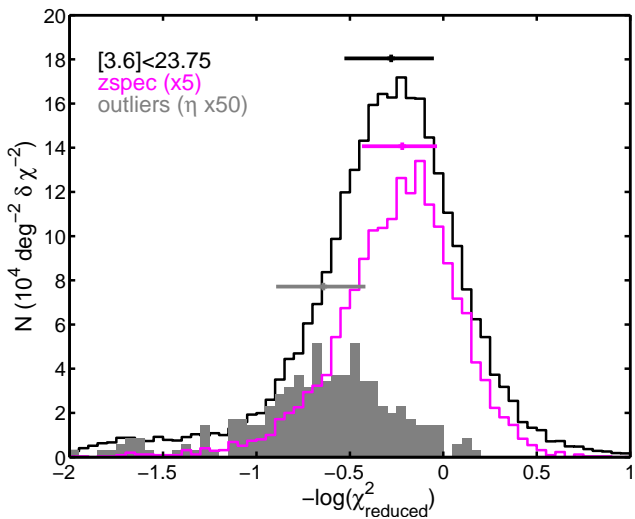}
\caption{\label{chi_zphot} Distribution of the normalized $\chi^{2}$
  resulting from the fit of the data to the templates during the
  calculation of $z_{\mathrm{phot}}$. The black line is for the full
  sample ([3.6]$<$23.75), and the magenta line and grey area are for
  the spectroscopic sample ($\times$5) and photometric redshift
  outliers ($\eta$$\times$50), respectively. The corresponding lines
  depict the median value and quartiles of each distribution.}
\end{figure}

Table~\ref{photozquality1} and ~\ref{photozquality2} quote the values
of $\Delta z_{\mathrm{phot}}$ as a function of redshift in the main
and flanking regions. Based on these results, we find that 62\%
(approximately 1$\sigma$) of the galaxies present values of $\Delta
z_{\mathrm{phot}}<|\Delta z|$. The median value and quartiles of
$\Delta z_{\mathrm{phot}}/(1+z)=0.036^{0.056}_{0.021}$ in the main
region are consistent with the statistics for
$\sigma_{\mathrm{NCMAD}}$ and also with $|\Delta
z|/(1+z)=0.027^{0.050}_{0.013}$. A similar agreement is found for the
sources in the flanking regions. Note that, as the $\Delta
z_{\mathrm{phot}}$ is computed from the zPDF its minimum value is
limited by the step size of the redshift grid ($\Delta$z$=$0.01), and
thus it tends to present larger values than $|\Delta z|$, specially
for very accurate $z_{\mathrm{phot}}$. Therefore, it is not surprising
that $\Delta z_{\mathrm{phot}}/(1+z)$ is on average larger than all
the other scatter estimates. In fact, this indicates that
$\Delta$$z_{\mathrm{phot}}$ provides a robust estimate of the
uncertainty in $z_{\mathrm{phot}}$, which can be underestimated if it
assumed to be equal to $\sigma_{\mathrm{NCMAD}}$ (see e.g.,
\citealt{2010ApJS..189..270C}).

In order to obtain a better characterization of the catastrophic
outliers caused by a poor fit to the data, we analyze the distribution
of sources as a function of the reduced $\chi^{2}$ of the SED
fitting. Figure~\ref{chi_zphot} shows the distribution of
$-log(\chi^{2})$ for the full photometric sample, the spectroscopic
sample and the catastrophic outliers. Approximately 83\% and 94\% of
the galaxies in the photometric and spectroscopic sample present
values of $\chi^{2}$ lower than the median, of the outlier
distribution ($-log(\chi^{2})<-0.6$), i.e., half of the outliers are
located within the $\sim$20\% and 5\% of the sources in the
photometric and spectroscopic samples with the worse values of
$\chi^{2}$.

Finally, we also find that 58$\%$ of the sources with significantly
different values of $z_{\mathrm{best}}$ and $z_{\mathrm{peak}}$
($|z_{\mathrm{best}}$-$z_{\mathrm{peak}}|$/(1+z)$>$0.2) are outliers.
These sources account for only 1\% of the spectroscopic sample, but
they represent $\sim$12\% of the outliers. Therefore, the difference
between both values is another useful indicator of possible outliers.

\subsection{Comparison of photometric redshift catalogs}

Here we compare the z$_{\mathrm{phot}}$ computed with {\it Rainbow} to
other previously published z$_{\mathrm{phot}}$ catalog and to the
estimates obtained with a different code. The alternative
z$_{\mathrm{phot}}$ are also included in the our data release (see
\S~\ref{datacatalogs}) in Table~\ref{dataredshift} .

\subsubsection{{\it Rainbow} vs. Ilbert et al. (2006a)}
\placetable{photozquality}
\begin{deluxetable*}{lcccccccc}
\centering
\tabletypesize{\footnotesize}
\setlength{\tabcolsep}{0.02in} 
\tablewidth{0pt}
\tablecaption{\label{rainbowi06}Rainbow z$_{\mathrm{phot}}$ versus I06 at i'$<$25 and [3.6]$<$23.75}
\tablehead{
& &  \multicolumn{3}{c}{{\it Rainbow}} & & \multicolumn{3}{c}{I06} \\
\cline{3-5}\cline{7-9}\\
\colhead{Redshift} &   \colhead{No.} & \colhead{$\sigma_{\mathrm{NCMAD}}$}  &  \colhead{$\eta$} &  \colhead{R($\eta$)} & &
\colhead{$\sigma_{\mathrm{NCMAD}}$}  &  \colhead{$\eta$} &  \colhead{R($\eta$)} \\
(1)&(2)&(3)&(4)&(5)&&(6)&(7)&(8)}
\startdata
All           & 5454&   0.034 &  2\% &  82\%&   & 0.036&  5\%& 55\%\\ 
\hline\\
0.$<$z$<$0.5  & 1444&   0.040 &  2\% &  83\%&   & 0.032&  5\%& 43\%\\ 
0.5$<$z$<$1.0 & 2787&   0.028 &  2\% &  80\%&   & 0.031&  3\%& 53\%\\ 
1.0$<$z$<$2.5 & 1143&   0.035 &  4\% &  80\%&   & 0.054&  8\%& 63\%\\ 
z$>$2.5 (LBGs)&   80&   0.063 &  9\% &  91\%&   & 0.345& 46\%& 42\%   
\enddata
\tablecomments{ \\
 Photometric redshift quality in the estimates with {\it Rainbow} and in I06.\\
(1) Spectroscopic redshift range.\\
(2) Number of sources in the redshift bin.\\
(3,6) Normalized median absolute deviation.\\
(4,7) Percentage of catastrophic outliers ($|\Delta z|/(1+z)>$0.20).\\
(5,8) Fraction of catastrophicc outliers in the other code presenting an accurate z$_{\mathrm{phot}}$. }
\end{deluxetable*}

We compare the $z_{\mathrm{phot}}$ presented in this paper with those
derived by \citet[][hereafter, I06]{2006A&A...457..841I} based on
optical data from the CFHTLS. These authors used a $i'$-band selected
sample with $i'<$24 and obtained photometric redshifts for the four
CFHTLS deep fields. The $z_{\mathrm{phot}}$ were computed using the
template fitting code {\it Le phare} (Arnouts \& Ilbert; e.g.,
\citealt{2009ApJ...690.1236I}) for $\sim500,000$ sources observed in 5
bands $u*,g',r',i',z'$. Their template library is based on an upgrade
of the empirical templates of \citet{1980ApJS...43..393C} and
\citet{1996ApJ...467...38K} computed by applying zero-point
corrections and interpolating between spectral types. Their
$z_{\mathrm{phot}}$ also include a Bayesian prior on the redshift
distribution. The accuracy of their results for the D3 field (the EGS)
is $\sigma_{\mathrm{NCMAD}}=0.035$ with $\eta=4\%$ for sources with
$i'<$24 and z$<$1.5. More recently, \cite{2009A&A...500..981C}
repeated essentially the same exercise using the latest data release
of the CFHTLS T004, obtaining $z_{\mathrm{phot}}$ of almost identical
quality.

\begin{figure*}
\centering
\includegraphics[width=8.5cm,angle=0.]{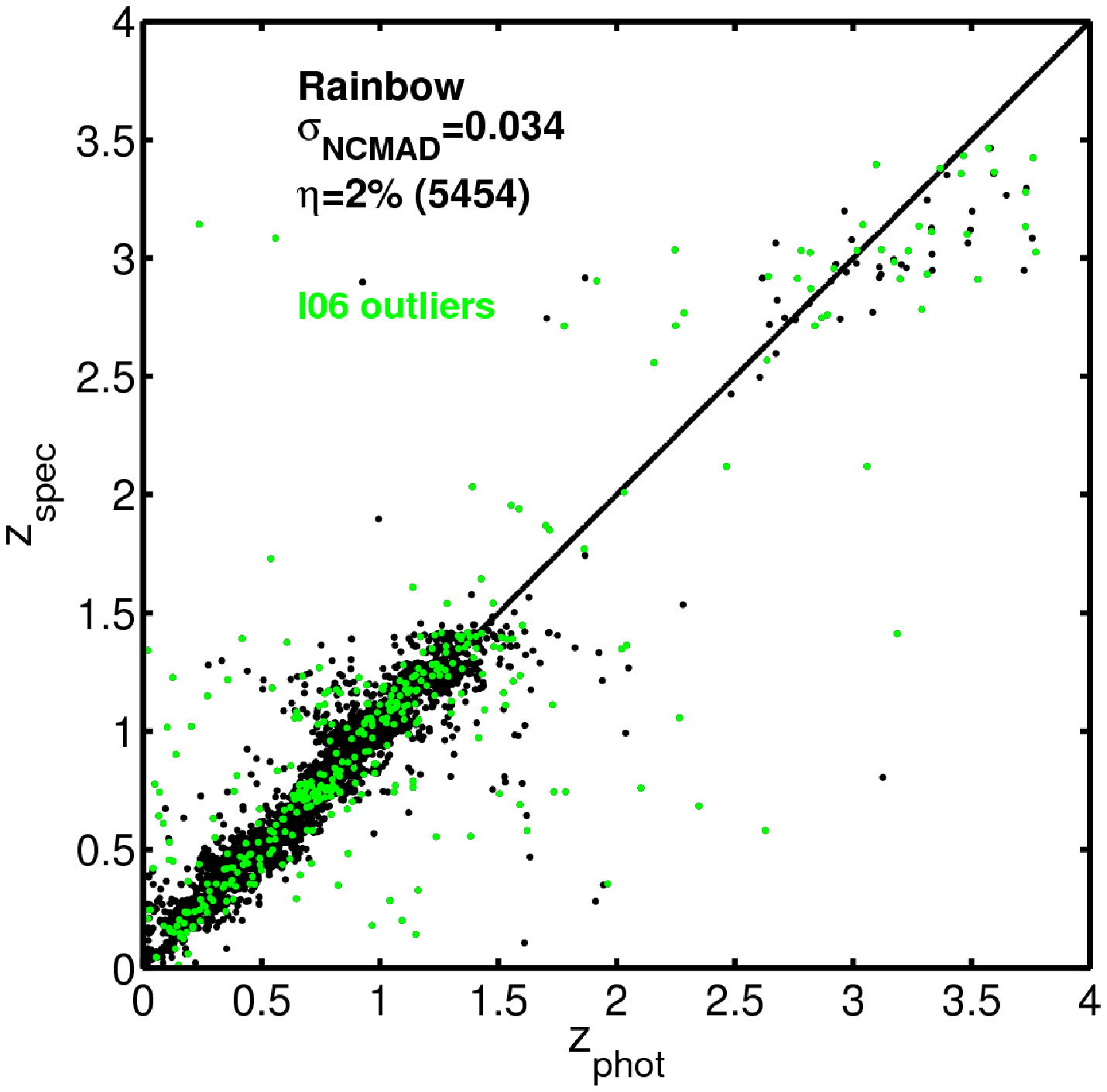}
\includegraphics[width=8.5cm,angle=0.]{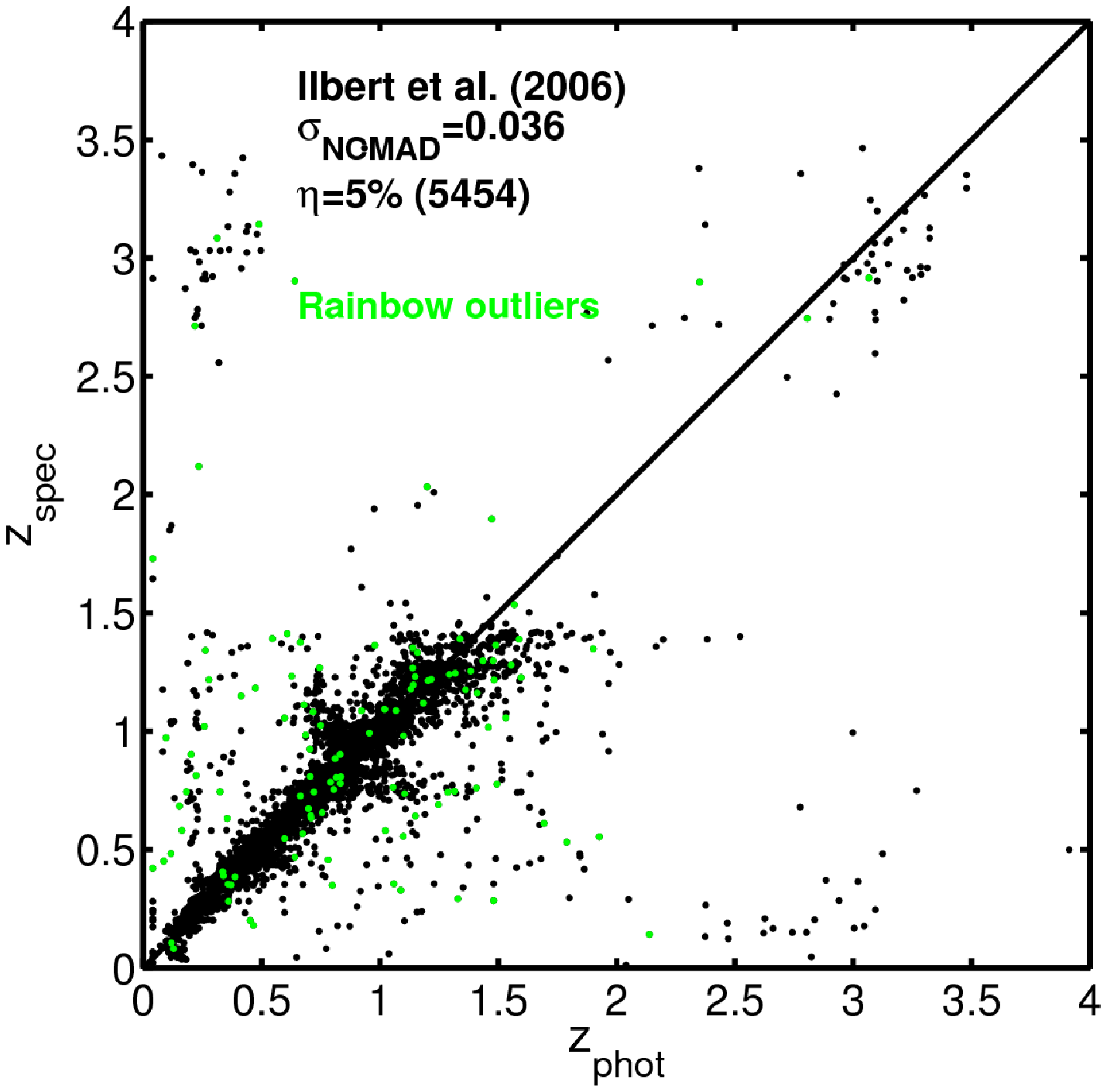}
\centering
\includegraphics[width=8.5cm,angle=0.]{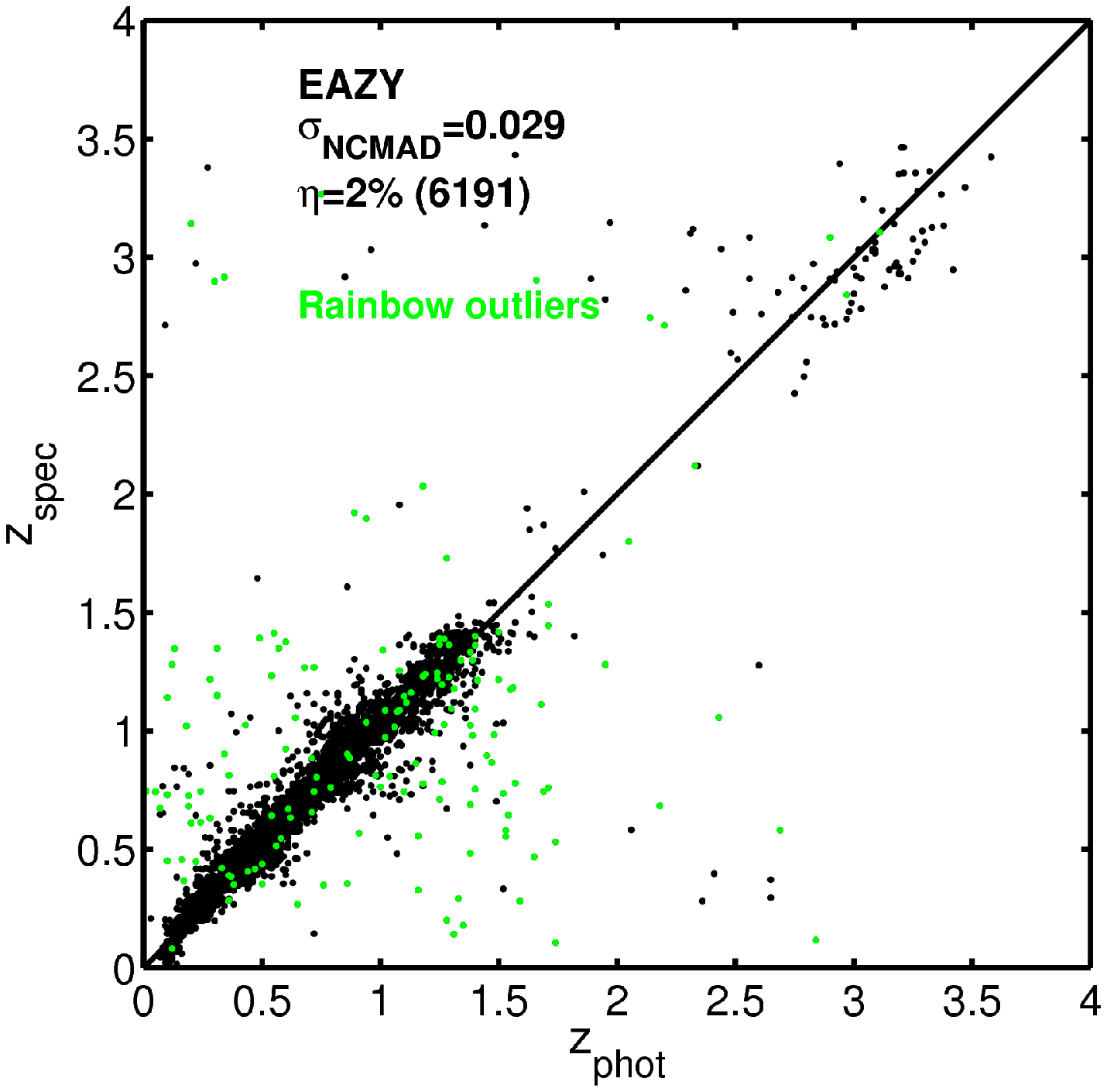}
\caption{\label{cfthls_pzeta} {\it Top:} Comparison of
  $z_{\mathrm{phot}}$ versus $z_{\mathrm{spec}}$ for the estimations
  presented in this paper (left) and the ones in I06 (right). The
  sample is drawn from the overlap region between the CFHTLS-D3 area
  and the IRAC mosaic (main region).  Both figures contain the same
  5454 sources simultaneously detected in both catalogs and in the
  DEEP2 sample at [3.6]$<$23.75 and $i'$$<$24.5. Green points depict
  $z_{\mathrm{phot}}$ in {\it Rainbow} for the outliers in I06 (left)
  and viceversa (right). {\it Bottom:} Comparison of
  $z_{\mathrm{phot}}$ computed with EAZY versus $z_{\mathrm{spec}}$
  for sources in the main region. This Figure is equivalent to the
  left panel of Figure~\ref{zaccuracy_c23} for estimates based in
  EAZY. Green dots depict catastrophic outliers in the
  $z_{\mathrm{phot}}$ estimated with {\it Rainbow}.}
\end{figure*}

The source density in the $i'$-band selected sample of I06 is 25, 42
and 96 sources/arcmin$^{2}$ up to limiting magnitudes of $i'=$24, 25
and 26.5 (the estimated SNR$\sim$5 level). The source density of the
IRAC selected catalog is $\sim$44 sources/arcmin$^{2}$ at
[3.6]$<$23.75.  This means that their source density at $i'<$25, which
is essentially the spectroscopic limit (R$=$25), is similar to ours.
However, at $i'<$24, the limiting magnitude for their best performing
$z_{\mathrm{phot}}$, the source density in $i'$ is approximately
50$\%$ that in IRAC.  At the faintest optical magnitudes, the source
density in the $i'$-band selected catalog is larger, although the
quality of these $z_{\mathrm{phot}}$ is worse than for the $i'$$<$24
sample, given that many of the galaxies will also be undetected in the
shallowest optical bands ($u^{*}$,$z'$).

Even presenting similar source densities, the nature of the galaxies
in an $i'$-band and an IRAC-selected samples is different, and some of
the sources in one selection will be missed by the other. We find that
the optically bright galaxies missed by the IRAC catalog
([3.6]$>$23.75) present a median and quartile redshifts
$z_{\mathrm{phot}}=$1.0$^{1.3}_{0.6}$, while the infrared bright
galaxies undetected in the optical ($i'$$>$26.5) present
$z_{\mathrm{phot}}=$1.8$^{1.1}_{2.3}$.  The high-z sources missed in
the IR selection are typically low-mass galaxies (similar to LBGs),
i.e., our catalog favors the detection of high-z massive galaxies, as
expected.

We cross-correlated the catalog of I06 to the IRAC selected sample
using a search radius of 1.5\arcsec. Due to small differences in the
extraction of the catalogs, the comparison is restricted to a slightly
smaller portion of the main region
(214.09$^{\circ}$$<$$\alpha$$<$215.72$^{\circ}$ and
52.20$^{\circ}$$<\delta<$53.16$^{\circ}$). Out of the 49605 IRAC
sources, 40\% and 88\% are detected in I06 to $i'<$24 and 26.5,
respectively.  The cross-match to the DEEP2 spectroscopic redshifts
contains 5454 galaxies simultaneously identified in all three catalogs
([3.6]$<$23.75, $i'$$<$26.5). Approximately 6\% of our spectroscopic
subsample is missed due to a more conservative source removal around
bright stars in I06.

The top panels of Figure~\ref{cfthls_pzeta} show the comparison of
$z_{\mathrm{phot}}$ versus $z_{\mathrm{spec}}$ for the galaxies in
common between the {\it Rainbow}, I06 and DEEP2 catalogs with
[3.6]$<$23.75 and $i'$$<$25, without any other requirement of band
coverage.  Table~\ref{rainbowi06} summarizes accuracy of the
$z_{\mathrm{phot}}$ in {\it Rainbow} and I06 for these sources as a
function of redshift. We also list the fraction of catastrophic
outliers in each catalog that is recovered in the other (shown as
green dots in Figure~\ref{cfthls_pzeta}).

The overall scatter in I06 for the sources in common with the IRAC
sample is consistent with their results for the whole D3 sample. The
comparison as a function of redshift indicates that I06 estimates at
lower redshift are slightly more precise, probably as a result of the
template optimization algorithm and the Bayesian prior (see Fig.6 of
I06), but also because of our slightly lower performance at
z$<$0.5. On the contrary, the fraction of catastrophic outliers in I06
is larger than in {\it Rainbow} for all redshifts, and particularly at
z$>2.5$. Moreover, the $z_{\mathrm{phot}}$ {\it Rainbow} is able to
recover $\sim$80\% of these outliers. At z$>$1, our larger band
coverage, mostly in NIR bands, provides more accurate estimates. Note
that the IRAC fluxes for the LBGs play a critical role on providing
more accurate redshifts (and stellar parameters) for these sources.

\subsubsection{{\it Rainbow} vs. EAZY}
\label{eazy}

Here we check again the quality and overall consistency of our SEDs
and $z_{\mathrm{phot}}$ by computing an independent estimation of the
$z_{\mathrm{phot}}$ with a different code. A successful result using a
different fitting code based on different template sets would certify
that the catalog reproduces accurately the observed SEDs and is
therefore suitable for galaxy population studies.

We computed alternative $z_{\mathrm{phot}}$ using the photometric
redshift code {\it EAZY} \citep{2008ApJ...686.1503B}. The advantage of
{\it EAZY} is that it was conceived to provide accurate photometric
redshift estimates for NIR selected samples in absence of a
representative calibration sample of spectroscopic redshifts. The code
makes use of a new set of templates computed from a $K$-limited
subsample of the Millennium Simulation (\citealt{2005Natur.435..629S},
\citealt{2007MNRAS.375....2D}) and modeled by fitting the synthetic
SEDs with PEGASE models and applying an optimization algorithm.  The
final result is set of 6 templates that essentially reproduces the
principal components of the catalog. Furthermore, a template error
function was introduced to account for systematic differences between
the observed photometry and the template photometry at different
wavelengths.  After trying different configurations for the input
parameters, we find that the best results in the
$z_{\mathrm{phot}}$-$z_{\mathrm{spec}}$ comparison are obtained using
the template error function and incorporating a Bayesian prior on the
redshift distribution similar to that of \citet[][see
\citealt{2008ApJ...686.1503B} for more
details]{2000ApJ...536..571B}. The use of the template error function
is decisive to weight the contribution of the IRAC bands at lower
redshifts as we have also verified in our own $z_{\mathrm{phot}}$ (see
\S~\ref{zeropoint}).

In addition, we find that a critical issue to avoid a severe
contamination from catastrophic $z_{\mathrm{phot}}$ is the use of the
purged photometric catalog produced by {\it Rainbow}. Prior to the
fitting procedure, {\it Rainbow} carries out a first pass on the
catalog where potential photometric outliers are removed.  If we use
the resulting catalog as input for {\it EAZY} the outlier fraction is
reduced by a factor $\sim$5, illustrating the relevance on the
photometric errors not only in the overall quality of the
$z_{\mathrm{phot}}$ but also in the catastrophic errors.

The bottom panel of Figure~\ref{cfthls_pzeta} depict the comparison
the z$_{\mathrm{phot}}$ with EAZY versus spectroscopic redshifts for
the galaxies in the main region (i.e., the same galaxies depicted in
the left panel of Figure~\ref{zaccuracy_c23}). As in the previous
section green markers indicate outliers in z$_{\mathrm{phot}}$ {\it
  Rainbow}. Table~\ref{photozquality1} and ~\ref{photozquality2}
summarizes the quality of z$_{\mathrm{phot}}$ for {\it Rainbow} and
EAZY in different redshift bins for the main and flanking regions,
respectively. The overall scatter and outlier fraction of both
estimates are roughly similar, with the estimates of EAZY presenting a
slightly higher accuracy in the main region, but lower in the flanking
fields.  We also note that z$_{\mathrm{phot}}$ with EAZY perform
better at z$<$0.5, being less sensitive to the mild broadening of the
zPDF present in {\it Rainbow}.  The tables also quote the 68\%
confidence intervals of z$_{\mathrm{phot}}$, which are similar for
both codes and are in good agreement with other results based on EAZY
(e.g., \citealt{2010ApJS..189..270C}). We show that the fraction of
sources with a quality parameter Q$_{z}$$\leq$1 in EAZY (good
photometric redshifts; \citealt{2008ApJ...686.1503B}) is typically
$>$90\% except for the highest redshift bin, where the photometry is
more uncertain, due to the intrinsic faintness of these sources, and
thus there is larger fraction of outliers. In addition, we find that,
roughly 40-50\% of the outliers in EAZY present a poor
z$_{\mathrm{phot}}$ estimate in {\it Rainbow} and
viceversa. Therefore, similarly to the galaxies with different values
of $z_{\mathrm{peak}}$ and $z_{\mathrm{prob}}$, galaxies with
significantly different estimates both catalogs are frequently
($\sim$50\%) outliers.

From the good agreement between the different z$_{\mathrm{phot}}$
estimates we conclude that the photometric catalog provides accurate
SEDs suitable for studies of galaxy populations irrespective of the
code used for the analysis.

\begin{figure*}
\centering
\includegraphics[width=7.5cm,angle=0.]{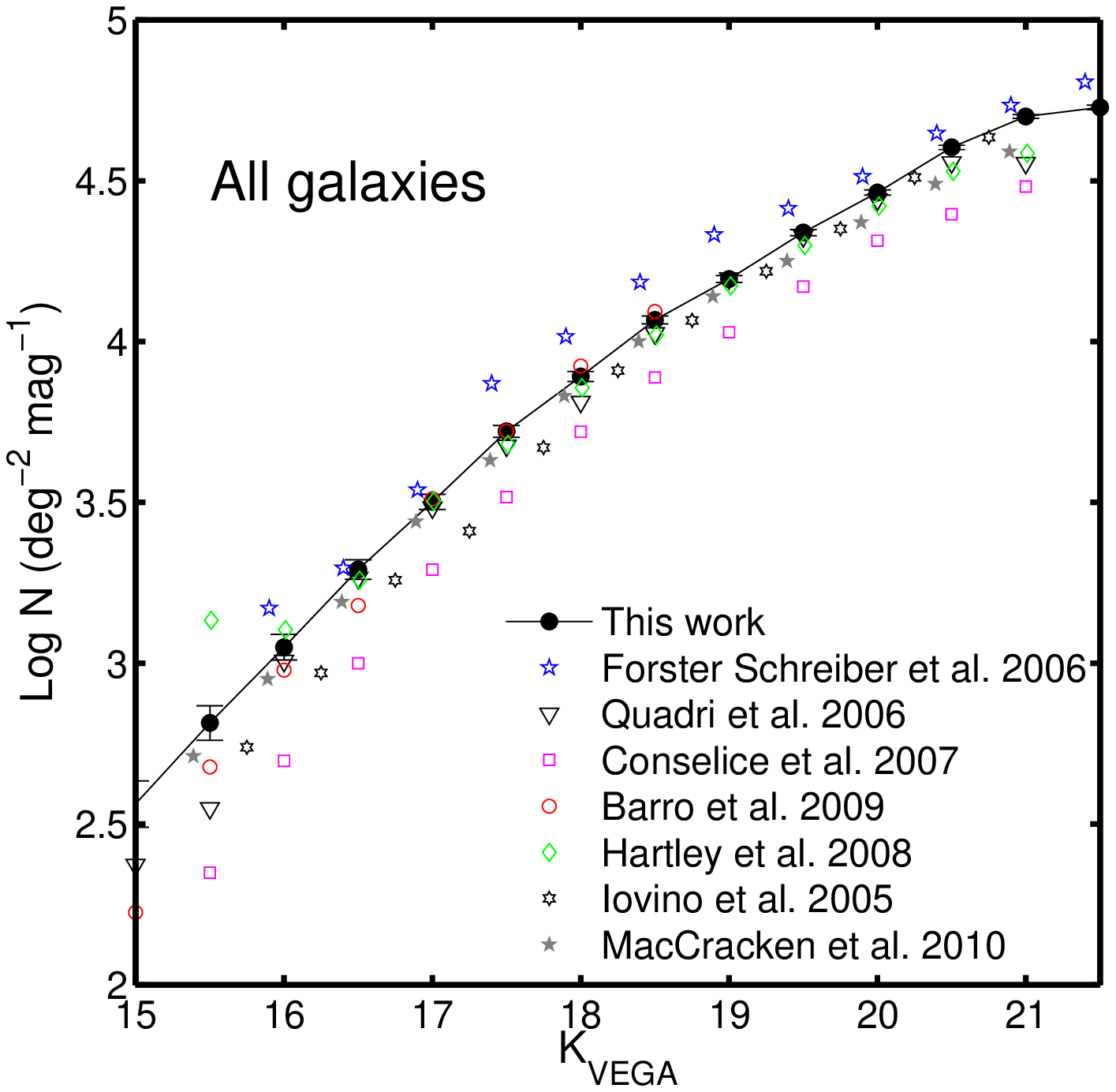}
\hspace{1cm}
\includegraphics[width=7.5cm,angle=0.]{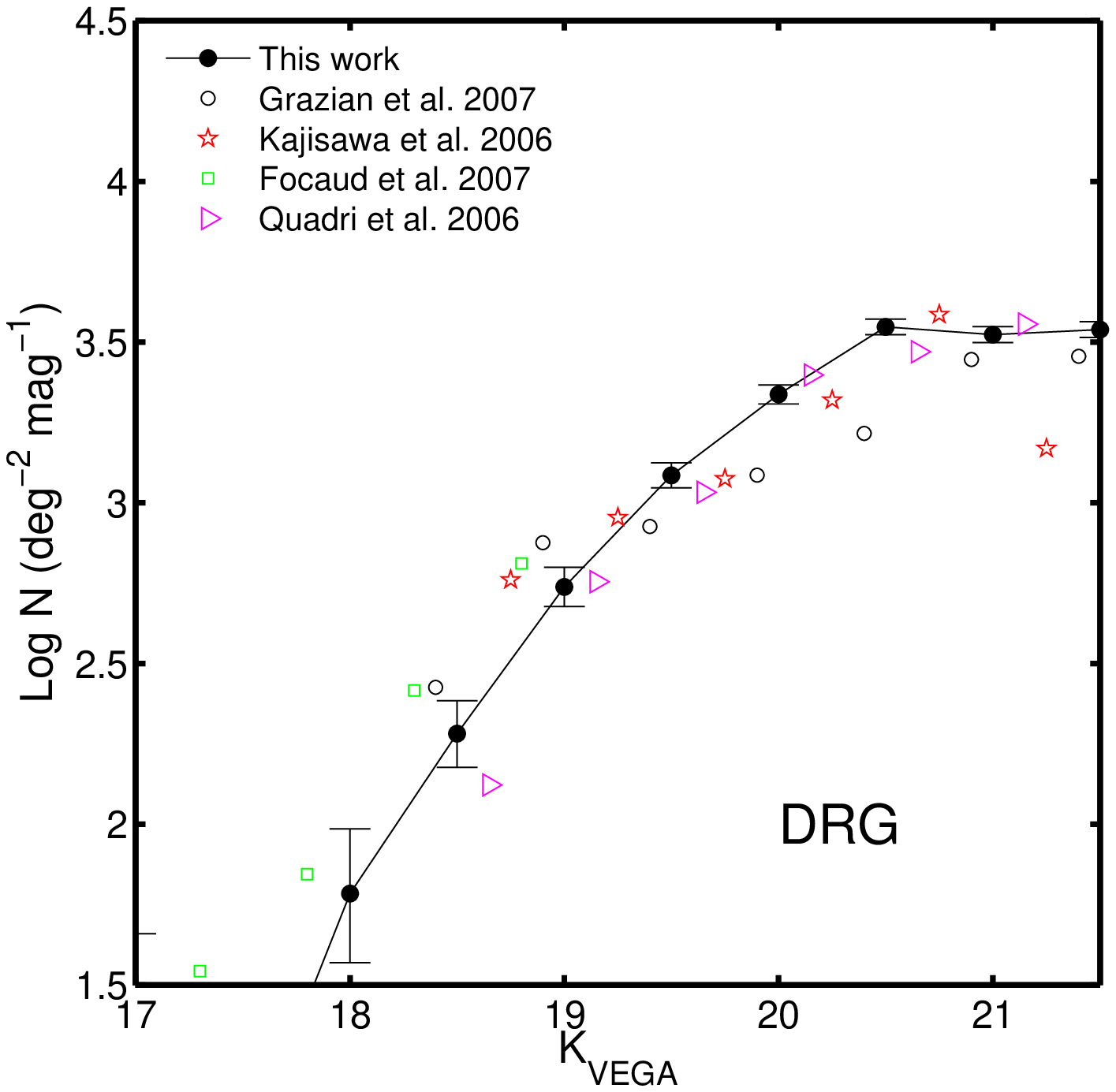}\\
\includegraphics[width=7.5cm,angle=0.]{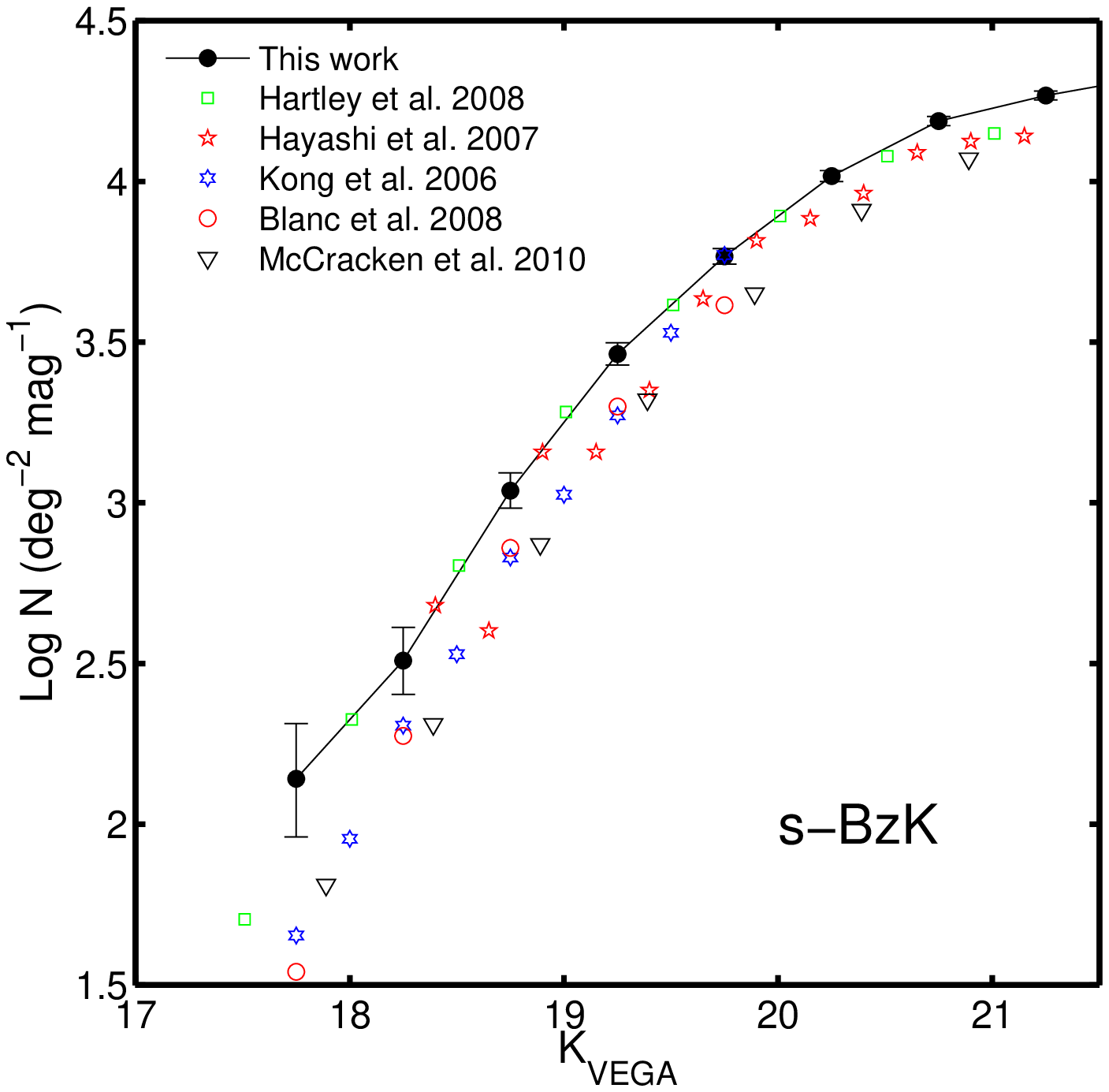}
\hspace{1cm}
\includegraphics[width=7.5cm,angle=0.]{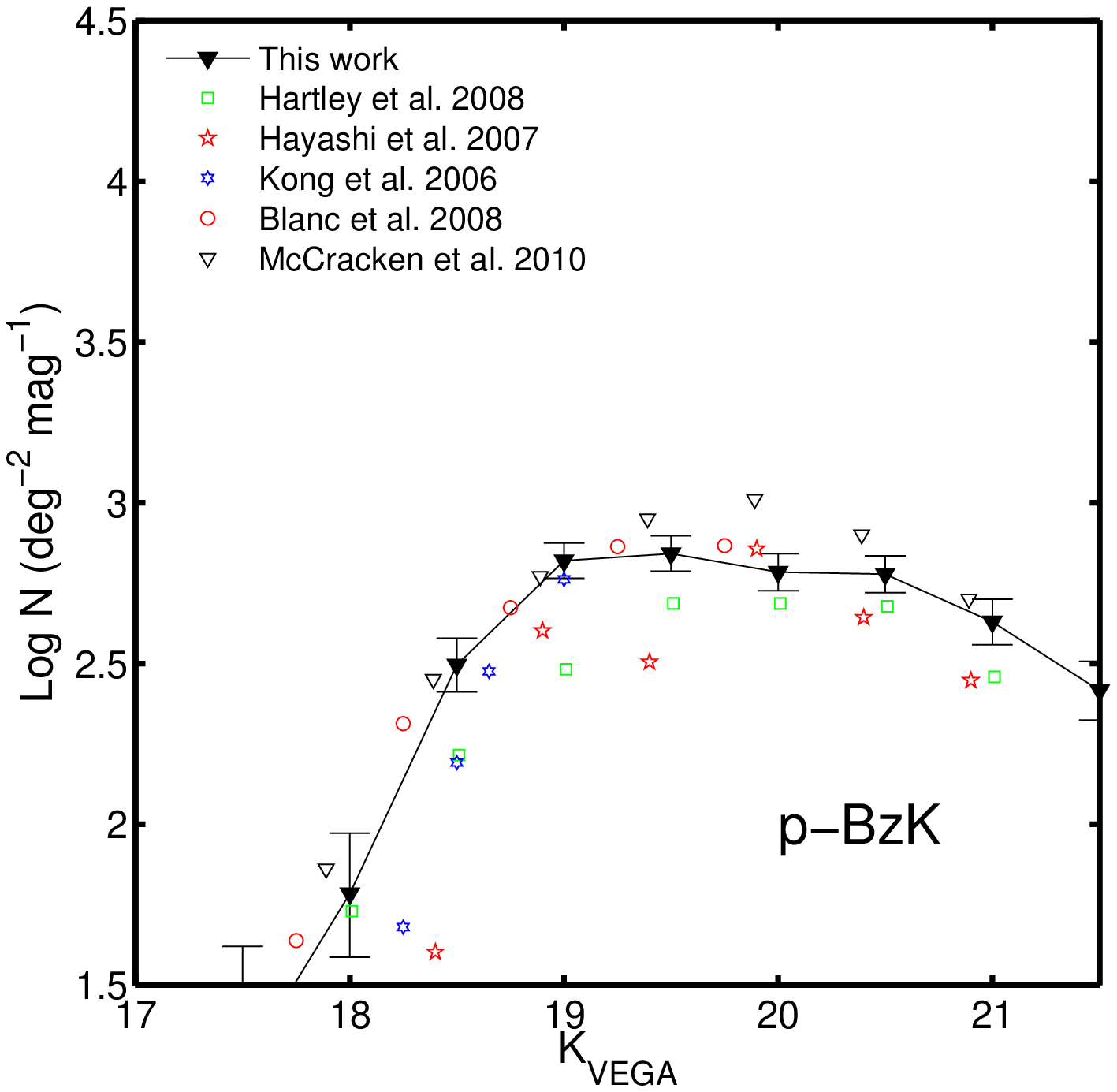}
\caption{\label{numbercounts} Galaxy number counts of our sample in
  the $K$-band, derived from synthetic observed magnitudes, compared
  to results from the literature. The black dots connected with a
  solid line depict the values derived in the present work. The other
  symbols show the results from other authors. {\it Top-left}; Number
  counts for the complete galaxy sample ([3.6]$<$23.75) in the main
  region.  {\it Top-right}; Number counts for DRG.  {\it
    Bottom-left}; Number counts for s-BzK galaxies.  {\it
    Bottom-right}; Number counts for p-BzK galaxies.}
\end{figure*}

\subsection{Number densities and redshift distribution of NIR-selected galaxies.}
\label{colorpop}

As an additional test of the accuracy of $z_{\mathrm{phot}}$ we
compare the number densities and redshift distributions of NIR
color-selected populations with the results from other authors. In
order to facilitate the comparison to the references, the magnitudes
in this section are given in Vega system.

Given the highly non-uniform band coverage of the field, we have
chosen to compute galaxy colors based on synthetic magnitudes. An
advantage of this method is that synthetic photometry behaves better
than directly observed values when deep data is not available in some
of the required bands, allowing us to assign robust fluxes for
undetected sources in the shallower bands.  This is the same procedure
that we used in PG08, and is similar to that presented in
\citet{2007A&A...465..393G} and \citet{2007AJ....134.1103Q}. We
restrict the analysis in this section to the 0.35\,deg$^{2}$ of the
main region which count with better photometry.

For obvious reasons, the success of this method depends critically on
the quality of the synthetic fluxes. In \S~\ref{zeropoint}, we showed
that these fluxes provide an accurate representation of the observed
values in the magnitude range covered by the observations. The median
offsets are very small and the scatter is consistent within a factor
$\sim$2 with the photometric errors at different magnitudes (see
Figure~\ref{zeropt}).

In order to avoid possible selection effects, we restrict the analysis
to NIR selected galaxies which would be fully represented in the
IRAC-selected sample. We selected Distant Red Galaxies (DRG;
\citealt{2003ApJ...587L..79F}) as galaxies with [$J$$-$$K$] $>$2.3,
and BzK galaxies, both star forming (s-BzK) and passively evolving
(p-BzK), following the equations in \citet{2004ApJ...617..746D}.  Both
criteria were proposed to target massive galaxies at z$\sim$2,
although DRG and p-BzK are best at selecting galaxies with a
significant fraction of evolved stars, whereas s-BzK select
star-forming galaxies similar to those found by the low-redshift
equivalent of the LBG criteria (LBG/BX;
\citealt{2004ApJ...604..534S}). For the DRG we convolved the templates
with the VLT/ISAAC $J$ and $K$ filters, whereas for the BzK we used
VLT/FORS B, HST/ACS $z$ and VLT/ISAAC $K$ which are the same filters
used in \citet{2004ApJ...617..746D}.

The top left panel of Figure~\ref{numbercounts} shows the $K$-band
\footnotemark[1] number counts for the IRAC-selected catalog compared
to other results from the bibliography. Our counts are in very good
agreement with the values of \citet{2007AJ....134.1103Q} for the MUSYC
survey, and with our previous results in Barro et al. (2009) for the
South region of EGS ($\sim$30$\%$ overlap with the IRAC sample).  The
overall agreement with the counts of the Palomar-WIRC catalog
\citep{2008MNRAS.383.1366C} is slightly worse.  However, the complete
Palomar-WIRC catalog covers a total area of 1.47\,deg$^{2}$ out of
which EGS is only a small fraction (0.20\,deg$^{2}$). At
$K_{\mathrm{VEGA}}$$\sim$20 our results are also consistent with those
in \citet{2008MNRAS.391.1301H}, \citet{2010ApJ...708..202M} and
\citet{2006AJ....131.1891F}, following the same trend as the latter up
to $K_{\mathrm{VEGA}}$=21. From this comparison, we conclude that the
IRAC catalog limited to [3.6]$=$23.75 is a good proxy of a $K$-limited
sample with at least $K_{\mathrm{VEGA}}$$<$21\,mag.

\footnotetext[1]{We used the following transformations when required 
$\Delta$K$_{\mathrm{VEGA-AB}}$(UKIRT,CFHT,SOFI)=1.90,1.85,1.87.}

The top-right and the bottom panels of Figure~\ref{numbercounts} shows
the comparison of the number counts for DRG and BzK (restricted to
z$>$1.4) galaxies with other values from the literature. There is good
agreement within the typical scatter (0.1-0.2\,dex), generally
associated with cosmic variance.  Our counts reproduce the most
representative features of the overall distribution, namely: the
plateau in DRG and p-BzK around $K_{\mathrm{VEGA}}$$\sim$20.5 and the
steep slope in the counts of s-BzK.  We note that our s-BzK counts are
slightly above those from \citet{2008ApJ...681.1099B} and
\citet{2010ApJ...708..202M} which count with very large surveyed areas
(0.71\,deg$^{2}$ and 2\,deg$^{2}$, respectively). On the contrary, our
results are in excellent agreement with
\citet[][0.63\,deg$^{2}$]{2008MNRAS.391.1301H}. In
\citet{2010ApJ...708..202M} the authors argue that their disagreement
with the counts of \citet{2008MNRAS.391.1301H} is the result of an
incorrect color correction in the filter system. However, for this
work we used the exact same filters as in \citet{2004ApJ...617..746D}
obtaining similar results to \citeauthor{2008MNRAS.391.1301H}. Thus,
the most plausible explanation is that there is an excess of galaxies
at z$\sim$1.5 in our region.

For p-BzK, our results are lie between those of
\citet{2008MNRAS.391.1301H} and \citet{2010ApJ...708..202M}. However,
the counts of p-BzK exhibit the largest scatter of the three
populations. This is not surprising given that p-BzK target a more
constrained population, prone to stronger clustering
(\citealt{2006A&A...453..507G},
\citealt{2006PASJ...58..951K},\citealt{2008MNRAS.391.1301H},
\citealt{2010ApJ...708..202M}) and hence significantly affected by
LSS.  Table~\ref{colorpop_densities} summarizes the accumulated
surface densities of DRG and BzK galaxies up to
$K_{\mathrm{VEGA}}$$=$20 and $K_{\mathrm{VEGA}}$=21. The values are
roughly consistent with the results of the studies shown in
Figure~\ref{numbercounts}, and with other values from the literature
(0.89 DRG/arcmin$^{2}$ in \citealt{2007AJ....134.1103Q}; 3.1
s-BzK/arcmin$^{2}$ and 0.24 p-BzK/arcmin$^{2}$ in
\citealt{2005ApJ...633..748R};3.2 s-BzK/arcmin$^{2}$, 0.65
p-BzK/arcmin$^{2}$ in \citealt{2007A&A...465..393G}). As mentioned
above, the excess of s-BzK by a factor of $\sim$1.5 could be caused by
a source over-density in the area. Note that this excess does not
necessarily affect s-BzK and p-BzK in the same manner due to the
different clustering properties of each population
(\citealt{2008MNRAS.391.1301H},
\citealt{2010ApJ...708..202M},\citealt{2010MNRAS.407.1212H}). In fact,
our density of p-BzK is not among the lowest values.

\begin{figure}
\includegraphics[width=8.7cm,angle=0.]{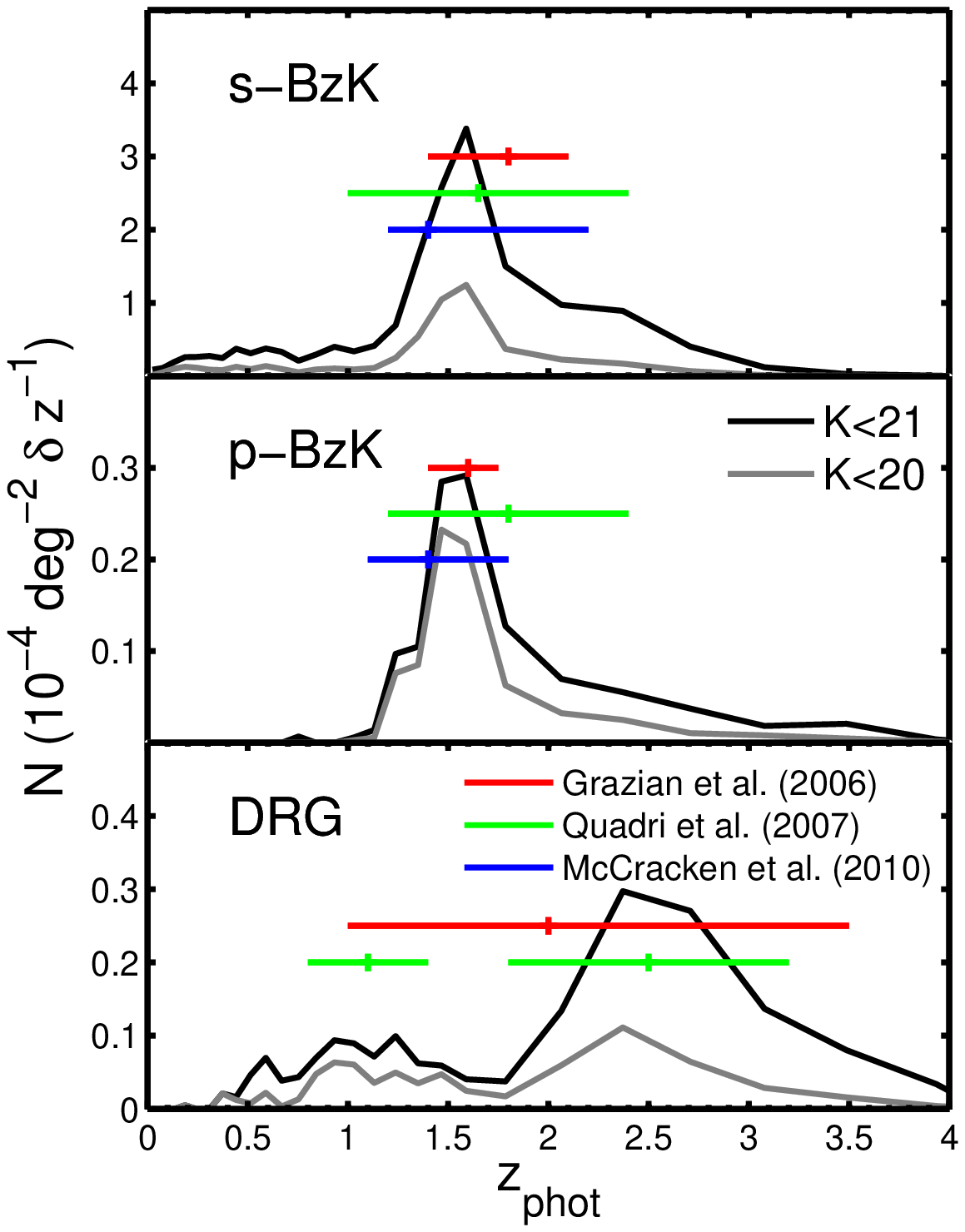}
\caption{\label{photozcolorpops} Photometric redshift distributions of
  s-BzK, p-BzK and DRG galaxies (from top to bottom) drawn from the
  IRAC-selected sample with [3.6]$<$23.75.  The black line and gray
  line show the distributions at $K_{\mathrm{VEGA}}$$<$21 and
  $K_{\mathrm{VEGA}}$$<$20, respectively. Our results are compared
  with the median (and quartile) of the redshift distributions in
  \citet[][red]{2007A&A...465..393G},
  \citet[][green]{2007AJ....134.1103Q} and
  \citet[][blue]{2010ApJ...708..202M} (also top to bottom). The two
  (green) intervals in the redshift distribution of DRG indicate the
  median values of the distribution at redshifts lower and higher than
  z$=$1.5 in the work \citet{2007AJ....134.1103Q}.}
\end{figure}

Figure~\ref{photozcolorpops} shows the $z_{\mathrm{phot}}$
distribution of DRG, s-BzK and p-BzK galaxies with
$K_{\mathrm{VEGA}}$~$<$20 and $K_{\mathrm{VEGA}}$~$<$21, compared to
some results from the literature. The distributions are convolved with
a $\delta z$=0.1 kernel in order to account for the
$z_{\mathrm{phot}}$ uncertainties. The redshift range spanned by the
different galaxy populations is in good agreement with the usual
distributions observed in other studies, i.e., z$>$2 for DRG and
1.4$<$z$<2.5$ for BzKs \citep{2004ApJ...617..746D}. DRG present a
secondary redshift peak around z$\sim$1, that accounts for a
significant fraction of the total population at bright
($K_{\mathrm{VEGA}}$$<$20) magnitudes (as already pointed out by other
authors, e.g., \citealt{2007AJ....134.1103Q},
\citealt{2007ApJ...660L..55C}).  Nevertheless, our surveyed area
(0.35\,deg$^{2}$) is not large enough to make (bright) low-z DRG the
dominant fraction, as in the 0.70\,deg$^{2}$ of the $K$-band Palomar
survey where $\sim$70$\%$ of these galaxies are found at z$<$1.4
\citet{2007ApJ...660L..55C}.  As expected, s-BzK and p-BzK present
almost identical redshift distributions, although the latter seems to
have a more extended high-redshift tail, being also less prone to low
redshift interlopers (probably as a consequence of the more
restrictive color criteria).  In summary, our results about the number
density and redshift distribution of color-selected z$>$1 galaxy
samples are consistent with previous studies
(\citealt{2005ApJ...633..748R}, \citealt{2007A&A...465..393G}; PG08),
indicating that the photometric redshift estimates are generally
robust at high-redshift.

\placetable{colorpop_densities}
\begin{deluxetable}{ccccc}
\setlength{\tabcolsep}{0.02in} 
\tablewidth{200pt}
\tabletypesize{\footnotesize}
\tablecaption{\label{colorpop_densities}Surface density of DRG and BzKs}
\tablehead{ & \multicolumn{2}{c}{$K_{\mathrm{VEGA}}$$<$21} & \multicolumn{2}{c}{$K_{\mathrm{VEGA}}$$<$20} \\
& \colhead{$\rho^{\mathrm{a}}$} & \colhead{$\tilde{z}^{\mathrm{b}}$} & \colhead{$\rho^{\mathrm{a}}$} & \colhead{$\tilde{z}^{\mathrm{b}}$}}
\startdata
DRG    &  1.4  &  2.47   &  0.5 & 2.24       \\

s-BzK  &  5.0  &  1.89   &  1.5 & 1.70       \\

p-BzK  &  0.5  &  1.85   &  0.3 & 1.73         
\enddata
\tablecomments{\\
$^{\mathrm{a}}$ Surface density of DRG and BzK in arcmin$^{-2}$.\\
$^{\mathrm{b}}$ Median photometric redshift of each sub-sample.}
\end{deluxetable}

\subsection{Photometric redshift distribution}
\label{zdistrib}

\begin{figure}
\includegraphics[width=8.7cm,angle=0.]{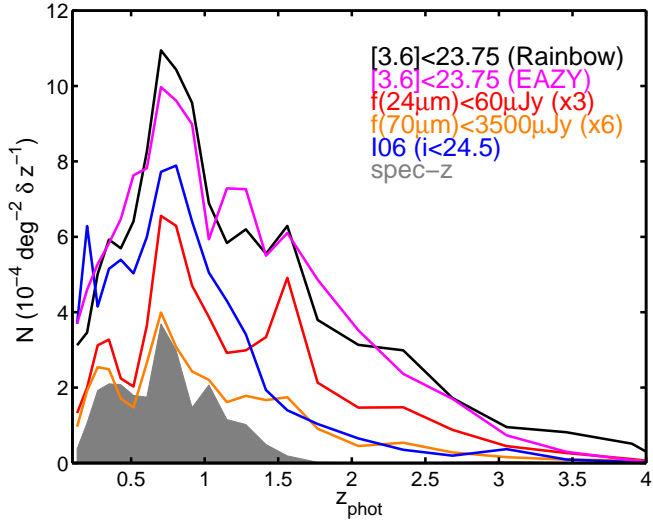}
\caption{\label{photoz_distrib} Photometric redshift distribution for
  the IRAC ([3.6]$<$23.75; black), MIPS 24~$\mu$m ($\times$3, red),
  MIPS 70~$\mu$m ($\times$6, orange) and spectroscopic (grey area)
  samples in the main region of the EGS.  The distribution of
  z$_{\mathrm{phot}}$ in I06 ($i'<$24.5) for the overlapping area with
  the IRAC mosaic is shown in blue for comparison.}
\end{figure}

Figure~\ref{photoz_distrib} shows the $z_{\mathrm{phot}}$
distribution for the IRAC selected sample in the main region, limited
to [3.6]$<$23.75. In addition, we also plot the redshift distributions
of the galaxies detected at 24~$\mu$m, 70~$\mu$m and the sub-sample with
spectroscopic redshifts.  In order to derive a realistic distribution,
accounting for the uncertainties in z$_{\mathrm{phot}}$, the
distribution was convolved with the typical width of the zPDFs. We
used a conservative upper limit of $\Delta z/(1+z)$=0.07. The shape of
the distribution is consistent with that expected for a magnitude
limited sample. At low redshift the number density increases as we
probe larger volumes, and then an exponential decay is observed as the
sources get fainter and the detection probability decreases.

The position of the minor prominences in the z$_{\mathrm{phot}}$
distribution are roughly consistent with the most remarkable peaks
observed in the spectroscopic redshift distribution at z$\sim$0.3,
z$\sim$0.7, z$\sim$1. The median redshift of the photometric redshift
distribution is $z$$=$1.2, 75$\%$ of the sources are below z$=$2.1,
and 90$\%$ below z$=$2.7.  The median of the distribution is
consistent with the results of \citet{2009ApJ...690.1236I} in the
COSMOS field for an i'+3.6$\mu$m selected sample (i'$<$25,
f(3.6)$>$1\,$\mu$Jy).  Although the IRAC S-COSMOS catalog is shallower
than ours, with a $\sim$50$\%$ completeness level at
f(3.6)$=$1\,$\mu$Jy, the median redshift limited to their faintest
magnitude bin (24.5$<$$i'$$<$25), $z=$1.06, is similar to ours. Note
that the small differences could arise from the presence of underlying
LSS in EGS, whereas this effect is largely reduced in the COSMOS
sample due the larger area of field ($\sim$1.73\,deg$^{2}$).  Finally,
the distribution is also in good agreement with our results in PG08
for the averaged redshift distribution of a combination of
IRAC-selected catalogs in the HDFN, CDFS and Lockman Hole fields. The
total combined area in PG08 is approximately that of the main region
of EGS, and the limiting magnitude of the catalog was slightly lower
(f(3.6)$<$1.6\,$\mu$Jy).  However, the redshift distribution also
peaks around z$=$0.8-1, consistently with ours.


\section{SED analysis: stellar masses}
\label{evalmass}

\begin{figure*}
\centering
\includegraphics[width=8.7cm,angle=0.]{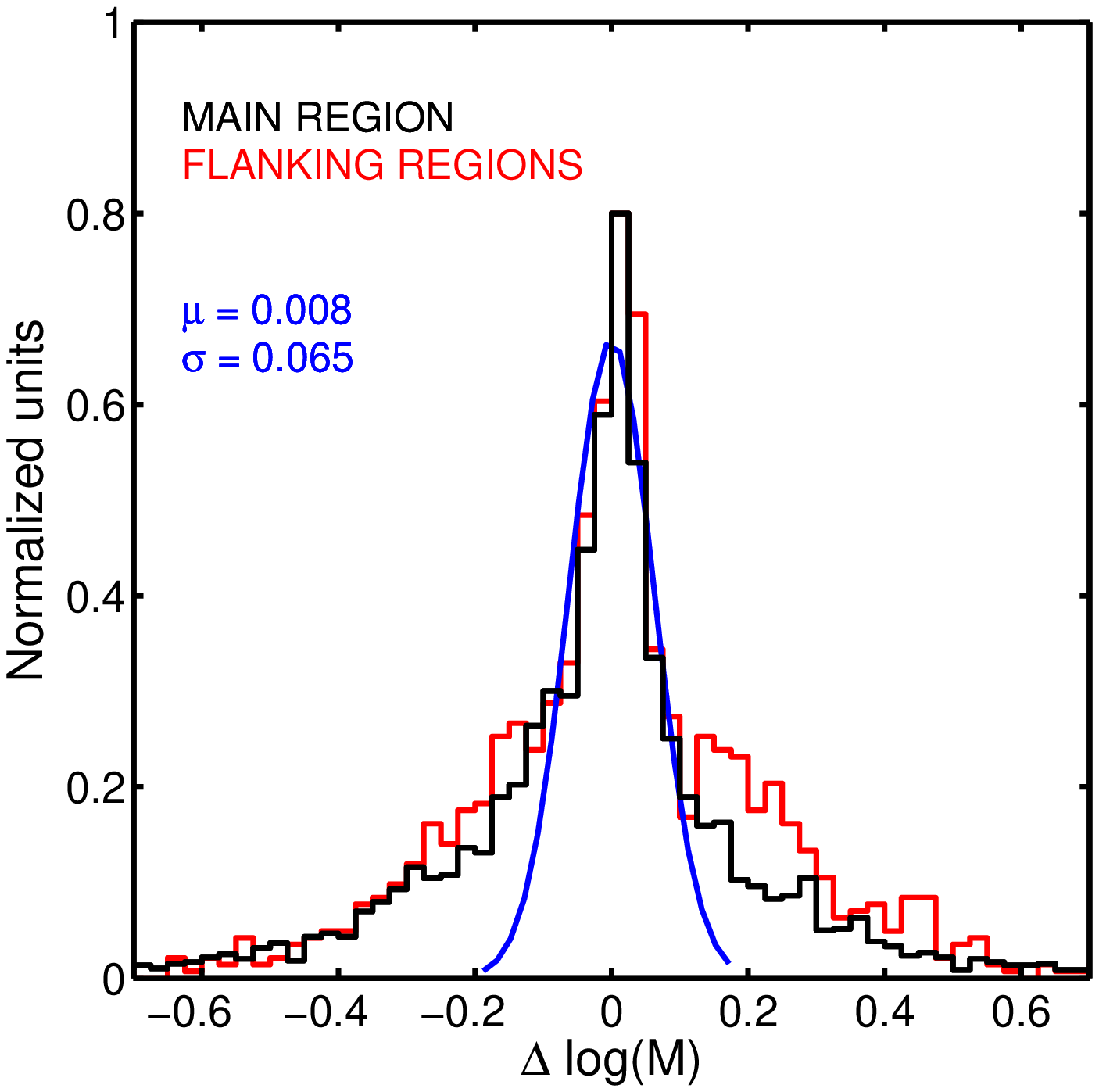}
\includegraphics[width=9cm,angle=0.]{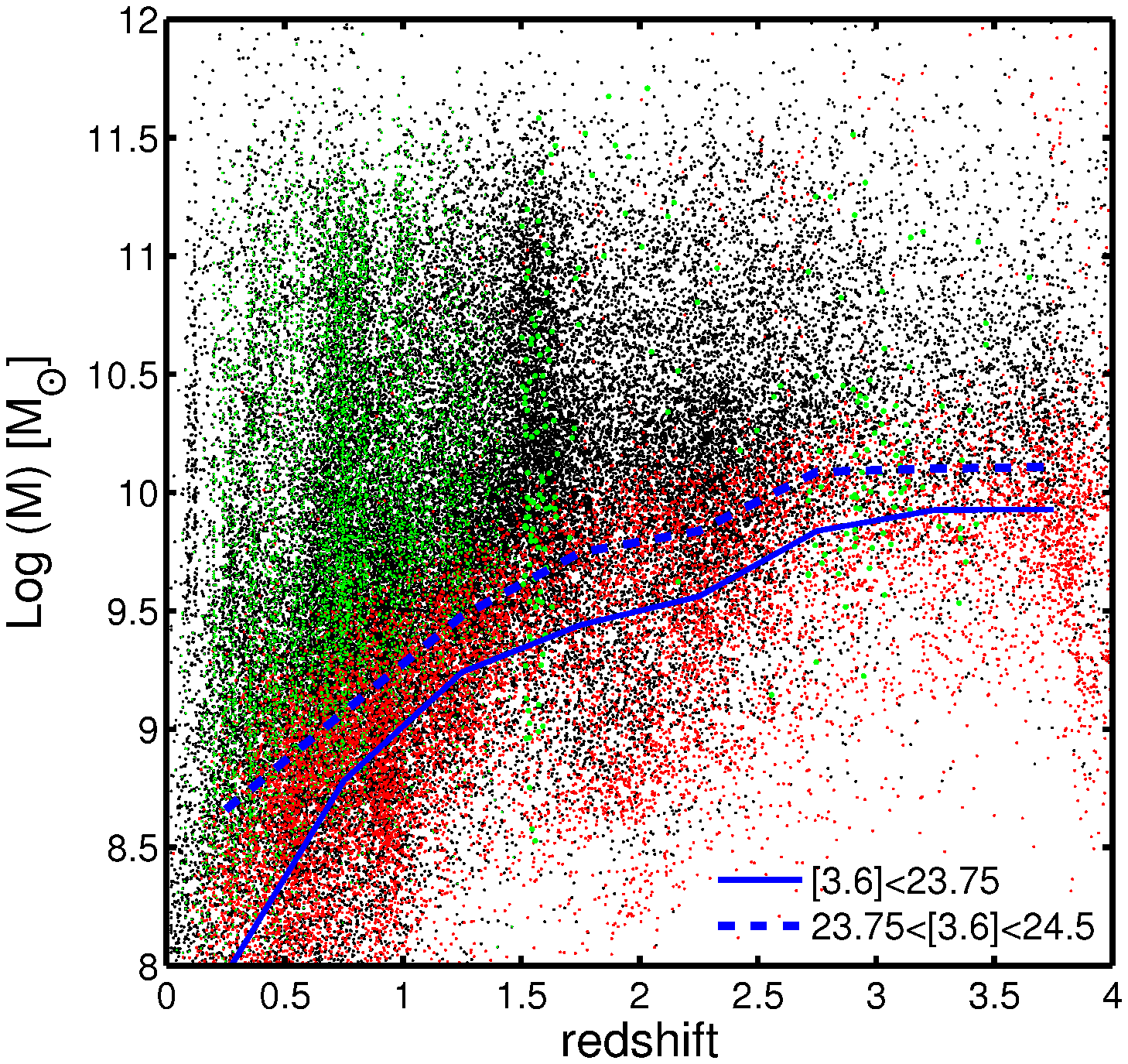}
\caption{\label{massesalone} {\it Left}: Comparison of the stellar
  masses estimated with [P01,SALP,CAL01] using spectroscopic and
  photometric redshifts in the main (black) and flanking (red)
  regions, respectively. The blue line indicates the best Gaussian fit
  to the central values of the main region distribution. {\it Right}:
  Stellar masses of the galaxies in the sample as a function of
  redshift.  The black dots depict galaxies with [3.6]$<$23.75 (85\%
  completeness level of the sample). The red dots depict galaxies
  23.75$<$[3.6]$<$24.75 (3$\sigma$ limiting magnitude). The green dots
  show galaxies with spectroscopic redshifts. The blue lines indicate
  the 90\% and 10\% percentiles of the mass distribution as a function
  of redshift for the galaxies with [3.6]$<$23.75 (solid) and
  23.75$<$[3.6]$<$24.75 (dashed), respectively.}
\end{figure*}

\begin{figure*}
\centering
\includegraphics[width=8.7cm,angle=0.]{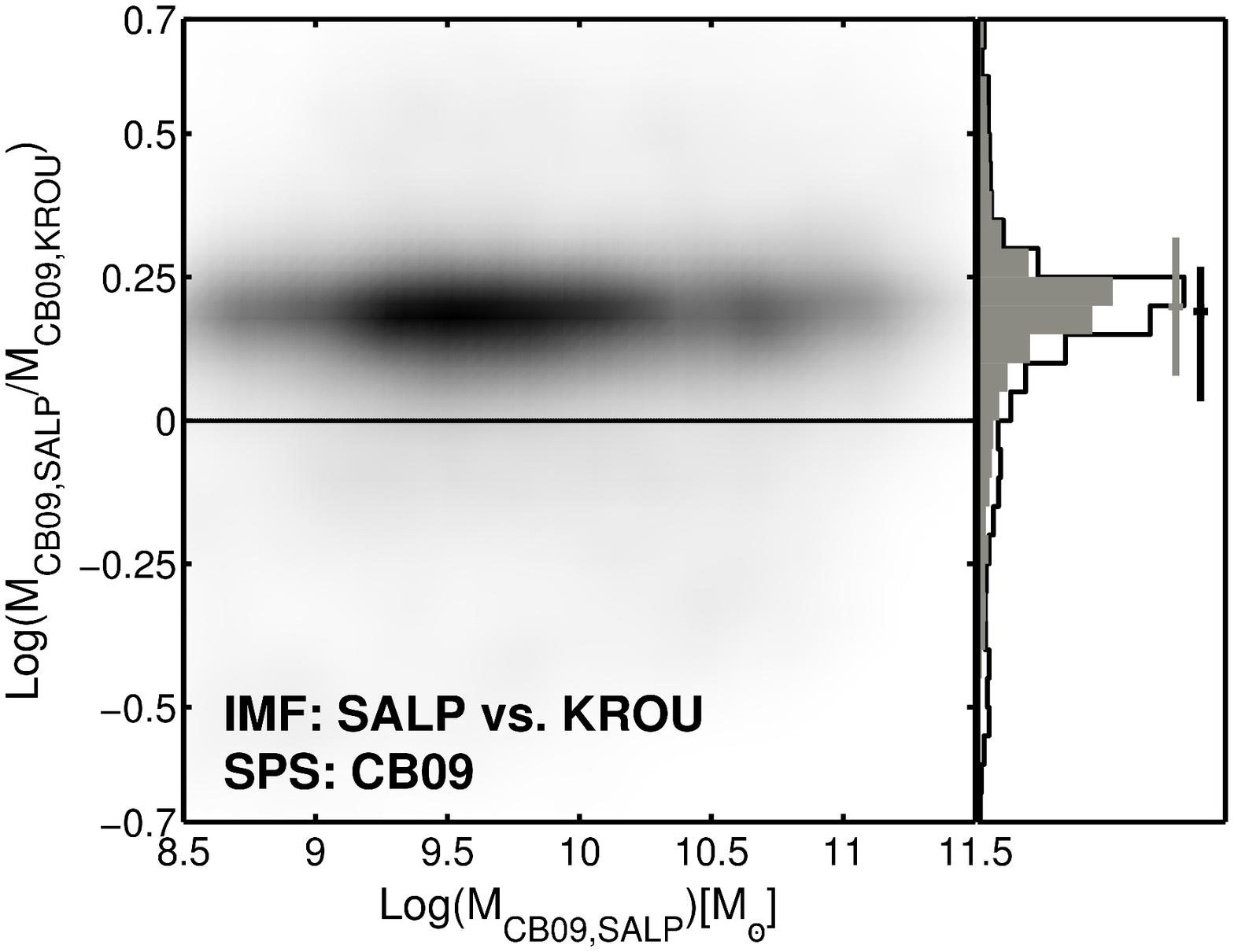}
\hspace{0.2cm}
\includegraphics[width=8.7cm,angle=0.]{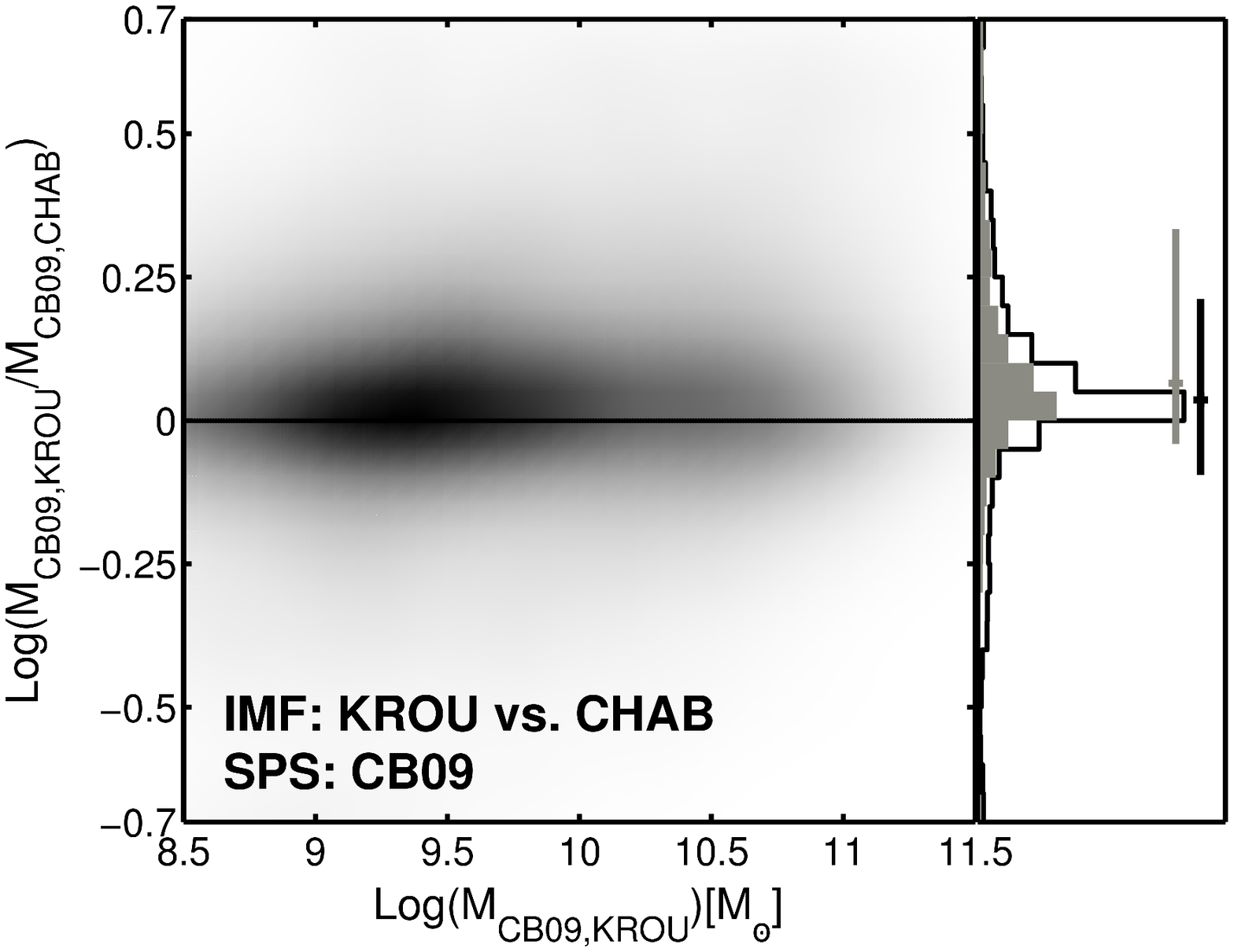}\\
\includegraphics[width=8.7cm,angle=0.]{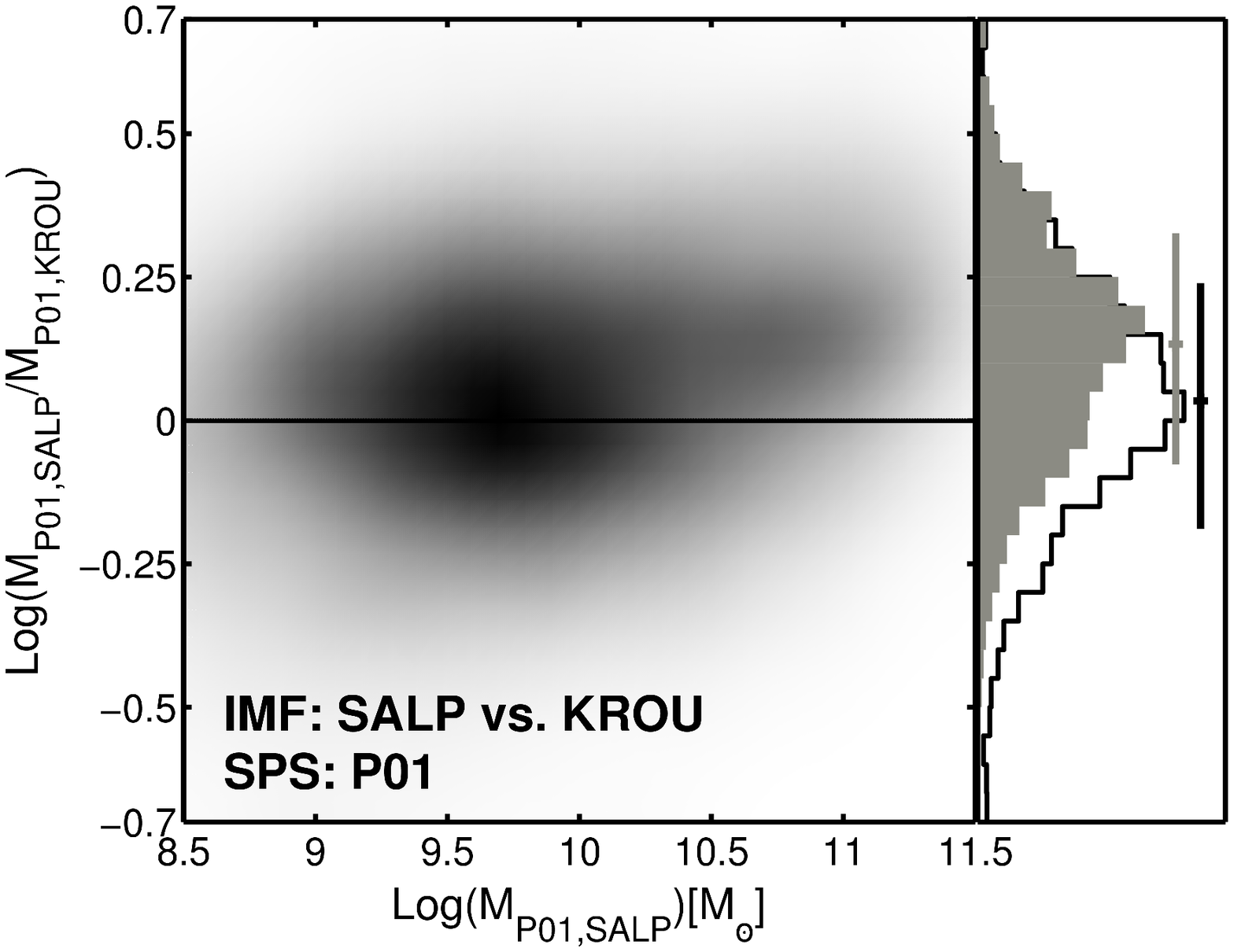}
\hspace{0.2cm}
\includegraphics[width=8.7cm,angle=0.]{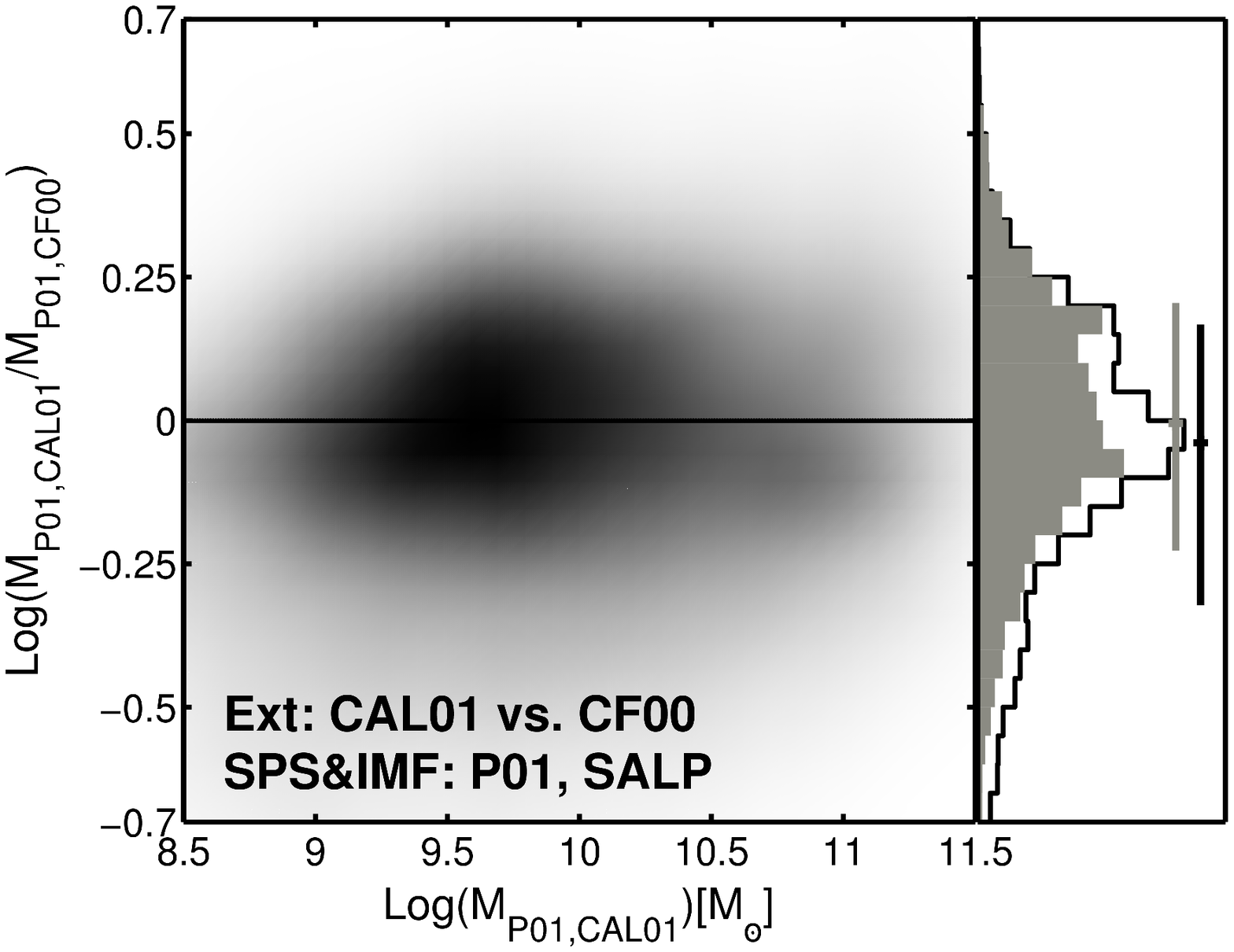}
\caption{\label{massesfig} Comparison of the stellar masses obtained
  using different IMFS and dust extinction laws for a given SPS
  library. Top-left: P01 models with a SALP and KROU IMFs. Top-right:
  CB09 models with a SALP and KROU IMFs. Bottom: CB09 models with a
  KROU and CHAB IMFs. Bottom-right: P01 models with a CAL01 and CF00
  extinction laws. The histograms in the right part of the plot depict
  the ratio of the stellar masses obtained with each IMF for galaxies
  with $M_{\mathrm{model}}$$<$10$\mathcal{M}_\sun$ (empty) and
  $M_{\mathrm{model}}$$>$10$\mathcal{M}_\sun$ (filled). The solid
  lines above the histogram show the median value and 1$\sigma$ of the
  distribution at both sides of the median.}
\end{figure*}

\begin{figure*}
\centering
\includegraphics[width=8.7cm,angle=0.]{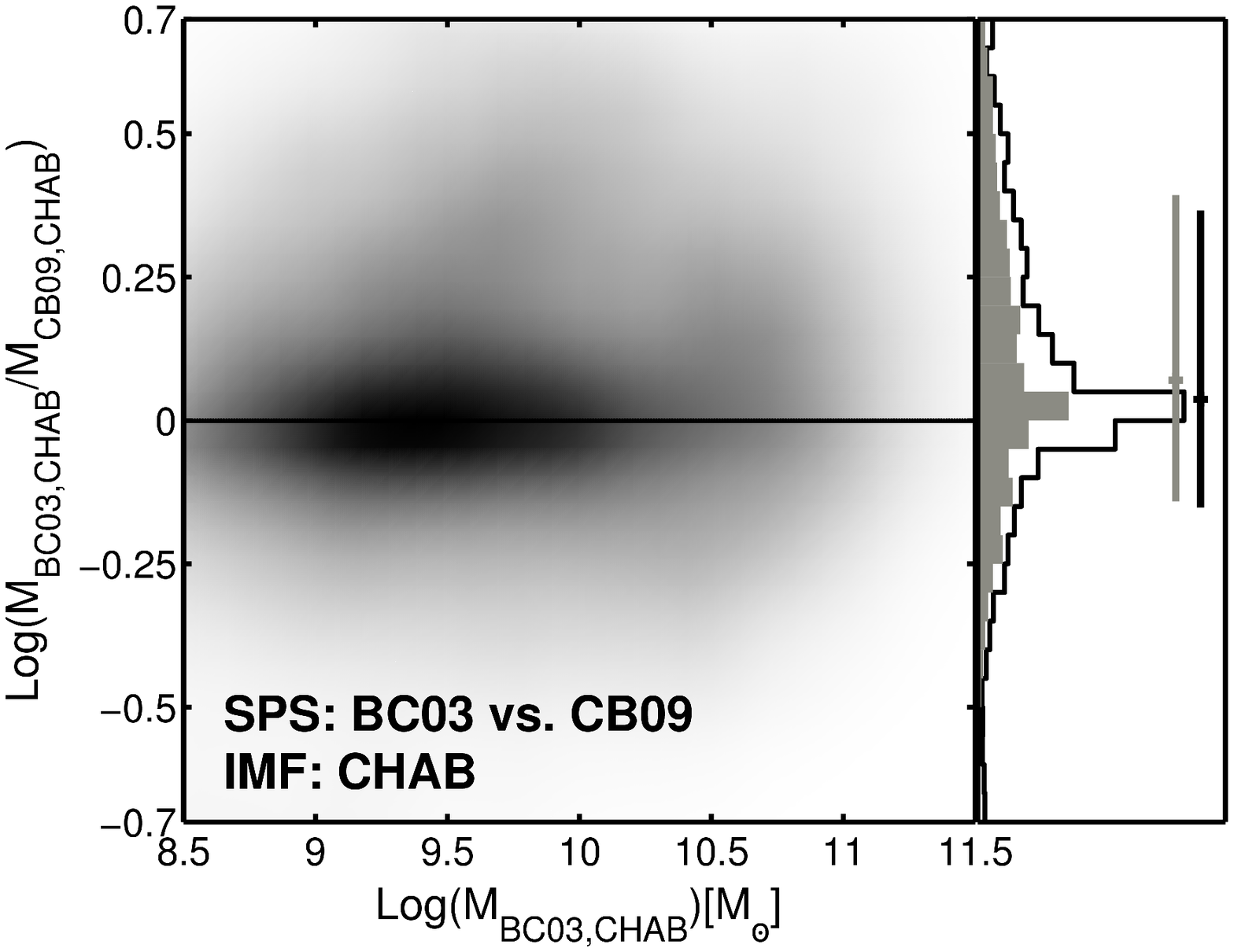}
\hspace{0.2cm}
\includegraphics[width=8.7cm,angle=0.]{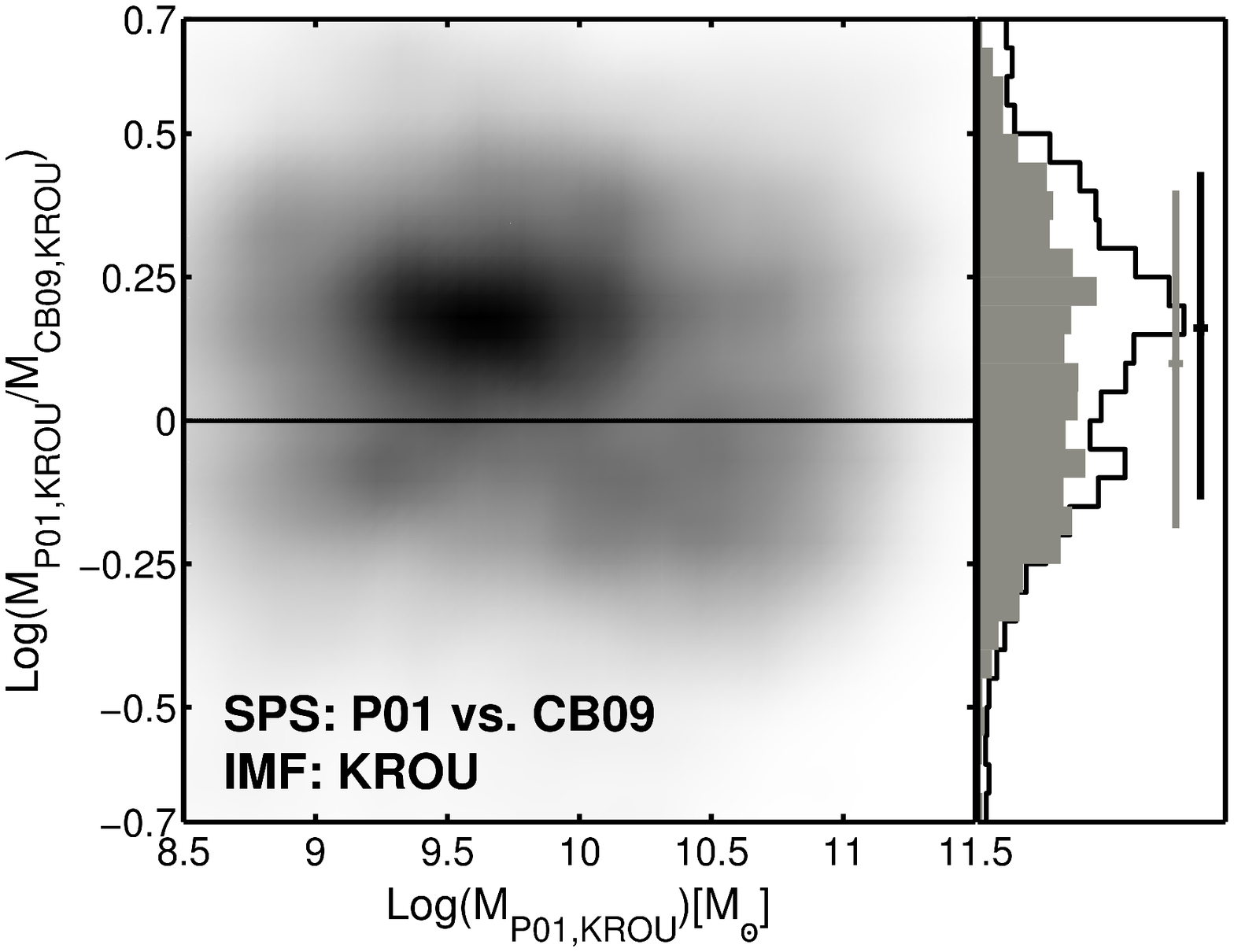} \\
\includegraphics[width=8.7cm,angle=0.]{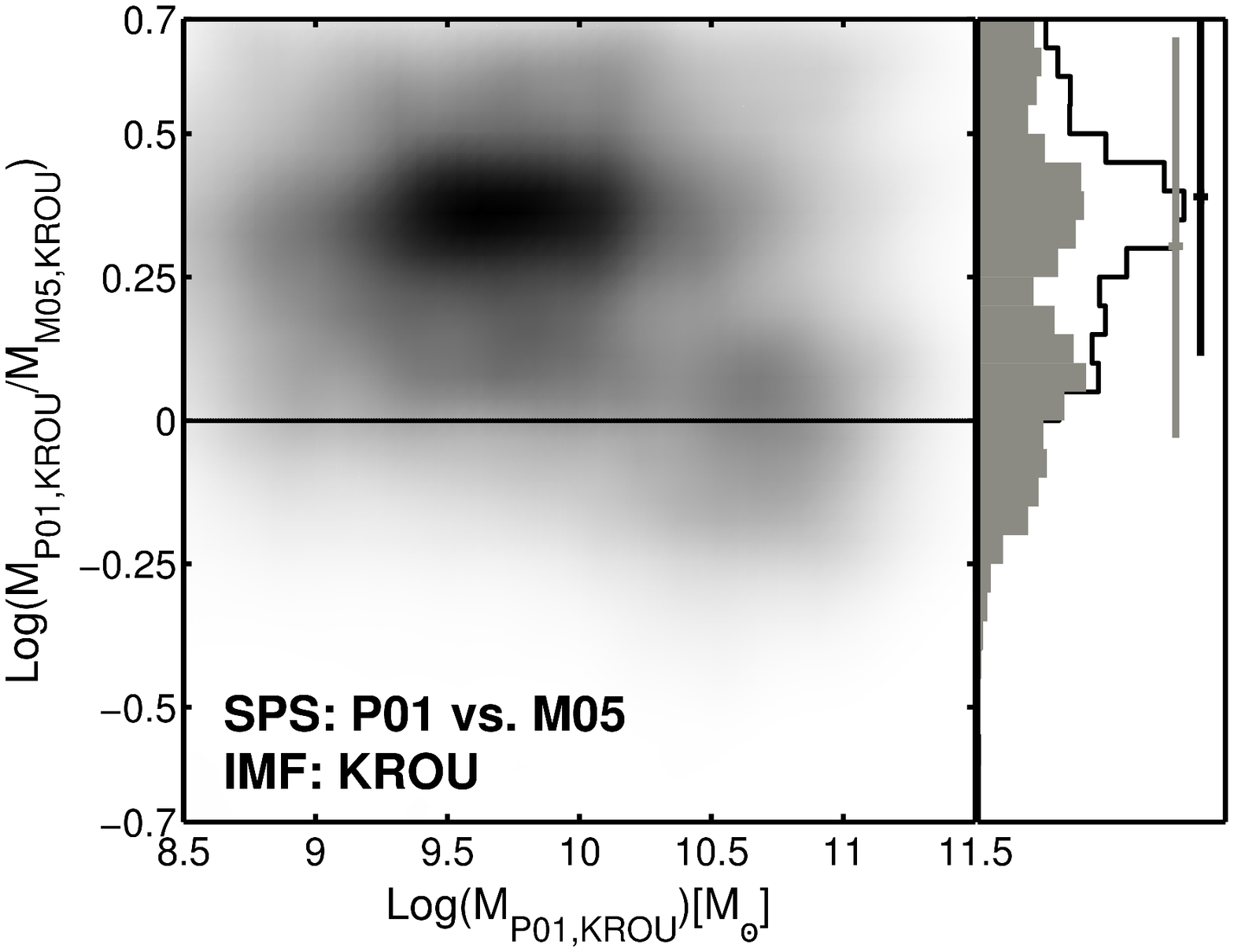}
\hspace{0.2cm}
\includegraphics[width=8.7cm,angle=0.]{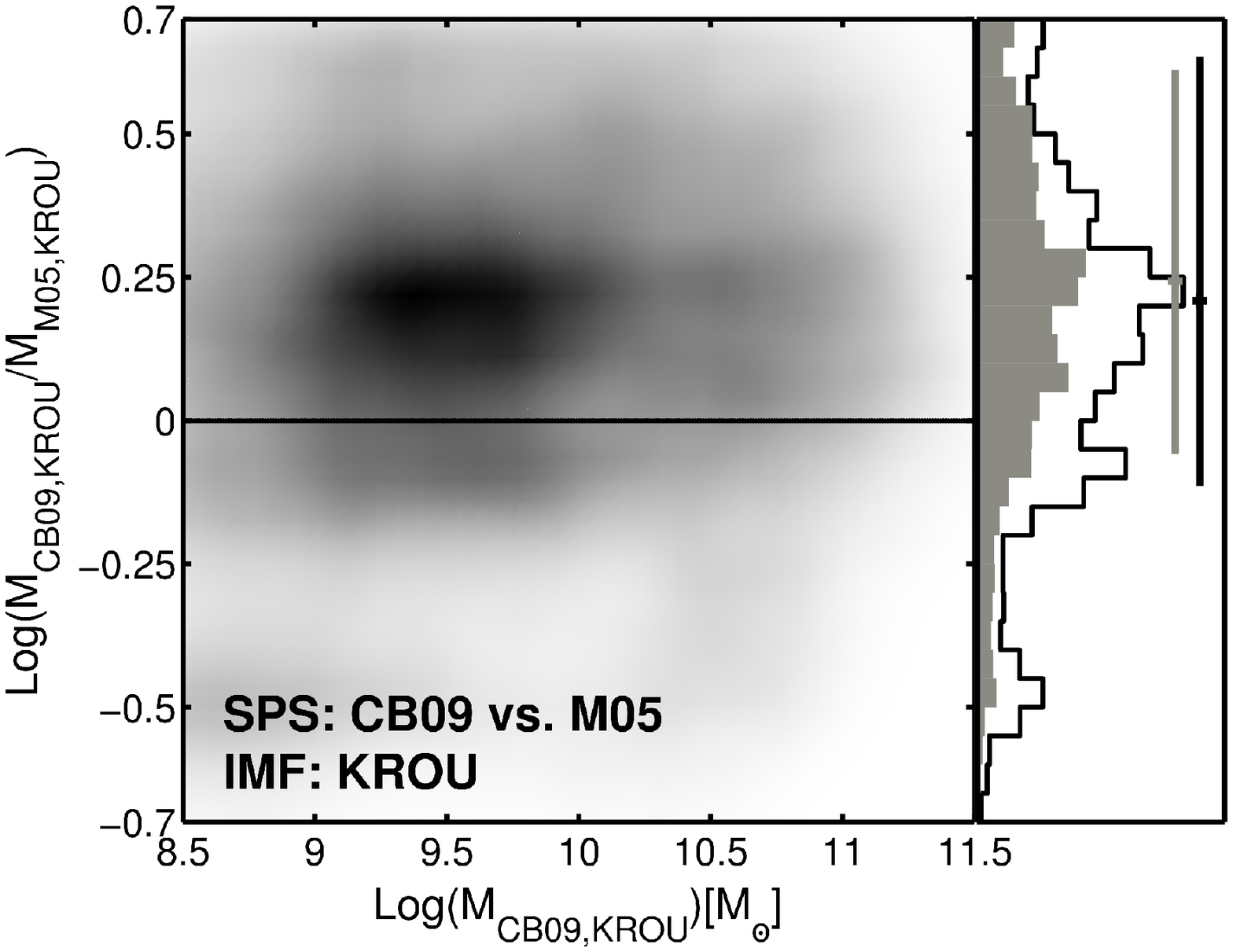}
\caption{\label{massesfig2} Comparison of the stellar masses obtained
  using different SPS models with the same IMF. Top-left: BC03 and
  CB09 models with a CHAB IMF. Top-right: P01 and CB09 models with a
  KROU IMF. Bottom-left: P01 and M05 models with a KROU
  IMF. Bottom-right: CB09 and M05 models with a KROU IMF.  The
  histograms in the right part of the plot depict the ratio of the
  stellar masses obtained with each model for galaxies with
  $M_{\mathrm{model}}$$<$10$\mathcal{M}_\sun$ (empty) and
  $M_{\mathrm{model}}$$>$10$\mathcal{M}_\sun$ (filled). The solid
  lines above the histogram show the median value and 1$\sigma$ of the
  distribution at both sides of the median.}
\end{figure*}

In this section we describe the method used to estimate stellar masses
based on the SED fitting. In addition, we analyze the goodness of our
stellar mass estimations quantifying the systematic and random errors
linked to assumptions in the input parameters for the stellar
population modeling. For the discussion in \S~\ref{masstest}, we use
only the spectroscopic sample in the main region (which count with
better photometry) and we force the z$_{\mathrm{phot}}$ to the
spectroscopic value.

\subsection{Stellar mass estimates}
\label{masses}

The stellar mass of each galaxy is estimated from the
wavelength-averaged scale factor required to match the template
monochromatic luminosities to the observed fluxes. This is possible
because our templates are obtained from stellar population synthesis
models which are expressed in energy density per stellar mass
unit. Note that the stellar mass estimate is not obtained from a
single rest-frame luminosity and its corresponding mass-to-light
ratio, which has been a typical procedure seen in the literature, but
from the whole SED. In our method, the fit to the multi-band data
implicitly constraints the mass-to-light ratio by determining the
most suitable template. Then, we estimate the mass from the averaged
template normalization, weighted with the photometric errors.  This
approach is less sensitive to the effects of the star-formation
history (SFH) or the photometric and template uncertainties in a
single band.  Objects fitted with pure AGN templates have no stellar
mass estimate, as their SED is dominated by non-stellar emission. The
random uncertainty of the stellar mass is estimated with a bootstrap
method by randomly varying the photometric redshift and observed
fluxes based on their quoted errors.

\subsection{Accuracy of the stellar masses}
\label{masstestintro}

\placetable{photozquality}
\begin{deluxetable*}{lcccc}
\centering
\tabletypesize{\footnotesize}
\setlength{\tabcolsep}{0.05in} 
\tablewidth{0pt}
\tablecaption{\label{masstable}Comparison of stellar masses computed with different modeling assumptions}
\tablehead{
\colhead{IMFs} &   \colhead{SPS model} & \colhead{Dust}  &  \colhead{ $\Delta\log$(M)} &   \colhead{ $\Delta\log$(M)}\\
(1)&(2)&(3)&(4)&(5)}
\startdata
SALP$-$KROU  & CB09&   CAL01 &  0.19$\pm^{0.07}_{0.12}$& 0.19$\pm^{0.12}_{0.10}$\\ 
SALP$-$KROU  &  P01&   CAL01 &  0.03$\pm^{0.20}_{0.17}$& 0.13$\pm^{0.19}_{0.21}$ \\ 
KROU$-$CHAB  & CB09&   CAL01 &  0.04$\pm^{0.11}_{0.09}$& 0.07$\pm^{0.24}_{0.10}$ \\ 
\hline
 SPS model & IMF & Dust & $\Delta\log$(M) & $\Delta\log$(M) \\
\hline\\
BC03$-$CB09  & CHAB&   CAL01 &  0.04$\pm^{0.28}_{0.15}$& 0.07$\pm^{0.30}_{0.21}$ \\ 
P01$-$CB09   & KROU&   CAL01 &  0.15$\pm^{0.23}_{0.29}$& 0.08$\pm^{0.28}_{0.27}$ \\ 
P01$-$M05    & KROU&   CAL01 &  0.39$\pm^{0.36}_{0.28}$& 0.30$\pm^{0.35}_{0.27}$ \\ 
CB09$-$M05   & KROU&   CAL01 &  0.16$\pm^{0.26}_{0.28}$& 0.20$\pm^{0.27}_{0.29}$ \\ 
\hline
     Dust  & IMF & SPS model & $\Delta\log$(M) & $\Delta\log$(M) \\
\hline\\
CAL01$-$CF00       & SALP&      P01&  -0.03$\pm^{0.20}_{0.23}$& 0.00$\pm^{0.20}_{0.21}$  \\
\enddata
\tablecomments{ Comparison of the stellar masses obtained under different combinations of the
modeling assumptions.\\
(1),(2),(3) SPS model, IMF and dust extinction law, alternatively. The first column indicate the parameters being compared.\\
(4) Log of median value and quartiles of the difference for galaxies with $\log$(M)$<$10$\mathcal{M}_\sun$.\\
(5) Same as (4) for galaxies with $\log$(M)$>$10$\mathcal{M}_\sun$ }
\end{deluxetable*}

In addition to the uncertainties inherited from the probabilistic
nature of $z_{\mathrm{phot}}$ and the intrinsic photometric errors,
there is another source of systematic uncertainty associated to the
assumptions in the SED modeling. Although significant effort has gone
into providing accurate SPS models, key ingredients of the theoretical
predictions are still poorly understood. As a result, there can be
substantial differences in the physical properties estimated with many
of the well-tested SPS models available in the literature. Most of
these differences arise from the different parametrizations of
potentially uncertain phases of the stellar evolution, such us the
asymptotic giant branch (AGB) or the thermally pulsating AGB
(\citealt{2005MNRAS.362..799M}, \citealt{2007ASPC..374..303B},
\citealt{2007ApJ...657L...5K}) Another critical aspect, is the choice
of an IMF. Although this is essentially assumed to introduce a change
in the overall normalization of the stellar mass, there are additional
effects attached, e.g., a change in the balance between low-mass and
high-mass stars varies the relative fraction of stars in different
points of the isochrones. Thus modifying the colors and M/L of the
modeled galaxies at different evolutionary stages
(\citealt{1998MNRAS.300..872M}, \citealt{2008ApJ...677L...5V}).

Apart from the choice of SPS models and IMF, additional effects might
arise from the assumed SFH, usually parametrized with $\tau$-models,
or the choice of a dust extinction law and metallicity. As recently
shown in \citet{2009ApJ...701.1839M} (also
\citealt{2008ApJ...677..219K} or \citealt{2009ApJ...701.1765M}), due
to all these effects, the physical properties of galaxies estimated
from broad-band photometry often presents large uncertainties
(typically within a scatter of 0.2\,dex for stellar masses), in
addition to systematic offsets. Moreover, these uncertainties can be
even larger (up to 0.6\,dex) for particularly sensitive galaxy
populations at high-z, such as bright red galaxies. See for example
the series of paper by
\citet{2009ApJ...699..486C,2010ApJ...708...58C,2010ApJ...712..833C}
for a detailed discussion of all these issues.

Taking these considerations into account, in the following Sections,
we analyze the accuracy of our stellar mass estimates quantifying the
uncertainty budgets associated with different effects.  First, we
study the effect of photometric redshift uncertainties. Then, we
evaluate the impact from the choice of SPS models, IMF and dust
extinction law restricting the analysis to the spectroscopic
sample. For the sake of clarity, we refer all comparisons to a default
choice of SED modeling parameters (as described in \ref{templates})
characterized by SPS models, IMF and extinction law
[P01,SALP,CAL01]. Finally, we verify that our stellar masses provide
realistic values by comparing them to other stellar mass catalogs
available in the literature.

Note that although the {\it a priori} assumptions on the SFH can also
introduce systematic effects in the estimated stellar masses, an in
depth analysis of these issues is clearly beyond the scope of this
paper (see e.g., \citealt{2010MNRAS.407..830M} for a detailed
discussion). Nonetheless, a comparison of the results obtained with a
single exponentially-declining stellar population (1-POP) and with a
single population plus a second burst (2-POP) is presented in PG08
(Appendix B) along with similar tests to the ones presented in the
next section.

The catalog of stellar masses presented in this paper (see
\S~\ref{datacatalogs}) contains the different values obtained with all
the modeling configurations discussed in the next sections.

\subsubsection{Effects of the Photometric redshifts, SPS models, IMF and
  extinction law}
\label{masstest}

The left panel of Figure~\ref{massesalone} shows the scatter in the
stellar masses estimated using $z_{\mathrm{phot}}$ and
$z_{\mathrm{spec}}$ for the 7,636 spectroscopic galaxies in the main
(black) and flanking regions (red). Approximately 68\% and 90\% of the
sources are confined within a {\it rms} of 0.16, 0.34\,dex and 0.20,
0.39\,dex in each region, respectively. Nonetheless, the distribution
shows a pronounced central peak that it is well reproduced by a
Gaussian distribution (blue line) with extended wings, indicating that
for the most accurate redshifts, the scatter is substantially reduced
($\sim$0.065~dex).  This is in good agreement with the results of
\citet[][see Figure 3]{2010ApJ...709..644I} scaled to the overall
accuracy of our photometric redshifts, which is slightly lower. The
right panel of Figure~\ref{massesalone} shows the range of stellar
masses as a function of redshift for the whole sample (black). In
order to illustrate the approximate limiting stellar mass inherited
from the magnitude limit ([3.6]$<$23.75; 85\% completeness), we also
depict the galaxies up to the 3$\sigma$ limiting magnitude
(23.75$<$[3.6]$<$24.75, red dots). Approximately 90\% of the galaxies
with [3.6]$<$23.75 present log(M)$>$10$M_{\odot}$ at z$\gtrsim$2.5
(blue line) in agreement with our results in PG08 for a similar
limiting magnitude. Similarly, $\sim$10\% of the faintest galaxies
(23.75$<$[3.6]$<$24.75), absent in our main sample, present stellar
masses larger than log(M)$>$10$M_{\odot}$ (blue dashed line). Note
however that the completeness in stellar mass can not be directly
extraploted from these limits because for any given redshift, galaxies
with different ages present different mass-to-light ratios. Hence, the
completeness is an age (or color) dependent value. In particular,
magnitude limited samples are known to be incomplete against the
oldest (red) galaxies (see e.g., \citealt{2006A&A...459..745F}). A
detailed analysis of the completeness limit as a function of the
galaxy type will be included in a forthcoming paper.

The first test on the effect of the SED modeling assumptions consist
on a comparison of the stellar masses computed with three different
choices of the IMF; a SALP , \citet{2001MNRAS.322..231K} and
\citet[][hereafter KROU,CHAB]{2003PASP..115..763C} IMFs. The naive
expectation is that, the stellar masses obtained with a SALP IMF are
on average larger than those obtained with the other two, as it
predicts a larger number of low-mass stars. On the contrary, the IMFs
of KROU and CHAB are quantitatively very similar and therefore the
differences are expected to be small and mass
independent. Table~\ref{masstable} summarizes the median value and
quartiles of the comparison of stellar masses obtained with each IMF
in combination with the P01 and Charlot \& Bruzual (2009; CB09) models
and a CAL01 extinction law against the reference values. The top
panels in Figure~\ref{massesfig} show this comparison for SALP or KROU
IMFs (left) and a KROU or CHAB IMFs (right) and the CB09 models. In
both cases the difference is essentially a constant value of factor
$\sim$1.6 and 1.2, respectively, consistently with the results of the
literature (e.g., \citealt{2009AIPC.1111..207S},
\citealt{2009ApJ...701.1839M}, \citealt{2009ApJ...701.1765M}). On the
contrary, the difference in the values obtained with a SALP or KROU
IMFs for the P01 models is significantly smaller than the for the CB09
models, showing also a larger scatter and a dependence on the stellar
mass. This effect seems to be related with a difference in the age
dependency of the mass-to-light ratio for each IMF in these particular
models (see e.g., \citealt{1998MNRAS.300..872M},
\citealt{2008ApJ...677L...5V} for a description of these effects).

\begin{figure*}
\centering
\includegraphics[width=8.cm,angle=0.]{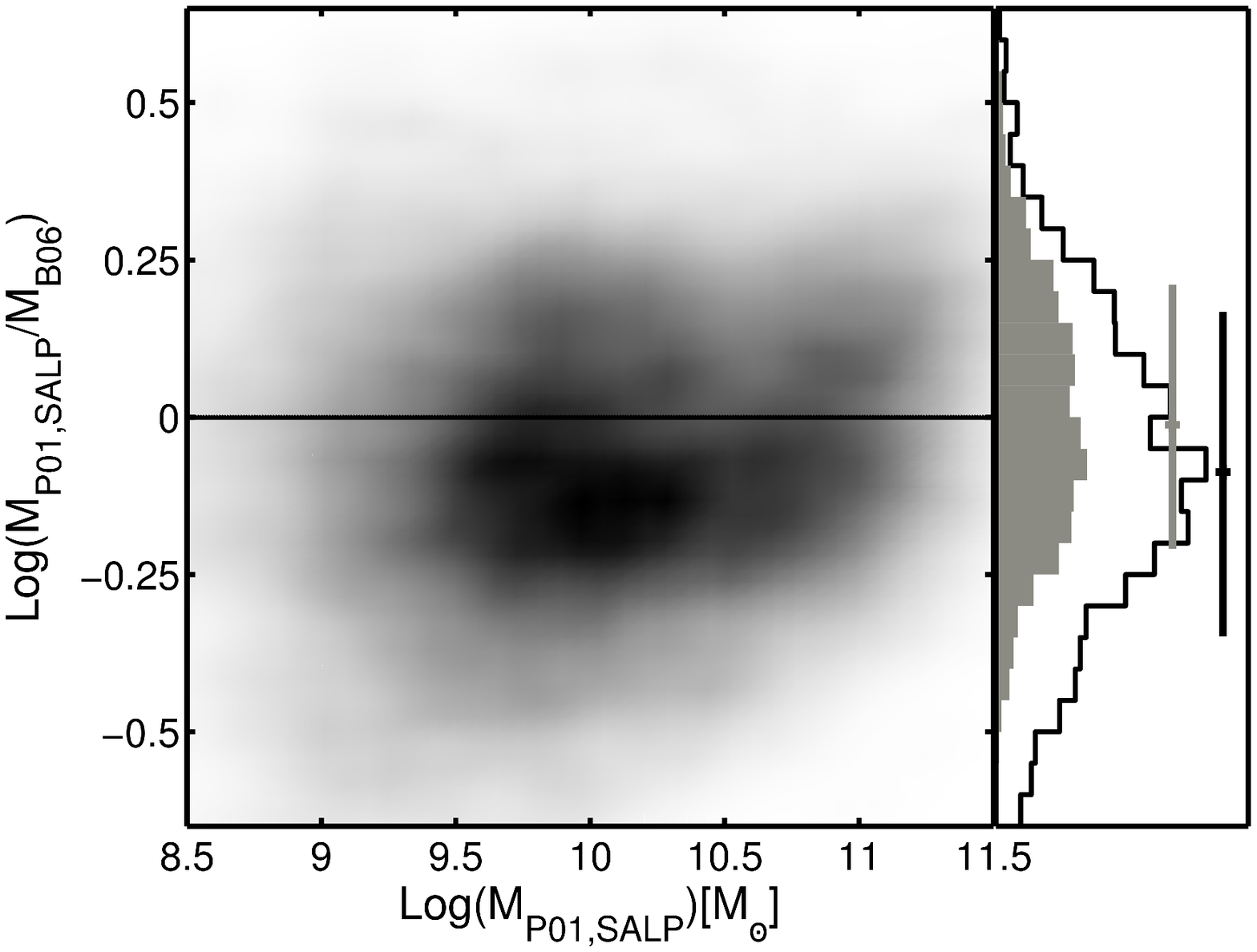}
\hspace{0.5cm}
\includegraphics[width=8.cm,angle=0.]{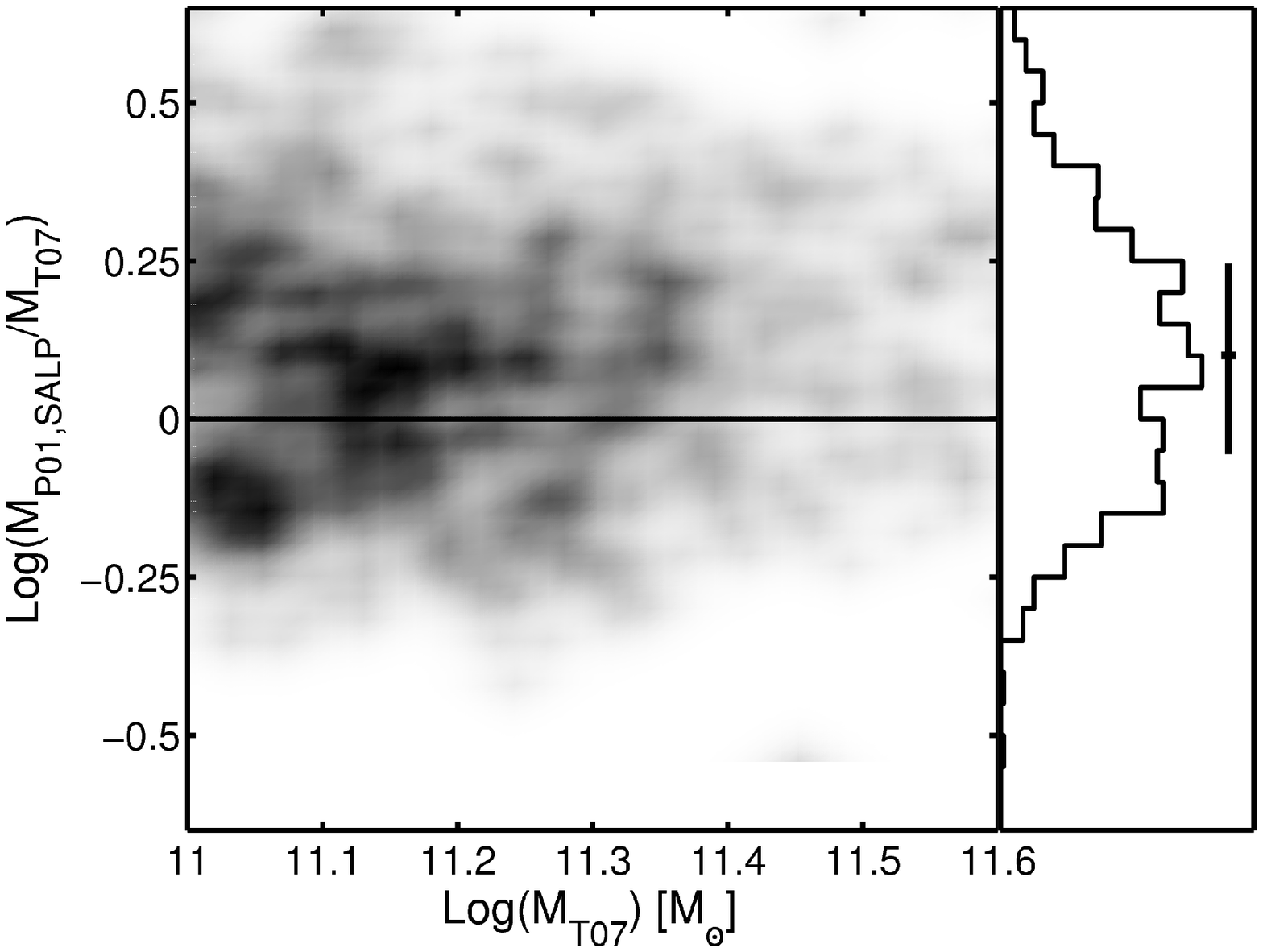}
\caption{\label{masses_bundy} Difference between our best-fit stellar
  mass using [P01,SALP] and the stellar masses of
  \citet{2006ApJ...651..120B}.  (left), and the mass limited sample
  (log~M$>$11\,$\mathcal{M}_\sun$) of \citet{2007MNRAS.382..109T}. The
  histograms in the right side are the same as in
  Figure~\ref{massesfig}. Note that the comparison to Trujillo et
  al. is limited to log~M$>$11\,$\mathcal{M}_\sun$.}
\end{figure*}

The second test on the modeling parameters is the comparison of
stellar masses obtained with the models of P01,
\citet[][BC03]{2003MNRAS.344.1000B},
\citet[][M05]{2005MNRAS.362..799M} and CB09 for the same IMFs. The
models of M05 were the first to account for the contribution of the
TP-AGB phase in the SPSs, a consideration that is expected to lead to
lower stellar masses compared to those obtained with P01 and
BC03. However, this difference should be reduced in the CB09 models,
the updated version of BC03, which include an improved treatment of
this particular phase.  The overall results of the comparison between
models is also summarized in Table~\ref{masstable} and in the panels
of Figure~\ref{massesfig2}.

Interestingly, we find that the difference between [BC03,CHAB] and
[CB09,CHAB] is relatively small, $\sim$0.04\,dex, and mostly
independent of the stellar mass. This suggests that, at least for the
present sample, taking into account the TP-AGB phase does not
introduce significant differences. A possible explanation could be
that, since the spectroscopic sample consist mostly on z$<$1 galaxies,
the available photometric coverage is not probing the rest-frame NIR
with sufficient detail. Only at higher redshifts (z$\gtrsim$1) the
IRAC bands would start probing the region of the SED that heavily
affected by the TP-AGB phases. Note also that the spectroscopic sample
analyzed here might not be a critical population to constraint the
effect of the TP-AGB, as for example the post-starburst galaxies
studied in \citet{2010ApJ...708...58C}.

In addition, we find that the estimates with [P01,KROU] are larger
than those obtained with [CB09,KROU] and [M05,KROU] with an average
offset of 0.15\,dex and 0.39\,dex, respectively. The difference with
respect to M05 is consistent with previous results (e.g.,
\citealt{2006ApJ...652...85M}, \citealt{2006ApJ...652...97V},
\citealt{2007ASPC..374..303B}) in spite of the slight dependence on
the mass. However, the 0.16\,dex offset between [CB09,KROU] and
[M05,KROU] (illustrated for completeness in bottom-right panel of
Figure~\ref{massesfig2}) is larger than expected revealing a more
complex relative difference between the two libraries beyond the
treatment of the TP-AGB phase.

Finally, the bottom-right panel of Figure~\ref{massesfig} shows the
comparison of the stellar masses estimates obtained with a CAL01 and a
Charlot\&Fall (2000; CF00) dust extinction laws for the P01 models and
a KROU IMF. The most relevant differences between a CAL01 and CF00
extinction laws is that the latter presents a larger attenuation of
the stellar component, which effectively leads to lower fluxes (mostly
in the UV) for similar values of the extinction. Furthermore, the
wavelength dependence of the attenuation in CF00 is greyer (i.e.,
shallower) than in CAL01. The overall result of the comparison is an
small offset of -0.03\,dex with a $\sim$0.2\,dex {\it rms}, similar to
what we found in PG08.  This is also in good agreement with the
results of \citet{2009ApJ...701.1839M}, indicating that the treatment
of the extinction law does not play a major role in the estimate of
the stellar mass (although it is more relevant for other estimated
parameters).

In summary, we find that after accounting for the different systematic
offsets, all models seem to be roughly consistent within a factor 2
($\sim$0.3\,dex). However, there are mass dependent systematics that
should be taken into account in the analysis of overall properties of
galaxy samples (e.g., \citealt{2009ApJ...701.1765M},
\citealt{2009ApJ...694.1171T}, \citealt{2010ApJ...709..644I}).

\subsubsection{Comparison to other stellar mass catalogs}
\label{masscatalogs}

In this section we compare our stellar masses with the estimates from
\citet{2006ApJ...651..120B} and \citet{2007MNRAS.382..109T}.  In the
former, the authors derived stellar masses for a large sample of
galaxies with spectroscopic redshifts from the DEEP2 survey in the
EGS. In the latter, the authors combined spectroscopic and photometric
redshifts to study the properties of a mass limited sample
(log~M$>$11\,$\mathcal{M}_\sun$).  Both works used the same
photometric dataset consisting on 5 bands: $BRI$ from the CFHT survey,
and $JK$ from the Palomar NIR survey.  The stellar masses in both
cases were essentially computed based on the fitting of the SEDs to a
grid of templates derived from BC03 models with a
\citet{2003PASP..115..763C} IMF and exponentially decreasing SFHs. In
particular, Bundy et al. used the rest-frame $K$-band luminosity and
mass-to-light ratio to scale the templates and compute the probability
distribution of the stellar mass and the most likely value. On the
contrary, Trujillo et al. (based on the results of Conselice et
al. 2007) did not renormalize the templates in a single band but used
the whole SED to scale the fluxes, similarly to our approach but
restricted to only 5-bands.

We cross-correlate the catalogs using a 2'' radius, and we double
check the validity of the match ensuring that the spectroscopic
redshifts (independently matched) are the same. The final sample
contains 4706 and 791 galaxies detected in the catalogs of Bundy et
al. and Trujillo et al., respectively.  For the comparison to Bundy et
al. the photometric redshifts were forced to the spectroscopic value
and for the comparison to Trujillo et al. the photometric redshifts
were forced to the values quoted in their paper.

The left panel of Figure~\ref{masses_bundy} shows the comparison of
the stellar masses with Bundy at al. for our default modeling
assumptions. Our estimates are slightly lower with an median
difference of $\Delta\log$(M)$=$-0.07$\pm$0.21\,dex. Also, we find
that the stellar masses computed with [P01,SALP] are in better
agreement with Bundy et al. than those obtained using the same
modeling configuration as in their work, [BC03, CHAB], which would
increase the difference in smaller masses to $\Delta\log$(M)$=$
-0.12\,dex.  We further investigate if this offset is caused by a
difference in the photometry by comparing our $K$-band magnitudes to
those of Bundy et al. that were computed using 2\arcsec radius
apertures (for the SED fitting). The sources in Bundy et al. are on
average $\Delta$$K=$0.12\,mag fainter than in our catalog, which would
imply a larger difference in the stellar masses if we simply scale
their magnitudes to our photometry. Thus, the most plausible
explanation for this small offset is the use of different techniques
for estimating the stellar masses, and specifically the use of IRAC
data in our study. The right panel of Figure~\ref{masses_bundy} shows
the comparison of the stellar masses with Trujillo at al.. The overall
comparison presents a good agreement with a median difference of
$\Delta\log$(M)$=$0.10$\pm$0.25\,dex, slightly larger than the offset
to masses of Bundy et al. However, the scatter of the distribution is
quite similar to that of the comparison to Bundy et al. for the
highest stellar masses $\log$(M)$>$11\,$\mathcal{M}_\sun$.

\section{SED analysis: Star Formation Rates}
\label{evalSFR}

In this section, we present the estimations of the SFRs of the
galaxies in our IRAC sample based on their UV-to-FIR SEDs. We also
discuss the quality of these estimates as well as their associated
systematic uncertainties.

The SFR of a galaxy is frequently computed from the UV and IR
luminosities through theoretical or empirical calibrations. As young
stellar populations emit predominantly in the UV, this wavelength
range is highly sensitive to recent events of star formation. However,
this UV emission is usually attenuated by dust, which re-emits the
absorbed energy in the thermal IR. Consequently, the ongoing SFR can
be estimated either by correcting the UV luminosity for extinction or
combining the IR emission and the unobscured UV flux.

Here we focus on the latter approach making use of the high quality
FIR fluxes observed with MIPS at 24 and 70~$\mu$m. Thus, assuming that
the total SFR of a galaxy can be estimated by summing up two
components (see e.g.,
\citealt{2005ApJ...625...23B}; \citealt{2007ApJ...670..279I}): the part
of the star formation that is probed by a tracer affected by dust
attenuation, so we only are able to observe directly a fraction of it
(i.e., the unobscured component), and the part of the star formation
that is hidden by dust (obscured component). The unobscured star
formation can be measured with the rest-frame UV emission, which can
be estimated from the optical/NIR SEDs for the galaxies in our
sample. The obscured component can be estimated from the total IR
thermal emission (thus, we will refer to it as IR-SFR or IR-based
SFR). However, its calculation is usually affected by the choice of
template libraries fitting the IR part of the SED and, more
significantly, by the photometric coverage in the MIR-to-mm spectral
range. In this section, we will focus on the analysis of the IR-based
SFR and the random and systematic uncertainties associated with the
different procedures used to estimate it.

The structure of this section is as follows. First, we describe how we
fit the IR part of the SEDs to dust emission models, and present the
different methods used to estimate and IR-based SFR from monochromatic
and integrated luminosities in the MIR-to-mm range. Then, we compare
these different methods and discuss the systematic and random
uncertainties inherent to the calculation of IR-based SFRs.

In this section, the reader must have in mind that the most useful
information to estimate IR-based SFRs comes from the MIPS 24~$\mu$m
fluxes. The reason is simple: these observations are the deepest in
the MIR-to-mm range, so we only have this SFR tracer for the vast
majority of sources in our sample. Ideally, it would be desirable to
have other fluxes in the IR to constrain the fits to dust emission
models, but this is only possible for a very small fraction of
galaxies which have MIPS 70~$\mu$m data, or other photometric points
in the (sub)-mm. Even with {\it Herschel} data, there will be a
significant population of galaxies that will only count with the MIPS
24~$\mu$m flux. Therefore, an important part of our discussion will be
assessing the reliability of IR-based SFRs based only on MIPS
24~$\mu$m data. For that purpose, we will take advantage of the very
deep observations carried out at 70~$\mu$m within the {\it Spitzer}
FIDEL Legacy Project, studying the variations in the estimated
IR-based SFRs fitting MIPS 24~$\mu$m and MIPS 70~$\mu$m
simultaneously.

\begin{figure}
\includegraphics[width=9cm,angle=0.]{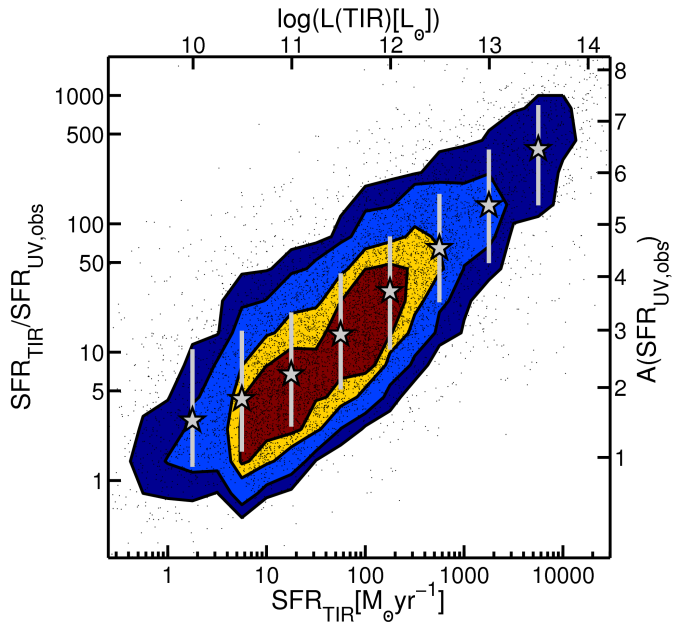}
\caption{\label{UV_TIR} Ratio of the two components of the total SFR
  (SFR$=$SFR$_{UV,obs}$+SFR$_{IR}$; Equation~\ref{sfrtir}) as a
  function of the SFR$_{\mathrm{TIR}}$ for a sub-sample galaxies
  detected in MIPS 24~$\mu$m (f(24)$>$60$\mu$Jy).Each contour contains
  (from the inside out) 25\%, 50\%, 75\% and 90\% of the sample,
  respectively. The underlying black dots depict the individual values
  of the ratio of SFRs. The grey stars with error bars depict the
  median and 1$\sigma$ of the ratio of SFRs in bins of
  SFR$_{\mathrm{TIR}}$.}
\end{figure}

\subsection{IR SED fitting}
\label{irfitting}

Typically, IR-based SFR are computed either from the total IR
luminosity, i.e., the integrated emission from 8 to 1000\,$\mu$m
[L(TIR)], or from monochromatic luminosities at different
wavelengths. Both methods require a detailed characterization of the
IR SED, which is usually obtained by fitting the observed fluxes to
dust emission templates. However, as mentioned above, these estimates
are largely dependent on the choice of templates. An issue that is
usually aggravated by the fact that typically the only measurement of
the MIR emission comes from the 24\,$\mu$m data, and occasionally
70\,$\mu$m, whereas the total IR luminosity is commonly dominated by
the emission at longer wavelengths $\lambda$$\sim$100\,$\mu$m.

Thus, in order to study in detail the intrinsic uncertainties in the
IR-based SFRs arising from these issues, we follow two different
approaches to fit the IR data to the dust templates: 1) we study the
galaxies detected at MIPS 24~$\mu$m fitting only this flux to models
of \citet[][CE01 hereafter]{2001ApJ...556..562C}, \citet[][DH02
  hereafter]{2002ApJ...576..159D}, which is a usual scenario in
studies of the IR-emission at high-z (see
e.g. \citealt{2009A&A...504..751S}, \citealt{2008ApJ...682..985W}). In
this case, we asses the differences between IR-based SFR (hereafter
SFR$_{i}$(24)) estimated with several methods, and the impact of using
different models; 2) we restrict the analysis to galaxies
simultaneously detected in IRAC and MIPS 24 and 70~$\mu$m, fitting all
fluxes at rest-frame wavelengths $\lambda$$>$~5~$\mu$m (where the
luminosity of a galaxy must present a significant non-stellar
contribution; see, e.g., \citealt{2006ApJ...648..987P};
\citealt{2007ApJ...656..770S} ) to the models of CE01, DH02 and also
\citet[][R09 hereafter]{2009ApJ...692..556R}. We refer to these
galaxies as the best-effort sample and their SFRs(8,24,70). The
notation indicates that the fit essentially includes 8, 24 and
70~$\mu$m data up to z$\sim$0.6, and 24 and 70~$\mu$m data at higher
redshift.  Based on this sample we can study the impact of having a
better constrained IR SED against the MIPS 24~$\mu$m only scenario
(e.g, as in \citealt{2010ApJ...709..572K}). In both cases, the fitting
is carried out by fixing the redshift to $z_{\mathrm{phot}}$ or
$z_{\mathrm{spec}}$ (if available).  Then, the excess resulting from
subtracting the predicted contribution from the stellar flux (given by
the best fitting optical template) to the MIR bands is fitted to each
set of models. In the case when only MIPS 24~$\mu$m data is used, the
templates are not fitted but rather scaled i.e., we obtain the
rest-frame monochromatic luminosity for that flux and redshift and we
select the most likely template based in their absolute normalization
in the total IR luminosity (as in e.g., \citealt{2006ApJ...640...92P},
\citealt{2009A&A...504..751S}). Moreover, for sources undetected in
the 24\,$\mu$m data, we set an upper limit of f(24)=60\,$\mu Jy$, the
approximate SNR=5 level of the MIPS data in EGS (see paper I for more
details), which allows us to provide an upper limit of the IR-based
SFR. Figure~\ref{example_sed} shows and example of the IR SED fitting
jointly with the optical template.

Based on the best fitting templates, we computed several IR
monochromatic and the integrated luminosities [L($\lambda)$ and
  L(TIR), respectively] as the median value of all the fitted template
sets. In the following, we describe various possibilities for IR-based
SFRs based on L($\lambda$) at different wavelengths. This relations
are calibrated from galaxy samples counting with extensive IR coverage
(at least more than 3-4 bands), and provides an alternative estimate
of the SFR based on milder template extrapolation than L(TIR), which
in principle makes them more robust when only few bands are available
for the fitting. Note however that the rest-frame wavelengths around
10-30~$\mu$m are wildly variable and thus extrapolating luminosities
in this region involves significant uncertainties, e.g., L(8) based on
24~$\mu$m data at z$\sim$2 (see \S~\ref{sfrs2470} for more details).

\subsection{Total SFR and IR-based SFRs estimates}
\label{sfrs}

Our method to estimate the total SFRs is based on a combination of the
IR emission and the unobscured UV flux (similarly to
\citealt{2006ApJ...648..987P}, \citealt{2009ApJ...703.1672K},
\citealt{2009A&A...504..751S}).In particular, we use the prescription
of \citet[][see also
  \citealt{2007ApJ...668...45P}]{2005ApJ...625...23B}, which is based
on the calibration for the total IR luminosity of
\citet{1998ARA&A..36..189K}, and parametrizes the contribution of
radiation that escapes directly in the UV.

\begin{equation}
SFR=SFR_{TIR}+SFR_{UV,\mathrm{obs}}
\end{equation}
\begin{equation}\label{sfrtir}
SFR(M_{\odot}\,\mathrm{yr}^{-1})=1.8\times10^{-10}[L(TIR)+3.3\times\,L(0.28)]/L_{\odot}
\end{equation}

\noindent where L(TIR) is the integrated total IR luminosity, and
L(0.28) is the rest-frame monochromatic luminosity at 0.28\,$\mu$m
(uncorrected for extinction).  The well sampled SEDs of our galaxies
at optical wavelengths allow a robust estimation of L(0.28) by
interpolating in the best fitting optical template. However, as
described in the previous section, the value of L(TIR) is strongly
model dependent, as it is based on an extrapolation from one or a few
MIR fluxes to the total emission from 8 to 1000\,$\mu$m.  An alternate
possibility is to obtain other IR-based SFRs based on L($\lambda$), thus
reducing the template dependence.  In the following, we will refer to
the IR-based SFR derived from L(TIR) as SFR$_{\mathrm{TIR}}$. In
addition, we compute four other IR-based estimates.

The first estimate is based on rest-frame monochromatic luminosity at
8\,$\mu$m (hereafter SFR$_{\mathrm{B08}}$). These estimate make use of
the empirical relation between L(8$\mu$m) and L(TIR) described in
\citet{2008A&A...479...83B} and the Kennicutt factor to transform to
SFR.

\begin{equation}
\label{bavouzet}
SFR_{\mathrm{B08}}(M_{\odot}~\mathrm{yr}^{-1})=1.8\times10^{-10}\times(377.9\times L(8)^{0.83})/L_{\odot}
\end{equation}

The second method in based on equation (14) of
\citet{2009ApJ...692..556R}, that relates the SFR (hereafter
SFR$_{\mathrm{R09}}$) to the observed flux in the MIPS 24~$\mu$m band
and the redshift. The redshift dependent coefficients of the relation
were computed using averaged templates derived from a set of empirical
IR-SEDs fitting local galaxies. This estimation of the SFR is
independent of the rest, as it is based on different templates.  The
conversion from IR luminosities to SFRs is also computed through the
Kennicutt factor. However, the authors scaled the factor to a
Kroupa-like (2002) IMF (the original factor is for a SALP IMF)
multiplying it by 0.66 (a similar conversion is obtained in
\citealt{2007ApJS..173..267S}). Here we undo that change for
consistency with the other methods that are computed using the factor
for a SALP IMF.

The last method is not strictly an IR-based SFRs but an estimate the global
SFR. It is based on the empirical relation given in
\citet{2006ApJ...650..835A} between the rest-frame monochromatic
luminosity at 24\,$\mu$m and the SFR (hereafter,
SFR$_{\mathrm{A-H06}}$),

\begin{equation}\label{aah}
SFR_{\mathrm{A-H06}}(M_{\odot}~\mathrm{yr}^{-1})=1.51\times10^{-8}\times L(24)^{0.871}/L_{\odot}
\end{equation}

This formula is based on the calibration of L(Pa${\alpha}$) versus
L(24~$\mu$m) obtained for a set of local ULIRGS using the Kennicutt
(1998) relation between L(Pa$\alpha$) and SFR. A similar result was
obtained by \citet{2007ApJ...666..870C} for resolved star-forming
regions in local starburst (see also \citealt{2009ApJ...703.1672K}).
This estimation refers to the global SFR, not the IR-based SFR,
because the empirical relation in \citet{2006ApJ...650..835A} already
takes into account the unobscured star formation (measured through the
observed Pa${\alpha}$ emission) and the extinction correction [applied
  to calculate L(Pa${\alpha}$) in that paper].

\subsection{Accuracy of the IR-based SFRs}
\label{accuracySFR}
 
In the following Sections we analyze the systematic uncertainties in
the IR-based SFRs associated with the use different models and indicators,
and also the number of photometric bands available for IR SED fitting.

First, we compare the values obtained with each of the methods
presented in the previous section for a sub-sample of MIPS 24~$\mu$m detected
galaxies. In this case, the SED is fitted to 24~$\mu$m data only
(\S~\ref{sfrs24}). Note that we have chosen several methods for
estimating the IR-based SFR that present intrinsically different
approaches, using either integrated and monochromatic luminosities or
observed fluxes. Here we also test the differences introduced by the
use of the CE01, DH02 or R09 models. For simplicity, in this case, the
comparison to the R09 models is done through the SFRs obtained with
their empirical relation (SFR$_{\mathrm{R09}}$) instead of fitting the
data to the three models.

Second, we study the differences in the SFRs obtained for MIPS 24~$\mu$m
sample and the best-effort sample, which count with a better IR
coverage based on IRAC-8.0 plus MIPS 24 and 70~$\mu$m data
(\S~\ref{sfrs2470}). With this test we quantify the systematic effects
associated with use of limited IR data. Finally, we repeat the
comparison of values obtained with each method for the best-effort
sample including also highly accurate SFRs drawn from other authors
based on a more detailed IR coverage (\S~\ref{sfrcomparison}). Based
on this comparison we asses the goodness of our best-effort SFRs and
the reliability of the different methods studied here.

For the sake of clarity, we will refer all the comparisons between the
SFRs estimated with each method to SFR$_{\mathrm{TIR}}$ which, as
explained in \S~\ref{irfitting}, is computed from the average total
infrared luminosity of all the fitted template sets. In addition, we
will refer to them just as SFRs (dropping the IR prefix). In the case
of SFR$_{\mathrm{A-H06}}$, the proper IR-based SFRs is obtained by
subtracting the contribution of the
SFR$_{UV,\mathrm{obs}}$. Nevertheless, as our working samples are
composed by strong IR-emitters, we are biased towards dust obscured
galaxies where this contribution is presumably small. For example,
Figure~\ref{UV_TIR}, which shows the ratio
SFR$_{UV,obs}$/SFR$_{\mathrm{TIR}}$, indicates that SFR$_{UV,obs}$ is
lower than SFR$_{\mathrm{TIR}}$ (in most cases), with a clear trend
for galaxies with intense star formation to present more and more
extincted starbursts.

\subsubsection{Analysis of IR-based SFRs: MIPS 24~$\mu$m sample}
\label{sfrs24}

\begin{figure*}
\includegraphics[width=9cm,angle=0.]{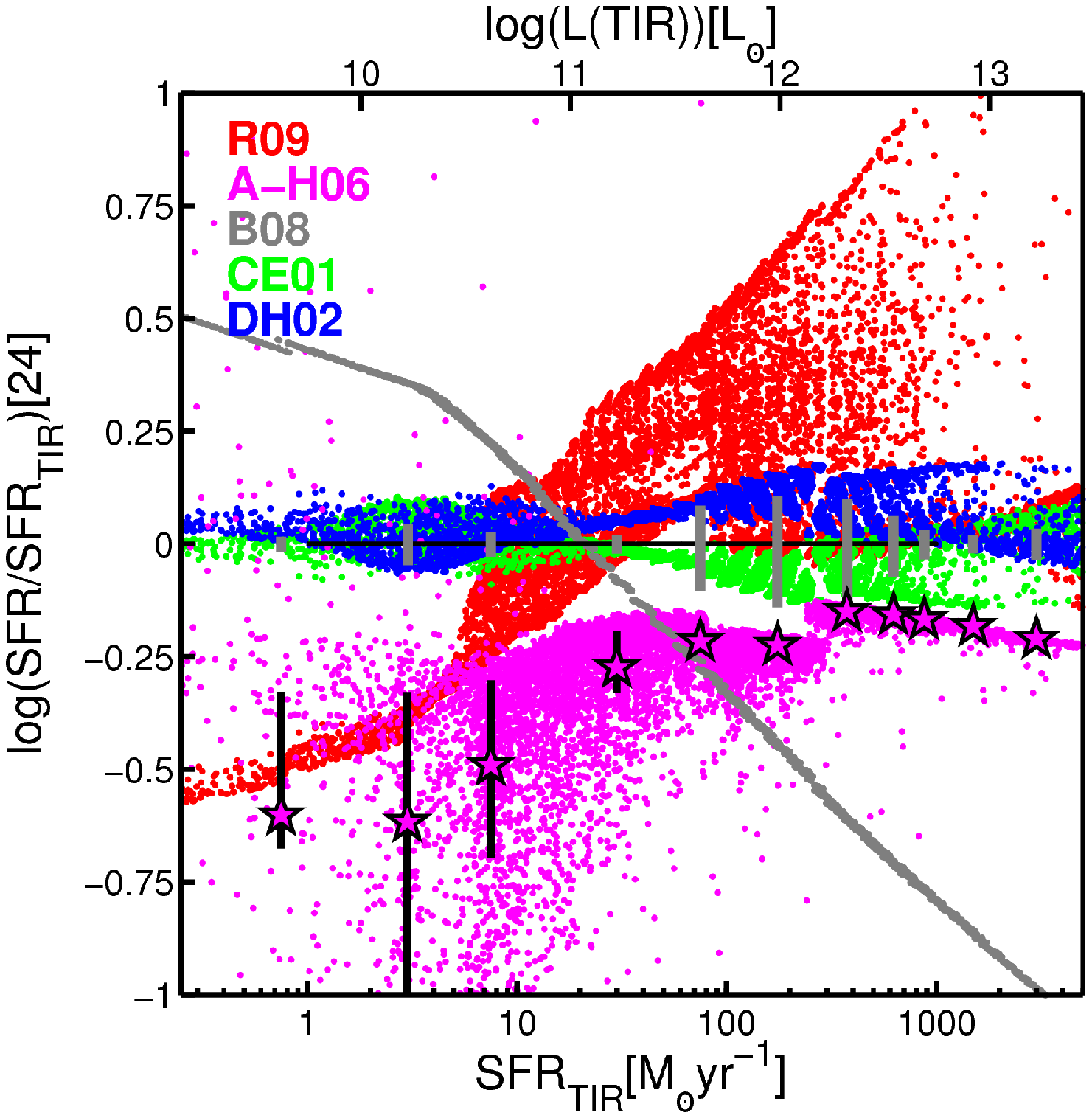}
\includegraphics[width=9cm,angle=0.]{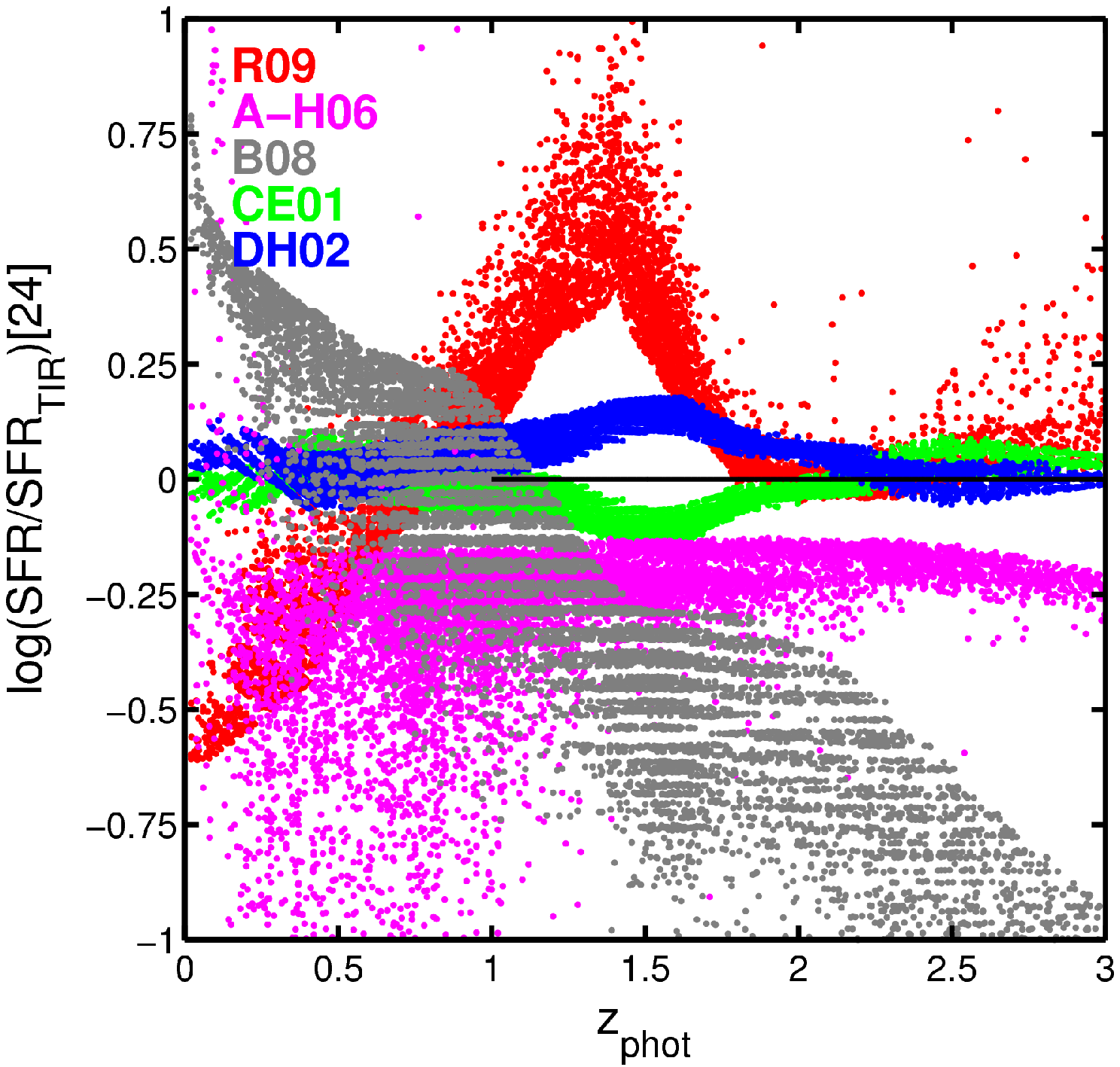}
\caption{\label{compare_sfrs} Comparison of different IR-based SFR
  indicators with respect to SFR$_{\mathrm{TIR}}$ as a function of
  SFR$_{\mathrm{TIR}}$(left) and redshift (right) for galaxies
  detected in MIPS 24~$\mu$m
  (f(24~$\mu$m)$>$60$\mu$Jy). SFR$_{\mathrm{TIR}}$ is computed from the
  average value of L$_{\mathrm{TIR}}$ in the templates of CE01 and
  DH02 fitted to the flux at 24~$\mu$m. The magenta points show
  SFR$_{\mathrm{A-H06}}$ estimated from the L(24) using the relation
  of \citet{2006ApJ...650..835A}; the magenta stars and error bars
  indicate the median value and 1$\sigma$ per SFR bin. The grey points
  show SFR$_{\mathrm{B08}}$ estimated from L(8) using the relation of
  \citet{2008A&A...479...83B}. The red points show
  SFR$_{\mathrm{R09}}$ estimated from MIPS 24~$\mu$m using the
  calibration of \citet{2009ApJ...692..556R}. The green and blue
  points depict SFR$_{\mathrm{TIR}}$ estimated from the templates of
  CE01 and DH02, respectively. The grey error bars depict the
  1$\sigma$ uncertainty in SFR$_{\mathrm{TIR}}$ per SFR bin.}
\end{figure*}

\begin{figure*}
\centering
\includegraphics[width=8.5cm,angle=0.]{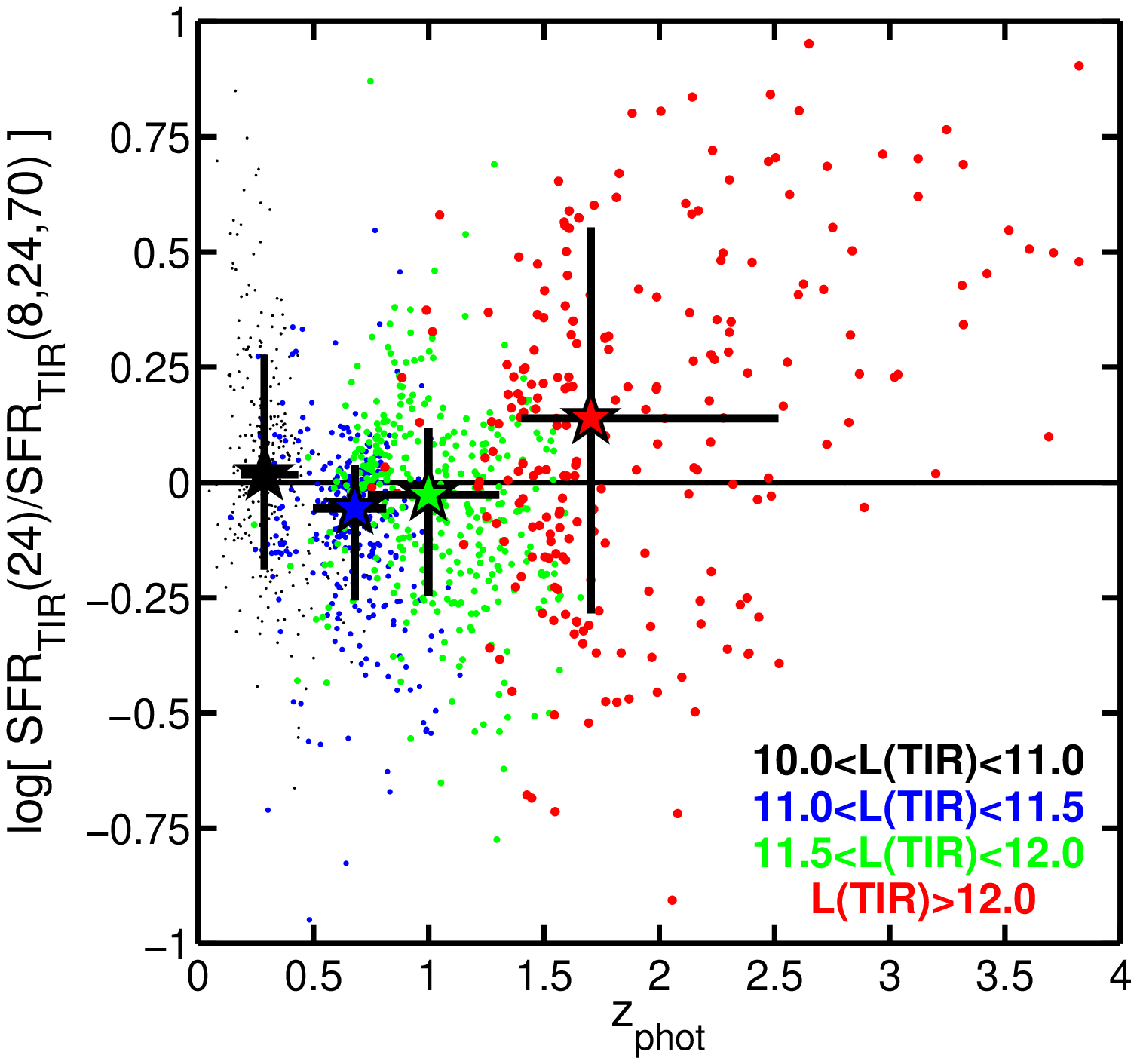}
\hspace{0.5cm}
\includegraphics[width=8.5cm,angle=0.]{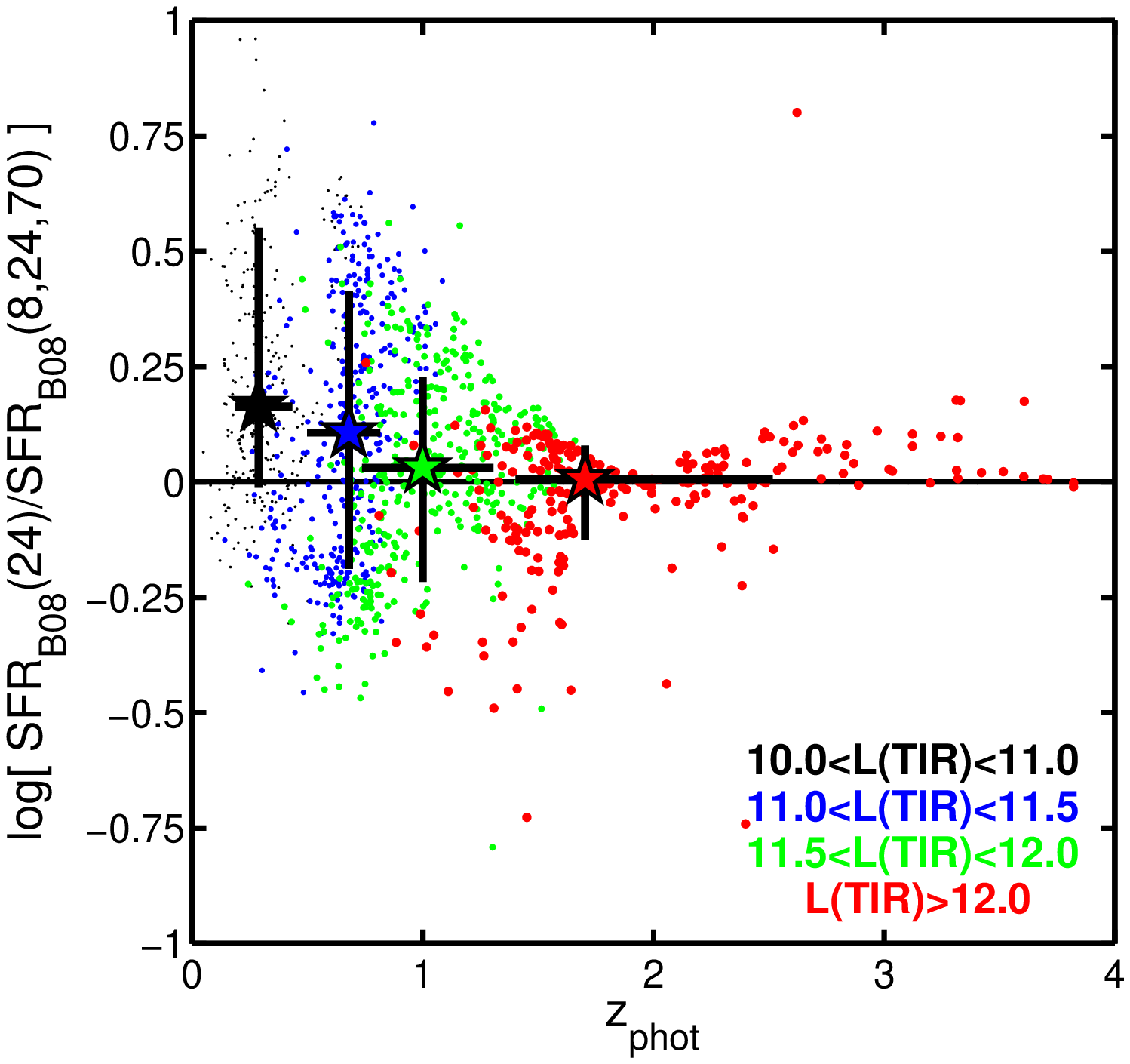}\\
\includegraphics[width=8.5cm,angle=0.]{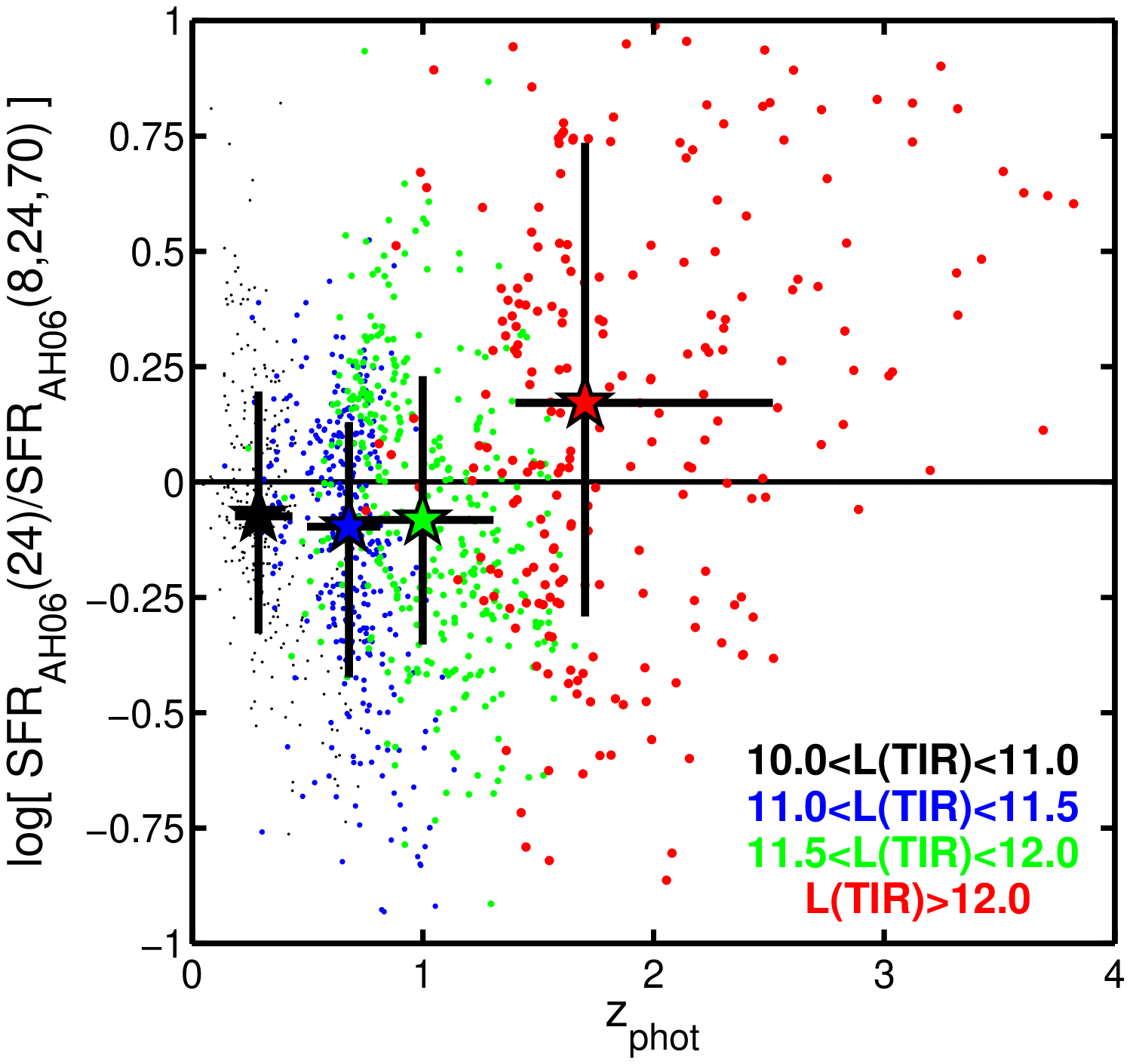}
\caption{\label{tracers_with70} Ratio of SFR$_{\mathrm{TIR}}$,
  SFR$_{\mathrm{B08}}$ (left and right) and SFR$_{\mathrm{A-H06}}$ (bottom)
  estimated with 24~$\mu$m data only and with 8, 24 and 70~$\mu$m data
  as a function of redshift. The color code indicate increasing bins
  of IR luminosity from {\it normal} galaxies to ULIRGS. The colored
  stars with error bars depict the median value and 1~$\sigma$ of the
  ratio of SFRs and the redshift per luminosity bin.}
\end{figure*}

Figure~\ref{compare_sfrs} shows the comparison of the IR-based SFRs
obtained with the different methods presented in Section~\ref{sfrs}
with respect to SFR$_{\mathrm{TIR}}$ as a function of SFR and
redshift.  All the estimates discussed in this section are based on
24~$\mu$m data only, i.e., SFR$_{i}$(24). We omit the parentheses for
simplicity.  The SFR$_{\mathrm{TIR}}$ estimated separately with the
models of CE01 (green dots) and DH02 (blue dots) are shown jointly
with the median and {\it rms} of both values in several SFR and
luminosity bins. The typical scatter of the SFRs estimated with both
libraries is smaller than $\sim$0.3\,dex, consistent with the results
by other authors (\citealt{2006A&A...451...57M};
\citealt{2007ApJ...668...45P}). When we compare the CE01 and DH02
libraries as a function of redshift, the maximum differences are
observed for galaxies at z$=$1-2. For these sources, the estimates
with the DH02 models are larger than those with CE01 models, as found
by \citet{2009A&A...504..751S}. In this redshift range, the 24\,$\mu$m
band is probing the spectral region where the 9.6\,$\mu$m silicate
absorption is found, jointly with the prominent PAHs around
7--9\,$\mu$m. The shape of the models in CE01 and DH02 template sets
is very different in this region, with the former presenting less
prominent PAH features than the latter. In fact, all the DH02 models
are identical below $\sim$9\,$\mu$m, while CE01 models present a wide
variety of spectral shapes, with a rising warm-dust continuum hiding
the PAH features between 6 and 20\,$\mu$m as we move to models with
higher IR luminosities.

The values of SFR$_{\mathrm{A-H06}}$ are systematically smaller than
SFR$_{\mathrm{TIR}}$. For
SFR$_{\mathrm{TIR}}$$>$100\,M$_{\odot}$yr$^{-1}$, we find
$\Delta$SFR$=$-0.18$\pm$0.05\,dex. For smaller values of the SFR, where
the unobscured and obscured star formation are comparable,
SFR$_{\mathrm{A-H06}}$ is down to a factor of 0.6\,dex smaller than
SFR$_{\mathrm{TIR}}$, with a larger scatter.

The comparison of SFR$_\mathrm{TIR}$ and SFR$_{\mathrm{B08}}$ clearly
indicates that the empirical relations L[8]-to-L$_{\mathrm{TIR}}$ in
\citet{2008A&A...479...83B} (Equation~\ref{bavouzet}) and in the models of
CE01 and DH02 (see e.g., Figure 8 of \citealt{2007ApJ...670..156D})
are substantially different. The ratio of the two SFRs as a function
of SFR$_\mathrm{TIR}$ is tilted with respect to the unity line, and
consequently, both estimates are only consistent within a narrow
interval around SFR$\sim$20\,M$_{\odot}$yr$^{-1}$ (or z$\sim$1).  For
SFR$_{\mathrm{TIR}}$$>$100 and 1000\,M$_{\odot}$yr$^{-1}$ (the latter
being the typical value for the z$\sim$2 galaxies detected by MIPS)
the SFR$_{\mathrm{B08}}$ values are 0.3 and 0.8\,dex lower than the
SFR$_{\mathrm{TIR}}$ estimates, respectively. In contrast, for
SFR$_{\mathrm{TIR}}$$\lesssim$10\,M$_{\odot}$yr$^{-1}$,
SFR$_{\mathrm{B08}}$ is larger than SFR$_{\mathrm{TIR}}$ by
$>$0.2\,dex.

The equation to calculate SFR$_{\mathrm{R09}}$
\citep{2009ApJ...692..556R} varies with redshift.  Consequently, the
SFR$_{\mathrm{R09}}$/SFR$_{\mathrm{TIR}}$ ratio presents different
trends as a function of both luminosity and redshift. In terms of
redshift, we distinguish three regions: 0$<$z$<$1.4,
1.4$<$z$\lesssim$1.75, and z$\gtrsim$1.75. At z$=$0-1.4, the ratio
increases with redshift from an average value of -0.5\,dex at z$=$0 to
0.5\,dex at z$\sim$1.4, being close to unity at z$\sim$0.75.  In the
interval from 1.4$<$z$\lesssim$1.75, the ratio decreases from 0.5\,dex
to nearly $\sim$0. Finally, at z$>$1.75, SFR$_{\mathrm{R09}}$ values
become roughly consistent with SFR$_{\mathrm{TIR}}$ with little
scatter up to z$=$3, $\Delta$SFR=0.02$^{0.06}_{0.04}$\,dex. These large
differences are related to the distinct shapes of the R09 and
CE01/DH02 templates. At z$\lesssim$0.5, the 24\,$\mu$m band probes a
spectral range dominated by warm dust and emission features found by
{\it Spitzer} at $\lambda$$\sim$17\,$\mu$m and identified with PAH or
nanoparticles \citep{2004ApJS..154..309W}. At these redshifts, our
sample is dominated by galaxies with L(TIR)$\sim$10$^{10}$\,L$_\sun$,
and the CE01 models for these luminosities differ from the
corresponding R09 templates by up to 0.5\,dex in the
$\lambda$$=$16--20\,$\mu$m. This explains the differences at low
redshift in the right panel of Figure~\ref{compare_sfrs}. At
z$=$0.5--1.0, our sample is dominated by LIRGs, and CE01 and R09
models for L(TIR)$\sim$10$^{10.75}$\,L$_\sun$ and
L(TIR)$\sim$10$^{11.25}$\,L$_\sun$ are very similar (up to
$\lambda$$=$1.5\,mm), resulting on very similar estimates of the SFR.
At z$\sim$1.4, the 24\mic\, band probes the spectral region around
10\mic, and the galaxies detected by MIPS in this range have
L(TIR)$\gtrsim$10$^{11.5}$\,L$_\sun$. For this luminosity, the CE01 and
R09 models differ considerably due to the relative strength of the
silicate absorption. For example, for a
L(TIR)$\gtrsim$10$^{12.25}$\,L$_\sun$, the R09 template predicts a
luminosity at 10\mic\, which is a factor of $\sim$0.7\,dex smaller than
the CE01 model corresponding to the same L(TIR). Below 8\mic,
rest-frame, the CE01 and R09 models are almost identical for LIRGs and
ULIRGs, explaining the good match between SFR$_{\mathrm{TIR}}$ and
SFR$_{\mathrm{R09}}$ at z$\gtrsim$2.

In summary, we conclude that whereas SFR$_{\mathrm{A-H06}}$ and
SFR$_{\mathrm{TIR}}$ are roughly consistent within $\sim$0.3~dex (
modulo a constant offset), regardless of the models used to fit the
IR-SED, the values of SFR$_{\mathrm{B08}}$ and SFR$_{\mathrm{R09}}$
present systematic deviations with respect to those that are not
consistent within the typical {\it rms}. Moreover, these differences
are not constant, but present a dependence of both redshift and
SFR. As a result, large systematic offsets (of $\pm$0.5~dex) with
respect to SFR$_{\mathrm{TIR}}$ are expected at certain redshifts,
e.g., $\Delta$SFR$\sim$+0.5 and -0.5~dex for SFR$_{\mathrm{R09}}$ and
SFR$_{\mathrm{B08}}$ at z$\sim$1.4, respectively.

\subsubsection{Analysis of IR-based SFRs: Best-effort vs. MIPS 24~$\mu$m }
\label{sfrs2470}

Here we study the impact on the IR-based SFRs of modeling the IR-SED
with limited photometric data. For that matter, we quantify the
differences in the SFRs estimated with each of methods compared in the
previous section using the sample characterized with MIPS 24~$\mu$m
data and with 8, 24 and 70~$\mu$m data, i.e, the best-effort sample
(note that $\lesssim$2\% of the sample in detected in MIPS 70~$\mu$m
for $\sim$20\% in MIPS 24~$\mu$m). In principle, the inclusion of
additional mid-IR fluxes must improve the quality of the estimates
given that there is a better sampling of the IR SED from which better
k-corrections to the monochromatic luminosities can be obtained.  On
the downside, the spectral range probed by the MIPS bands gets
narrower with redshift, and the 24~$\mu$m channel shifts progressively
into to PAH region, where models are more uncertain and different
libraries differ significantly. Also, as the observed 70~$\mu$m moves
further away from the tip of the IR-emission ($\sim$100$\mu$m), the
uncertainty in the extrapolated L(TIR) increases.

Figure~\ref{tracers_with70} shows the ratio SFR(24)/SFR(8,24,70) as a
function of redshift for each of the different methods to estimate the
SFR, except for SFR$_{\mathrm{R09}}$, that only depends on the
observed flux at 24~$\mu$m and the redshift) . The color code
indicates four different bins of infrared luminosity. The colored
stars with error bars depict the median value and 1~$\sigma$ of the
redshift and the ratio of SFRs(24/8,24,70) for different luminosity
bins.

\begin{figure*}
\centering
\includegraphics[width=8.5cm,angle=-90.]{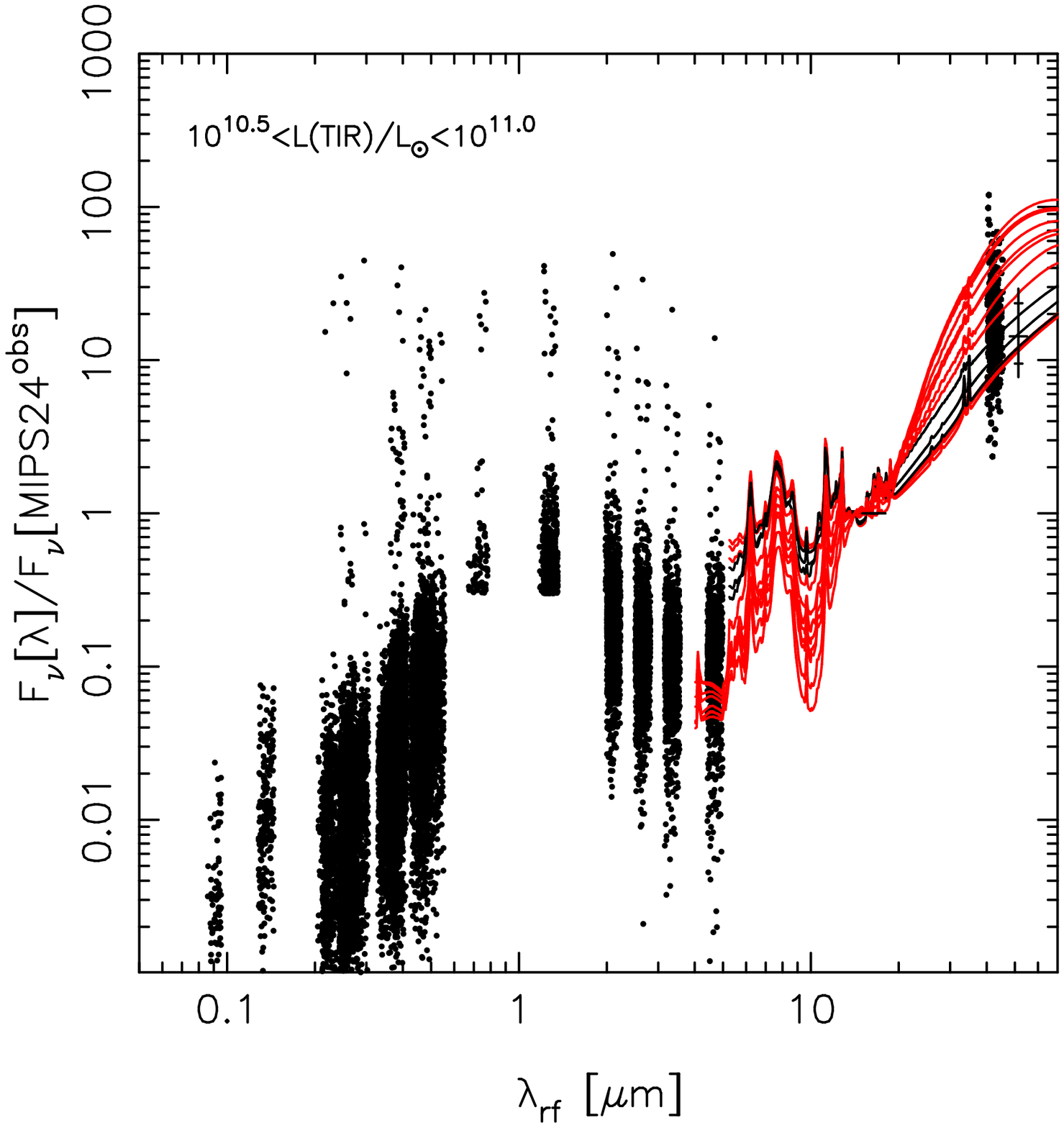}
\hspace{0.7cm}
\includegraphics[width=8.5cm,angle=-90.]{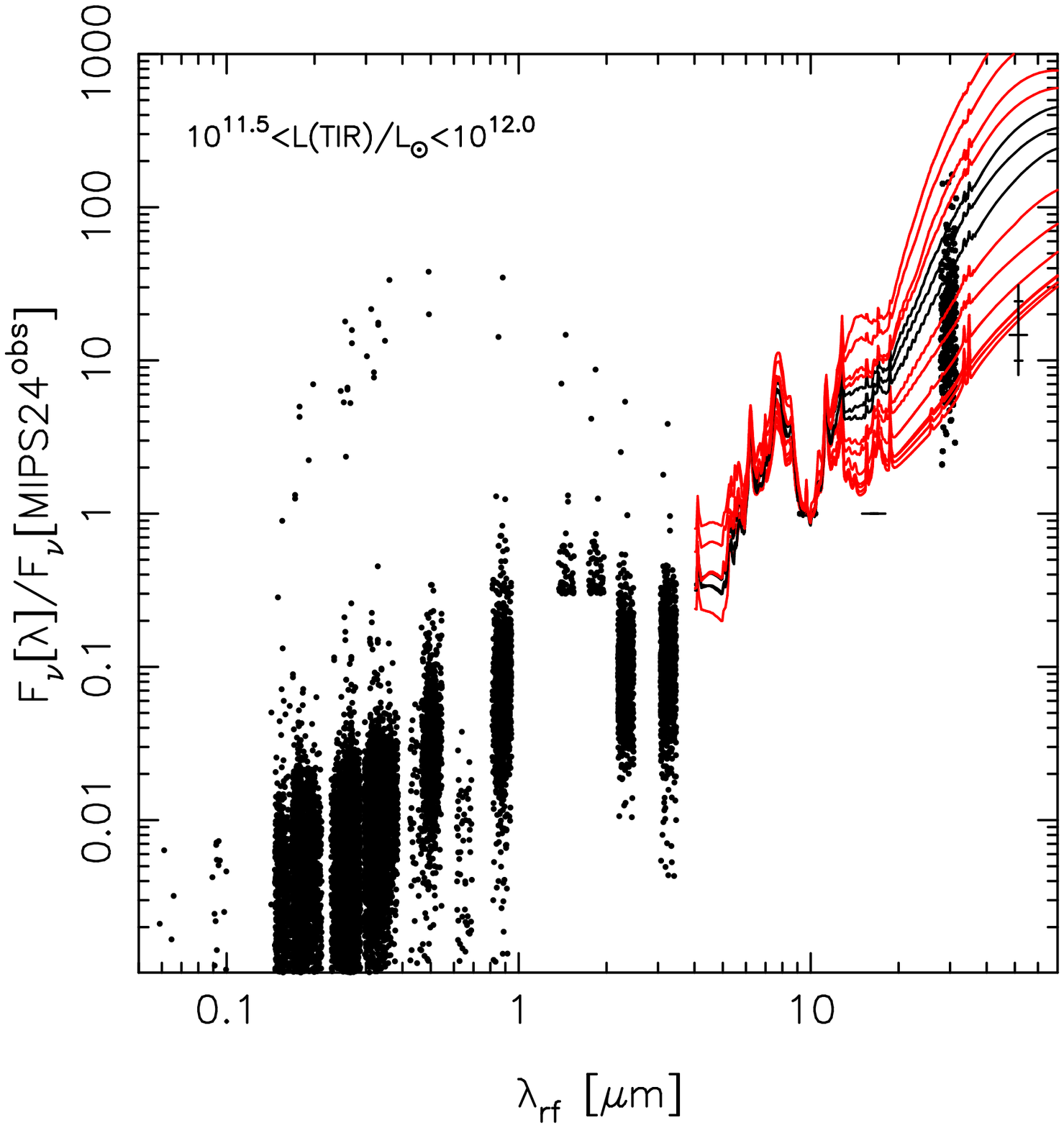}
\caption{\label{rieke_24_70} Rest-frame SEDs of galaxies detected in
  MIPS 24 and 70~$\mu$m at 0.6$<$z$<$0.8 with
  10$^{10.5}$$<$L(TIR)$<$10$^{11.0}$ (left panel) and 1.3$<$z$<$1.6
  with 10$^{11.5}$$<$L(TIR)$<$10$^{12}$ (LIRGS) (right-panel).  The
  fluxes are normalized to the flux in MIPS 24~$\mu$m. The solid lines
  depict the templates of \citet{2009ApJ...692..556R}: black lines are
  for templates with L(TIR) within the corresponding IR-luminosity
  range, red lines are for the rest of the templates.  The vertical
  line and marks on the right of the MIPS 70~$\mu$m fluxes show the
  median, quartiles and 1~$\sigma$ of the distribution of observed
  fluxes.}
\end{figure*}

The upper-left panel of Figure~\ref{tracers_with70} shows that
SFR$_{\mathrm{TIR}}$(24) and SFR$_{\mathrm{TIR}}$(8,24,70) are
$\sim$1$\sigma$ consistent within $\sim$0.20\,dex up to ULIRG
luminosities, showing a small offset (mostly at z$\gtrsim$0.5) in
SFR$_{\mathrm{TIR}}$(24) towards underestimating the SFR by
$\Delta$SFR=-0.05$\pm$0.20\,dex.  On the other hand, ULIRGs (typically
at z$\gtrsim$1.5) present values of SFR$_{\mathrm{TIR}}$(24) larger
than SFR$_{\mathrm{TIR}}$(8,24,70) with an average difference of
$\Delta$SFR=0.15$\pm$0.40\,dex. This is consistent with the results
found for ULIRGs at this redshift by several authors, who report
excesses of a factor of 2-10 in the SFRs estimated from MIPS 24~$\mu$m
only (\citealt{2007ApJ...670..156D}, \citealt{2007ApJ...668...45P},
\citealt{2008ApJ...675..262R}). Note that estimating the
IR-luminosities for these galaxies based on MIPS 24~$\mu$m data alone
is intrinsically difficult as this band is probing the most variable
region of the IR-SED, featuring emission from PAHs and Silicate
absorptions. In fact, further motive for these discrepancies could
associated to a change in the relative strength of these components in
high-z galaxies with respect to the local templates, particularly for
the ULIRG templates.
 
Figure~\ref{rieke_24_70} present further evidence of this issue. The
left and right panels of the Figure show the rest-frame SED normalized
to the flux at 24\,$\mu$m for galaxies at $\tilde{z}$$=$0.7 and
$\tilde{z}$$=$1.5 in different L(TIR) ranges. The red and black lines
are the dust-emission templates of \citet{2009ApJ...692..556R}. The
black templates are those corresponding to the IR-luminosity range
shown in the legend. Note that we have selected the redshift ranges
and template normalization with the specific aim of stressing the
differences in warm-to-cold dust colors between the R09 models and the
actual observations. The same differences apply to other template
sets.  The vertical line to the right of the MIPS 70~$\mu$m data
depicts the median and 1~$\sigma$ of the distribution of MIPS
70~$\mu$m fluxes (normalized to 24\,\mic, i.e., the S$_{70}$/S$_{24}$
color). At z$\sim$0.7, the models for a
L(TIR)$=$10$^{10.5-11.0}$\,L$_\sun$ nicely predict the actual colors
observed for galaxies (the templates plotted in black match the median
and 1$\sigma$ range of observed colors). However, at z$\sim$1.5,
galaxies present smaller colors than what the models for the
appropriate luminosity range predict. This suggest that the excess in
SFR$_{\mathrm{TIR}}$(24) could be related to a difference in spectral
shapes for ULIRGs at high redshift in comparison with local
ULIRGs. Either due to the strength of the PAH and the Silicate
features (\citealt{2010A&A...518L..29E};
\citealt{2010A&A...518L..15P}) or due to additional continuum emission
by a obscured AGN \citep{2007ApJ...670..156D}.

The lower panel of Figure~\ref{tracers_with70} shows
SFR(24)/SFR(8,24,70) for the B08 recipe. The ratio of SFRs presents
just the opposite trend of what we find for SFR$_{\mathrm{TIR}}$,
i.e., the offset and {\it rms} of the comparison are larger at lower
redshifts and almost non-existent ($<$0.1\,dex) at high-z. This is not
surprising considering that MIPS 24~$\mu$m shifts towards 8~$\mu$m
with increasing redshifts, reducing the impact of the
k-corrections. As a result, SFR$_{\mathrm{B08}}$ is nearly insensitive
to the inclusion of 70~$\mu$m data at z$\sim$2. Note however that this
does not mean that it is a better estimation of the SFR. At
z$\lesssim$0.5, SFR$_{\mathrm{B08}}$(24) is larger than
SFR$_{\mathrm{B08}}$(8,24,70) by $\Delta$SFR$=$0.18$\pm$0.23\,dex.

Finally, the right panel of Figure~\ref{tracers_with70} shows the
comparison of SFR(24)/SFR(8,24,70) for the A-H06 recipe. The overall
trends are analogous to those observed for SFR$_{\mathrm{TIR}}$ but
with a larger scatter ($\sim$0.30\,dex), i.e., the offset and {\it
  rms} increases with redshift up from
$\Delta$SFR$=$-0.10$\pm$0.26\,dex at z$<$1 to
$\Delta$SFR$=$0.19$\pm$0.47\,dex at z$=$2--4.

\begin{figure*}
\includegraphics[width=9cm,angle=0.]{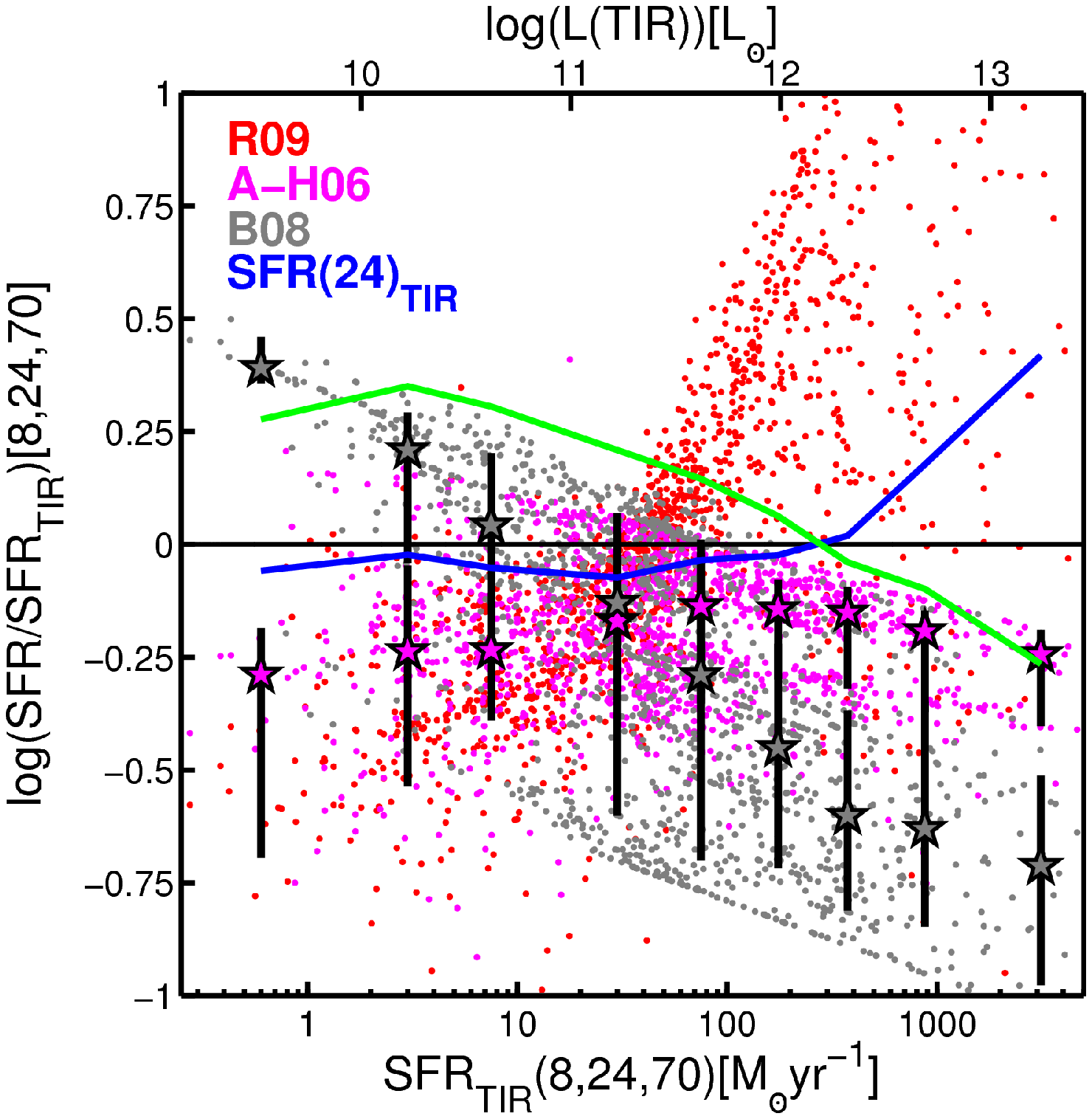}
\includegraphics[width=9cm,angle=0.]{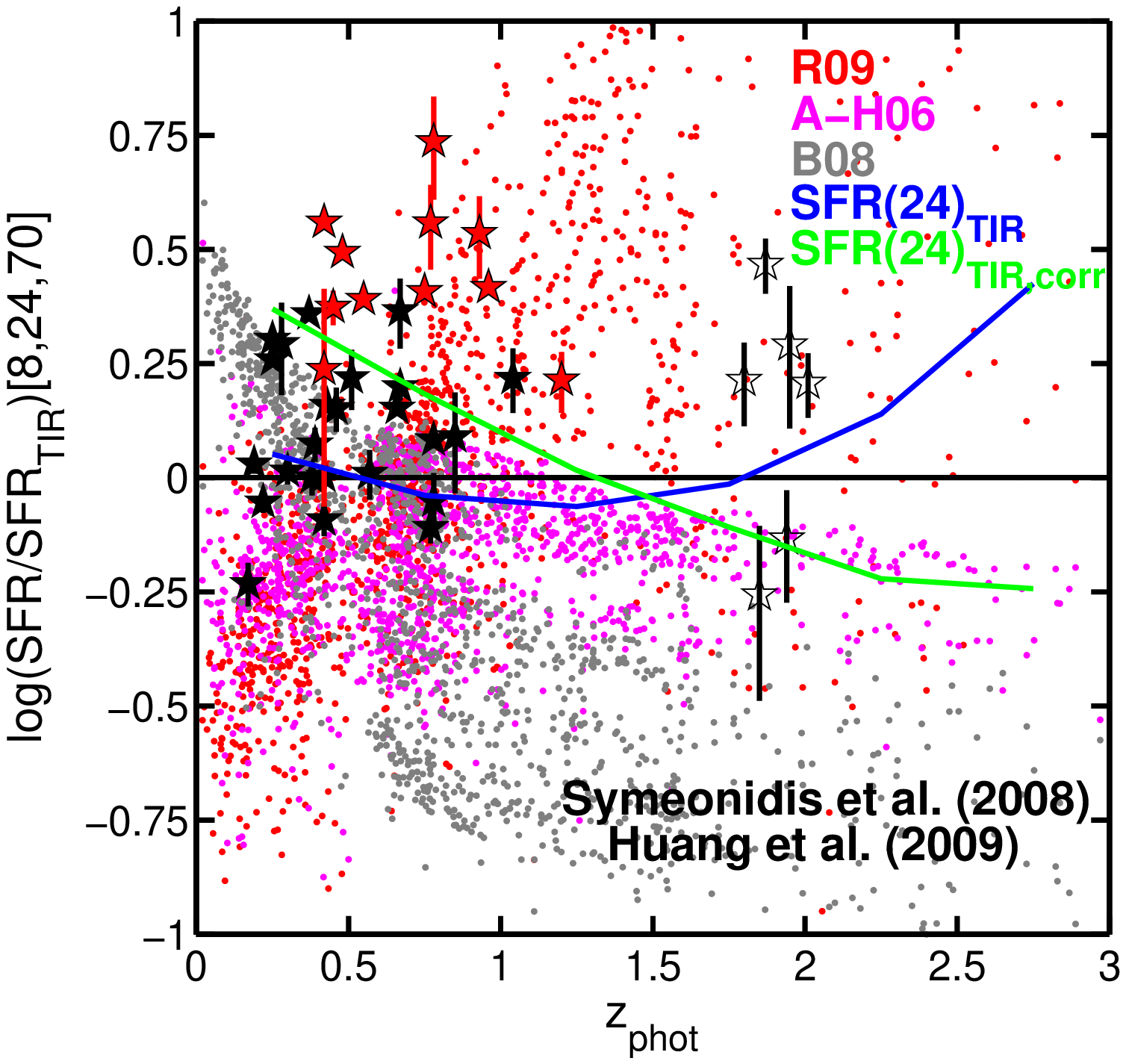}
\caption{\label{compare_sfrs_70} Comparison of different IR-based
  SFR(8,24,70) indicators with respect to
  SFR$_{\mathrm{TIR}}$(8,24,70) as a function of
  SFR$_{\mathrm{TIR}}$(left) and redshift (right) for galaxies
  detected in MIPS 24~$\mu$m (f(24~$\mu$m)$>$60$\mu$Jy) and
  MIPS 70~$\mu$m (f(70~$\mu$m)$>$3500$\mu$Jy) . The color code is the
  same as in Figure~\ref{compare_sfrs}. Here in also show the median
  values and 1$\sigma$ of the comparison to SFR$_{\mathrm{B08}}$ (grey
  stars with error bars). The blue line joins the median values of
  SFR$_{\mathrm{TIR}}$(24)/SFR$_{\mathrm{TIR}}$(8,24,70) per
  luminosity bin (similarly to the colored stars in the top-left panel
  of Figure~\ref{compare_sfrs}).  The black stars depict the ratio of
  SFRs for galaxies in common with the sample of \citet[][filled stars
    at z$<$1.2]{2008MNRAS.385.1015S} and \citet[][open stars at
    z$\sim$2]{2009ApJ...700..183H}. The IR-based SFR for these sources
  is one of the most accurate available at the moment, and it is in
  overall good agreement with our estimates with
  SFR$_{\mathrm{TIR}}$.}
\end{figure*}

\subsubsection{Analysis of IR-based SFRs: Best-effort sample}
\label{sfrcomparison}

In this Section we compare again the IR-based SFRs obtained with
different methods, but this time for the best-effort sample, i.e, with
estimates based on 8, 24 and 70~$\mu$m data. We also present a
comparison of IR-based SFRs to galaxies in common with other authors
counting with better IR SED coverage (e.g., MIPS-160 or
IR-spectroscopy) and therefore more reliable SFRs.

Figure~\ref{compare_sfrs_70} shows a comparison of the SFRs(8,24,70)
obtained with each method with respect to
SFR$_{\mathrm{TIR}}$(8,24,70) (our reference value), as a function of
SFR (left-panel) and redshift (right-panel). The color code is the
same as in Figure~\ref{compare_sfrs}.  To simplify the comparison to
the results of the previous section, the Figure also shows the ratio
of SFR$_{\mathrm{TIR}}$(24) to SFR$_{\mathrm{TIR}}$(8,24,70) i.e.,
basically the values shown in the upper-left panel of
Figure~\ref{tracers_with70}. The blue line joins the median values of
the ratio of SFRs per bin of luminosity (and redshift).

The relative differences with respect to SFR$_{\mathrm{TIR}}$(8,24,70)
remains mostly unchanged with respect to what is shown in
Figure~\ref{compare_sfrs} for estimates based on 24~$\mu$m data.  For
SFR$_{\mathrm{A-H06}}$ the most noticeable differences is that the
overall {\it rms} increases by $\sim$50\% for the highest SFRs and the
median ratio for
SFR$_{\mathrm{TIR}}$(8,24,70)$>$100M$_{\odot}$yr$^{-1}$ decreases to
$\Delta$SFR$=$-0.12$\pm$0.12\,dex. For SFR$_{\mathrm{B08}}$, the
median ratio SFR$_{\mathrm{B08}}$/SFR$_{\mathrm{TIR}}$ presents a
smaller tilt and a significant increment of the {\it rms} with respect
to the values in the MIPS 24~$\mu$m sample. Finally,
SFR$_{\mathrm{R09}}$, which only depends on the observed flux in MIPS
24~$\mu$m, presents the same trend at low-z as in
Figure~\ref{compare_sfrs}. However, at z$\sim$2 it tends to
overestimate SFR$_{\mathrm{TIR}}$(8,24,70) because the former present
similar values to SFR$_{\mathrm{TIR}}$(24) that, as shown in the
previous section, overpredicts SFR$_{\mathrm{TIR}}$(8,24,70) at
z$\sim$2.

As the relative trends between the estimates of the SFR(8,24,70) has
barely changed, our conclusions from Section~\ref{sfrs24} still apply,
i.e, the values of SFR$_{\mathrm{A-H06}}$ are roughly consistent with
those of SFR$_{\mathrm{TIR}}$, but the SFR based on the calibrations
of B08 and R09 present systematic deviations with respect to these
that can be significant (up to 0.75~dex) in certain SFR and redshift
intervals. In addition, we find that the typical {\it rms} of the
comparison of SFR(8,24,70) estimates is 20\% larger with respect to
the previous comparison based on 24~$\mu$m data. This is not
surprising considering that Figure~\ref{compare_sfrs} shows only
functional relation of each method but none of the uncertainties
attached to the fit of data.

Note that the comparisons shown in Figure~\ref{compare_sfrs} and
~\ref{compare_sfrs_70} only illustrates the expected uncertainty
budget associated to the use of different SFR estimates, but they do
not demonstrate that any of them provides intrinsically more accurate
results. Nonetheless, having MIPS 70~$\mu$m data to constrain the
shape of the IR SED, it is reasonable to assume that the values of
SFR$_{\mathrm{TIR}}$(8,24,70) would provide more reliable values than
the other 3 methods. In order to verify this statement and to asses
the accuracy of SFR$_{\mathrm{TIR}}$(8,24,70), we compare the SFRs to
the results from other authors based on better photometric datasets.
In particular, we compare our SFRs against the values of
\citet[][S08]{2008MNRAS.385.1015S} and
\citet{2009ApJ...700..183H}. The latter studied the SFRs of a
spectroscopic sample of high-z (z$\sim$1.9) galaxies with strong
IR-emission (f(24)$>$0.5~mJy). For these galaxies, the authors provide
accurate SFRs estimated from a very detailed coverage of the IR SED
including {\it Spitzer}/IRS spectroscopy and data at 24, 70 and
160$\mu$m, 1\,mm and 1.4\,GHz. In S08 the authors describe the IR
properties of a 70~$\mu$m selected sample restricted to galaxies
detected at 160~$\mu$m and having reliable spectroscopic redshifts
ranging from 0.1$<$z$<$1.2 (z$=$0.5).

The sources in common with S08 and \citet{2009ApJ...700..183H} are
shown in the right panel of Figure~\ref{compare_sfrs_70} as black
stars (open and closed, respectively). In addition, we show as red
stars those sources which were poorly fitted to the models of CE01 and
DH02 in the work of S08. We find that our values of
SFR$_{\mathrm{TIR}}$(8,24,70) for the majority of the z$<$1.2 galaxies
tend to underestimate the SFRs of S08 with a median (considering only
black stars) difference and scatter of
$\Delta$SFR$=$0.09$\pm^{0.20}_{0.14}$\,dex. However, there is small
group of sources for which the SFRs are systematically underestimated
by $\sim$0.5\,dex or more (red stars). In S08, the authors showed
that, for these galaxies, the IR SED fitting to the models of CE01 and
DH02 severely under-fitted the data at 160$\mu$m, whereas the models
of \citet[][SK07]{2007A&A...461..445S} allowed a better fit to the
data (see also \citealt{2010MNRAS.403.1474S}). As a result, the values
of L(TIR) obtained from the fit to models of CE01 and DH02 would be
systematically lower than the estimates for SK07. These strong
discrepancies in the fitting of CE01 and DH02 models do not seem to be
the usual scenario (e.g., \citealt{2010ApJ...709..572K}), although
some issues fitting the MIPS 160~$\mu$m fluxes of local galaxies with
the models of CE01 has been reported \citep{2009A&A...507.1793N}.

A possible explanation for this issue could be related to fact that
S08 makes use of the 4 IRAC bands in the SED fitting. As a result,
these bands contribute significantly to the $\chi^{2}$ (more than the
MIPS bands), whereas they only represent a minimum fraction of the
total IR luminosity.  Nonetheless, some intrinsic differences in the
IR-SED of MIPS 160~$\mu$m selected samples are expected, given that
these are usually biased towards {\it cold} galaxies, i.e., galaxies
with a relatively large (and probably not very frequent) cold dust
content in comparison with the amount and emission of the warm dust
featured in the models of CE01 and DH02. In summary, the differences
in the SFR(8,24,70) with respect to the values of S08 for these
sources are most likely the result of combined SED modeling issues and
selection effects.

For the rest of the sources, the $\sim$0.10\,dex offset in
SFR(8,24,70) towards underestimating the values of S08 is in good
agreement with the results of \citet[][K10]{2010ApJ...709..572K} for a
sample of galaxies selected at 70~$\mu$m (and counting with 160~$\mu$m
data for $\sim$20\% them). The authors indicate that the estimates of
L(TIR) for 160~$\mu$m detected sources computed without fitting that
flux can be underestimated by up to 0.20\,dex at z$<$1 and
$\sim$0.3\,dex at higher redshifts. The authors also point out that
these effect could be related to a bias in 160$\mu$m selected samples
towards selecting intrinsically cooler objects (as opposed to
70~$\mu$m selections). A similar statement is made in S08 based on the
160/70 flux ratios of their sample.

The comparison to the SFRs of \citet{2009ApJ...700..183H} for galaxies
at z$\sim$2 present too few sources to provide a significant
result. However, the overall results are in relatively good agreement
within 0.3\,dex. We find again that SFR$_{\mathrm{TIR}}$(8,24,70)
slightly under-predicts the SFRs of some galaxies, consistently with
the conclusions of K10 for 160$\mu$m detected galaxies. Moreover, we
find that the values of SFR$_{\mathrm{TIR}}$(24) for the galaxies of
\citet{2009ApJ...700..183H} overestimates SFR(8,24,70) by a factor of
$\sim$4, following the trend shown by the blue line.

Finally, we further check the accuracy of our estimates at high-z,
comparing them to the values of SFR$_{\mathrm{TIR}}$(24) corrected
with the empirical relation of \citet{2006ApJ...640...92P}. This
correction was conceived to mitigate the excess in the IR-SFRs of
high-redshift galaxies estimated from 24~$\mu$m data. The correction
was computed by matching the SFR(24) to the SFRs estimated from the
average stacked fluxes in MIPS 24, 70 and 160~$\mu$m of a sample of
z$\sim$2 galaxies. The green line in Figure~\ref{compare_sfrs_70}
joins the median ratios of
SFR$_{\mathrm{TIR,CORR}}$(24)/SFR$_{\mathrm{TIR}}$(8,24,70) as a
function of redshift.  The overall results are that the values of
SFR$_{\mathrm{TIR,CORR}}$(24) are a factor of $\sim$3-4 lower than
SFR$_{\mathrm{TIR}}$(24) at z$\sim$2. As a result these estimates are
also slightly lower than our predictions for
SFR$_{\mathrm{TIR}}$(8,24,70). Nonetheless, the values of
SFR$_{\mathrm{TIR,CORR}}$(24) also agree within 0.3\,dex with the SFRs
of the galaxies in common with \citet{2009ApJ...700..183H}.

\subsection{Summary of the SFRs}

The accuracy of the SFRs estimated from IR tracers up to intermediate
redshifts has been demonstrated by the good agreement with the
estimates based on other tracers such us dust corrected UV/optical
indicator (\citealt{2007ApJ...670..279I};
\citealt{2007ApJS..173..267S,2009ApJ...700..161S}). On the other hand,
the systematics effects in the IR-based SFRs of the most luminous
galaxies (ULIRGS) at high redshift are quite significant. Some of
these issues arise from the assumptions made in the estimation of
IR-based SFRs, such as the validity of the local templates at high
redshift or the contribution of obscured AGNs to the IR
luminosity. However, the most relevant issues arise from the lack of
enough data to constrain the full IR SED. Particularly for studies
based on 24~$\mu$m data alone. Nonetheless, the breadth and quality of
the MIPS 24~$\mu$m data ensures that it will continue leading multiple
studies of IR-based SFR for the foreseeable future. Thus, quantifying
the systematic effects between the SFRs(24) computed with different
methods, and the differences in the SFRs(24) with respect to the SFRs
computed from more IR data, provides a useful information.

Our analysis shows that, although the values of
SFR$_{\mathrm{TIR}}$(24) are consistent with those of
SFR$_{\mathrm{A-H06}}$(24) within 0.3~dex (the usual uncertainty
quoted for IR-based SFRs) the values of SFR$_{\mathrm{B08}}$(24) and
SFR$_{\mathrm{R09}}$(24) can be significantly deviated (up to
$\pm$0.5~dex) with respect to SFR$_{\mathrm{TIR}}$(24) for certain
redshift and luminosity ranges.  The differences in the SFRs obtained
with these methods remain mostly unchanged for SFR(8,24,70), and we
find that the discrepancies in SFR$_{\mathrm{B08}}$ or
SFR$_{\mathrm{R09}}$ with respect to SFR$_{\mathrm{TIR}}$(24) do not
provide a better agreement to the SFRs of other authors computed from
very detailed IR photometric data. Therefore, out of the four methods
to estimate the IR-SFR discussed here, SFR$_{\mathrm{TIR}}$ present
(after accounting for intrinsic systematics) the more accurate
results.

From the analysis of sample of MIPS 70~$\mu$m detected galaxies, we
find that SFR$_{\mathrm{TIR}}$(24) is reasonably consistent with the
values of SFR$_{\mathrm{TIR}}$(8,24,70) up to ULIRG luminosities
(typically at z$\lesssim$1.4) showing only a small deviation towards
underestimating SFR(8,24,70) by $0.05$\,dex with an {\it rms} of
0.2~dex. However, at z$>$1.5 the agreement is significantly worse. The
values of SFR$_{\mathrm{TIR}}$(24) tend to overestimate SFR(8,24,70)
by a median value of 0.15$\pm$0.40~dex. As already pointed out by
other authors, the best approach to solve this issue is to apply a
correction factor that reduces the estimated values at high-z
(\citealt{2006ApJ...640...92P} or \citealt{2009A&A...504..751S}).

The comparison of SFR$_{\mathrm{TIR}}$(8,24,70) (our best-effort SFRs)
to the SFRs computed by other authors based on a better IR photometric
coverage (including MIPS 160~$\mu$m) also shows an excellent
agreement, proving that these estimates are robust.  The overall
results are consistent within 0.3\,dex presenting only a small
systematic deviation in SFR$_{\mathrm{TIR}}$(8,24,70) towards
underestimating the values including MIPS 160~$\mu$m data by
-0.09\,dex (mostly z$<$1.2 galaxies). Note that since this comparison
is restricted to MIPS 160~$\mu$m detected sources the could be some
selection effects, and thus this offset might not apply for all
galaxies (see e.g., the results of K10 based on stacked fluxes in MIPS
160~$\mu$m for a 70~$\mu$m selected sample).


\section{Data access}
\label{datacatalogs}

All the data products for the 76,936 IRAC 3.6+4.5\,$\mu$m-selected
([3.6]$<$23.75) sources in the EGS are presented here. These include:
(1) the photometric redshift catalog containing the estimates with
{\it Rainbow}, {\it EAZY} and from I06, when available
(Table~\ref{dataredshift}); (2) the stellar mass catalog containing
the values estimated with each of the different modeling
configurations described in \S~\ref{evalmass} (Table~\ref{datamass});
(3) the SFR catalog containing the UV- and IR- based SFRs obtained
with the different methods and calibrations discussed in
\S~\ref{evalSFR} (Table~\ref{datasfr}). A table containing the
UV-to-FIR SEDs for all these sources is presented in Paper I. The
number of objects and unique identifier of this table and the tables
presented in the following is the same.

A larger version of these catalogs containing all galaxies down
[3.6]$<$24.75 (3$\sigma$ limiting magnitude) are available through the
web utility {\it Rainbow Navigator}\footnotemark[2] (see paper I for a
more detailed description), that provides a query interface to the
database containing all the data products of the multiple {\it
  Rainbow} tasks that we have used in the papers. {\it Rainbow
  Navigator} has been conceived to serve as a permanent repository for
future versions of the data products in EGS, and also to similar
results in other cosmological fields (such as GOODS-N and GOODS-S,
presented in PG08).

\footnotetext[2]{http://rainbowx.fis.ucm.es}

\subsection{Table~\ref{dataredshift}: Photometric redshift catalog}

These are the fields included in Table~\ref{dataredshift}:

\begin{itemize}
\item {\it Object}: Unique object identifier starting with irac000001.
  Objects labeled with an underscore plus a number (e.g,
  irac000356\_1) are those identified as a single source in the IRAC
  catalog built with SExtractor, but deblended during the photometric
  measurement carried out with the {\it Rainbow} software (see
  \S~\ref{photometry}). Note that, although the catalog contains
  76,936 elements, the identifiers do not follow the sequence
  irac000001 to irac076185. This is because the catalog is extracted
  from a larger reference set by imposing coordinate and magnitude
  constraints. The table is sorted according to this unique
  identifier.
\item $\alpha,\delta$: J2000.0 right ascension and declination in
  degrees.
\item {\it zphot-peak}: Maximum likelihood photometric
  redshift.
\item {\it zphot-best}: Probability weighted mean photometric
  redshift. This is the value of z$_{\mathrm{phot}}$ used along the
  paper.
\item {\it zphot-err}: 1$\sigma$ uncertainty in the photometric
  redshift as estimated from the zPDF.
\item {\it zphot-EAZY}: Photometric redshift estimated using the EAZY
  code \citep{2008ApJ...686.1503B} on our SEDs with the default
  templates and including the $K$-band luminosity prior.
\item {\it Qz}: Estimate of the quality of the photometric redshifts
  computed with EAZY. Reliable photometric redshifts present values of
  {\it Qz}$\leq$1 \citep{2008ApJ...686.1503B}.
\item {\it zphot-I06}: Photometric redshift from
  \citet{2006A&A...457..841I}. These are only available for galaxies
  in the main region.
\item {\it zspec}: Spectroscopic redshift (set to -1 if not
  available).
\item {\it qflag}: Spectroscopic redshift quality flag from 1 to 4.
  Sources with {\it qflag}$>$3 have a redshift reliability larger than
  80$\%$.
\item {\it N(bands)}: Number of photometric bands used to derive the
  photometric redshift.
\item {\it Stellarity}: Total number of stellarity criteria satisfied.
  A source is classified as a star if it satisfies 3 or more criteria.
  A description of all the stellarity criteria and the accuracy of the
  method is given in \S~5.4 of Paper I.
\end{itemize}
\noindent

\subsection{Table~\ref{datamass}: stellar mass catalog}

The stellar masses are estimated from the same templates used to
compute the photometric redshifts. These templates were computed using
several combination of stellar population synthesis library, IMFs and
extinction laws. Our reference stellar masses are those obtained with
[P01,SALP,CAL01] (see \S{\ref{analyzesed}}).  We provide 2 different
stellar mass estimates based on these templates depending on the
redshift used during the fitting procedure, namely: {\it zphot-best}
and {\it zspec}.  In addition, we additional estimates obtained with:
1) the stellar population models of
\citet[][BC03]{2003MNRAS.344.1000B},
\citet[][M05]{2005MNRAS.362..799M} and Charlot \& Bruzual (2009;
CB09), 2) The IMFs of \citet[][KROU]{2001MNRAS.322..231K} and
\citet[][CHAB]{2003PASP..115..763C}, and 3) the dust extinction law of
\citet[][CF00]{2000ApJ...539..718C}. In Table~\ref{datamass}, we give
6 additional stellar mass estimates obtained under different stellar
population modeling assumptions, namely: [P01,KROU], [BC03,CHAB],
[M05,KROU], [CB09,CHAB], [CB09,SALP] and [P01,SALP,CF00].  The
extinction law in all cases except for the last is CAL01.  For these
stellar mass estimates we use {\it z-fit}, which is equal to {\it
  zphot-best} unless {\it zspec} is available.

These are the fields included in Table~\ref{datamass}:

\begin{itemize}
\item {\it Object}: Unique object identifier as in the photometric
  catalog.
\item $\alpha,\delta$: J2000.0 right ascension and declination in
  degrees.
\item {\it Mass(best)}: Stellar mass [log~M$_{\odot}$] with the
  associated uncertainty estimated with {\it zphot-best} using our
  default modeling parameters [P01,SALP,CAL01].
\item {\it Mass(zspec)}: Stellar mass [log~M$_{\odot}$] with the
  associated uncertainty, estimated with {\it zspec} using our default
  modeling parameters [P01,SALP,CAL01].
\item {\it z-fit}: Value of the photometric redshift used during the
  SED fitting with the [P01,KROU], [BC03,CHAB], [M05,KROU],
  [CB09,CHAB], [CB09,SALP], [P01,SALP,CF00] models.  It is equal to
  {\it zphot-best} unless {\it zspec} is available.
\item {\it Mass(P01,KROU)}: Stellar mass [log~M$_{\odot}$] with the
  associated uncertainty, estimated with the modeling parameters
  [P01,KROU,CAL01] and {\it zphot-fit}.
\item {\it Mass(BC03,CHAB)}: Stellar mass [log~M$_{\odot}$] with the
  associated uncertainty, estimated with the modeling parameters
  [BC03,CHAB,CAL01] and {\it zphot-fit}.
\item {\it Mass(M05,KROU)}: Stellar mass [log~M$_{\odot}$] with the
  associated uncertainty, estimated with the modeling parameters
  [M05,KROU,CAL01] and {\it zphot-fit}.
\item {\it Mass(CB09,CHAB)}: Stellar mass [log~M$_{\odot}$] with the
  associated uncertainty, estimated with the modeling parameters
  [CB09,CHAB,CAL01] and {\it zphot-fit}.
\item {\it Mass(CB09,SALP)}: Stellar mass [log~M$_{\odot}$] with the
  associated uncertainty, estimated with the modeling parameters
  [CB09,SALP,CAL01] and {\it zphot-fit}.
\item {\it Mass(P01,CF00)}: Stellar mass [log~M$_{\odot}$] with the
  associated uncertainty, estimated with the modeling parameters
  [P01,SALP,CF00] and {\it zphot-fit}.
\end{itemize}
\noindent

\subsection{Table~\ref{datasfr}: SFR catalog}

The unobscured UV-SFR is obtained from the best-fitting optical
template modeled with [P01,SALP,CAL01]. The rest-frame IR luminosities
and IR-based SFRs are computed either from the average value of the
best fitting templates from the dust emission models of
\citet[][CE01]{2001ApJ...556..562C} and
\citet[][DH02]{2002ApJ...576..159D} to MIPS 24~$\mu$m data only, or
the average value of the best fitting dust emission models of CE01,
DH02 and \citet[][R09]{2009ApJ...692..556R} to IRAC-8.0, MIPS 24 and
70~$\mu$m data. Only fluxes at rest-frame $\lambda>$5~$\mu$m are
considered in this method. In both cases we use {\it zphot-fit}, which
is equal to {\it zphot-best} unless {\it zspec} is available. For
sources un-detected in MIPS 70~$\mu$m at z$\gtrsim$0.6 both methods
provide similar results modulo the effect of the R09 templates.

Note that the SFRs has been computed for all the MIPS 24~$\mu$m and  70~$\mu$m
detections, but only sources with f(24)$>$60~$\mu$Jy and
f(70)$>$3.5$m$Jy (the 5$\sigma$ detection limit) are discussed in
\S~\ref{evalSFR}. In addition, sources un-detected in MIPS 24~$\mu$m are fitted
using an upper limit value of f(24)$=$60~$\mu$Jy. In this cases the
quoted L(TIR) and SFRs are negative values.

\begin{itemize}
\item {\it Object}: Unique object identifier as in the photometric
  catalog.
\item $\alpha,\delta$: J2000.0 right ascension and declination in
  degrees.
\item {\it f(24),f(70)}: Observed flux [$\mu$Jy] and uncertainties in
  MIPS 24 and 70~$\mu$m.
\item {\it z-fit}: Value of the redshift used during the IR SED
  fitting. It is equal to {\it zphot-best} unless {\it zspec} is
  available.
\item L(TIR,24): Total IR luminosity [log~M$_{\odot}$yr$^{-1}$],
  calculated by integrating the (average) dust emission model from
  8\mic\ to 1000\mic\ . This value is computed by fitting the observed
  flux in MIPS 24~$\mu$m to the models of CE01\&DH02.
\item SFR$_{0.28}$: Unobscured UV-based SFR [M$_{\odot}$yr$^{-1}$]
  estimated from the rest-frame luminosity at 0.28\mic\, interpolated
  in the best-fit optical template, $\nu L_{\nu}$(0.28), using the
  Kennicutt (1998) calibration.
\item SFR$_{\mathrm{TIR}}$(24): IR-based SFR [M$_{\odot}$yr$^{-1}$],
  estimated from L(TIR) using the calibration of Kennicutt
  (1998). This value is computed by fitting the observed flux in MIPS
  24~$\mu$m to the models of CE01\&DH02.
\item SFR$_{\mathrm{CE01}}$(24): Same as SFR$_{\mathrm{TIR}}$(24) but
  fitting the MIPS 24~$\mu$m data to the models of CE01 only.
\item SFR$_{\mathrm{B08}}$(24): IR SFR [M$_{\odot}$yr$^{-1}$]
  estimated from the rest-frame monochromatic luminosity at 8$\mu m$
  using the calibration of \citet{2008A&A...479...83B}.  This value is
  computed by fitting the flux in MIPS 24~$\mu$m to the models of CE01\&DH02.
\item SFR$_{\mathrm{A-H06}}$(24): Total SFR [M$_{\odot}$yr$^{-1}$]
  estimated from the rest-frame monochromatic luminosity at 24\,$\mu
  m$ using the calibration of \citet{2006ApJ...650..835A}. This value
  is computed by fitting the observed flux in MIPS 24~$\mu$m to the
  models of CE01\&DH02. Note that to obtain the IR-SFR part of this
  value, the unobscured UV-SFR must be subtracted according to
  Equation~\ref{sfrtir}.
\item SFR$_{\mathrm{R09}}$(24): IR SFR [M$_{\odot}$yr$^{-1}$]
  estimated from the observed flux in MIPS 24~$\mu$m and the redshift
  using the formula of \citet[][eq~14]{2009ApJ...692..556R}.
\item
  L(TIR),SFR$_{\mathrm{TIR}}$,SFR$_{\mathrm{CE01}}$,SFR$_{\mathrm{B08}}$,SFR$_{\mathrm{A-H06}}$(8,24,70).
  Same as the previous values but fitting the IR SED with IRAC-8.0,
  MIPS 24 and 70~$\mu$m data to the models of CE01, DH02 and R09.
  Note that SFR$_{\mathrm{R09}}$ has been omitted because its value is
  independent of the flux in IRAC-8.0 nor MIPS 70~$\mu$m.
\end{itemize}
\noindent

\section{Summary}

In this paper, and the companion (Barro et al. 2010a; Paper I), we
have presented an IRAC-3.6+4.5$\mu$m selected sample in the Extended
Groth Strip characterized with UV-to-FIR SEDs.  The photometric
catalog includes the following bands: far-UV and near-UV from GALEX,
$u^{*}g'r'i'z'$ from the CFHTLS, $u'gRiz$ from MMT and Subaru
observations, $BRI$ from CFHT12k, $V_{606}$, $i_{814}$, $J_{110}$ and
$H_{160}$ from \HST\ , $JK$ from Palomar Observatory, CAHA and Subaru
data, and [3.6]-to-[8.0], 24\mic, and 70\mic\, data from {\it Spitzer}
IRAC and MIPS surveys. Our catalog contains 76,936 sources down to a
85\% completeness level ([3.6]$<$23.75) over an area of
0.48\,deg$^{2}$.  In addition, we have cross-correlated our sample
with the redshift catalog from DEEP2, and with X-ray and VLA-20cm
radio data.

Paper I presented the data, the procedure to measure consistent
UV-to-FIR photometry using our own dedicated software ({\it Rainbow}),
and the analysis of the multi-band properties of the sample.  We
showed that the SEDs present the level of consistency required to
characterize the intrinsic stellar populations of the galaxy. In this
paper, we have presented a galaxy-by-galaxy fitting of the UV-to-FIR
SEDs to stellar population and dust emission models. From the best
fitting optical and IR templates, we have estimated: (1) photometric
redshifts, (2) stellar masses, and (3) SFRs. Then, we have analyzed in
detail their accuracy and reliability with respect to different
parameters. In the following we present the summary of the most
important results of this analysis, organized by parameter.

Photometric redshifts ($z_{\mathrm{phot}}$) were estimated from the
comparison of the UV-to-NIR SEDs to stellar population and AGN
templates. This comparison was carried out with our own dedicated
software (within the {\it Rainbow} package) using $\chi^2$
minimization algorithm (see \citealt{2008ApJ...675..234P} for more
details), and with the EAZY code \citep{2008ApJ...686.1503B}. These
are our main results about photometric redshifts:

\begin{itemize}

\item Two new features have been included in the {\it Rainbow}
  photometric redshift code over the previous implementation in
  \citet{2008ApJ...675..234P} to improve the quality of the estimates:
  (1) a zero-point re-calibration of the observed photometry, and (2)
  the use of template error function as a weight term in the SED
  fitting procedure. Both features are computed simultaneously and
  iteratively based on the comparison of observed and synthetic
  photometry in a spectroscopic control sample. The results show an
  overall good agreement between observations and templates. The
  zero-point corrections are typically $\lesssim$0.1~mag and converge
  after a few iterations. The overall {\it rms} in the residual is a
  factor of $\sim$2 the median photometric uncertainty. The most
  noticeable discrepancies present at $\lambda$$>$3~$\mu$m and (to a
  lesser extent) around 250nm. These are the result of limitations in
  the stellar templates in the NIR range, and a possible excess in the
  strength of the dust attenuation with respect to a
  \citet{2000ApJ...533..682C} extinction law, respectively.

\item The comparison of our photometric redshifts to 7636 secure
  spectroscopic redshifts from DEEP2 and \citet[][LBGs at
    z$>$3]{2003ApJ...592..728S} shows an overall accuracy of
  $\sigma_{\mathrm{NCMAD}}$$\equiv1.48\times\mathrm{median}\left(\left|\frac{\Delta
    z-\textrm{median}(\Delta
    z)}{1+z_{\mathrm{spec}}}\right|\right)$$=$0.034 (where $\Delta
  z$$=$$z_{\mathrm{phot}}-z_{\mathrm{spec}}$) and
  $\sigma_{\mathrm{NCMAD}}$=0.046, with $\eta$$=$2\% and 3\%
  catastrophic outliers ($\eta$ defined as the fraction of galaxies
  presenting $\sigma_{\mathrm{NCMAD}}$$>$0.2) in the EGS main region
  (covered by the CFHTLS) and flanking regions (covered with fewer and
  shallower bands), respectively.  The overall scatter in
  $z_{\mathrm{phot}}$ does not depend strongly on the redshift,
  presenting a minimum value around z$=$0.5-1
  ($\sigma_{\mathrm{NCMAD}}$=0.028 and 0.040 in the main and flanking
  regions, respectively), and increasing by a factor of $\sim$1.3 at
  lower and higher redshifts (up to z$<$1.5). At z$\sim$3, the
  $z_{\mathrm{phot}}$ accuracy for 91 LBGs with secure spectroscopy is
  reduced to $\sigma_{\mathrm{NCMAD}}$=0.063 and $\eta$=10\%.

\item The accuracy of the $z_{\mathrm{phot}}$ is mostly independent of
  the [3.6] magnitude. However, it decreases with the optical
  magnitude from $\sigma_{\mathrm{NCMAD}}$$=$0.030 at R$=$22 to
  $\sigma_{\mathrm{NCMAD}}$$=$0.060 at R$=$25. Approximately 50\% of
  the catastrophic outliers have R$>$23.5 and $\log(\chi^{2})$$>$0.6.
  Approximately 60\% of the sources with significantly different
  values of $z_{\mathrm{best}}$ and $z_{\mathrm{peak}}$
  ($|z_{\mathrm{best}}$-$z_{\mathrm{peak}}|$/(1+z)$>$0.2) are
  catastrophic outliers.

\item The $z_{\mathrm{phot}}$ statistics for the 1995 and 262
  spectroscopic galaxies detected in MIPS 24~$\mu$m
  (f(24)$>$60~$\mu$Jy) and 70~$\mu$m (f(70)$>$3.5~mJy) in the main
  region are similar to the rest of the sample with
  $\sigma_{\mathrm{NCMAD}}$=0.033, $\eta$$=$3\%, and
  $\sigma_{\mathrm{NCMAD}}$=0.045, $\eta$$=$1\%, respectively. The
  accuracy for the 142 X-ray sources is similar
  ($\sigma_{\mathrm{NCMAD}}$=0.038) although with larger fraction of
  outliers ($\eta=$10\%), probably as a result of some degree of
  contamination by the AGN, for which reliable $z_{\mathrm{phot}}$ are
  difficult to estimate based on stellar templates. The worst results
  are found for a very few (12) power-law galaxies (PLGs, identified
  as obscured AGNs): $\sigma_{\mathrm{NCMAD}}$=0.052 ($\eta=$17\%). We
  also note that sources with increasing fluxes in the IRAC bands
  (f$_{[3.6]}$$<$f$_{[4.5]}$$<$f$_{[5.8]}$$<$f$_{[8.0]}$) makes up for
  up to 15\% of the total number of outliers.

\item The $z_{\mathrm{phot}}$ {\it Rainbow} are in good agreement with
  those from the $i'$-band selected catalog of
  \cite[][I06]{2006A&A...457..841I}, which overlaps with our sample in
  the main region. For the 5454 galaxies in common between the two
  catalogs with [3.6]$<$23.75 and $i'<$24.5, the accuracy of the
  $z_{\mathrm{phot}}$ at z$<$1 is roughly the same,
  $\sigma_{\mathrm{NCMAD}}$$=$0.035. At higher redshifts, our larger
  band coverage (mostly in the NIR) provides more accurate results and
  less severe systematic errors and uncertainties. In particular, for
  galaxies at z$\sim$3 (the LBG sub-sample), the outlier fraction in
  I06 is 46\% for only 9\% in {\it Rainbow}. Our $z_{\mathrm{phot}}$
  catalog and the one presented in I06 are complementary: whereas the
  NIR-selected sample detected more galaxies at high-z, which are too
  faint in the optical to be included in the I06 catalog, the IRAC
  catalog misses a population of low-mass galaxies at z$<$1 which are
  recovered by the $i'$-band selection in I06.

\item We showed that the photometric catalog provides robust SEDs by
  obtaining a different realization of the $z_{\mathrm{phot}}$ catalog
  with similar quality using the code EAZY
  (\citealt{2008ApJ...686.1503B}). In particular, these alternative
  photometric redshifts are slightly more accurate for the sources in
  the main region, particularly at z$<$0.5, whereas they present a
  larger scatter in the flanking regions. Moreover, these
  $z_{\mathrm{phot}}$ exhibit a slightly larger systematic deviation
  ($\Delta$z/(1+z)$=$0.019 and 0.027 in the main and flanking regions,
  respectively) than the $z_{\mathrm{phot}}$ computed with {\it
    Rainbow}.

\item We further tested the accuracy of our $z_{\mathrm{phot}}$ by
  checking the number densities and z$_{\mathrm{phot}}$ distributions
  of a sub-sample of (NIR selected) s-BzK ($\rho=$5.0~arcmin$^{2}$;
  $\tilde{z}=$1.89), p-BzK ($\rho=$0.5~arcmin$^{2}$; $\tilde{z}=$1.85)
  and DRG ($\rho=$1.4~arcmin$^{2}$; $\tilde{z}=$2.47). These are in
  relatively good agreement with the results from the literature down
  to $K_{\mathrm{VEGA}}$$<$21. The most significant difference is an
  excess of $\sim$1.5 in the density s-BzKs, which could be caused by
  an overdensity of galaxies at z$\sim$1.5.

\item The median redshift of the ([3.6]$<$23.75) sample,
  $\tilde{z}$$=$1.2, is consistent with that of the flux limited
  samples of \cite{2008ApJ...675..234P} and
  \cite{2009ApJ...690.1236I} in different fields.
\end{itemize}

Stellar masses for the whole sample were obtained in a
galaxy-by-galaxy basis by fitting the optical-to-NIR SEDs to stellar
population synthesis models. In addition, we analyzed the effects of
the choice of different stellar population synthesis (SPS) libraries,
IMFs and dust extinction laws on our estimations. For that, we
considered a reference set of assumptions to which several
combinations of input parameters were compared. This reference stellar
masses were obtained with the PEGASE 2.0 \citep{1997A&A...326..950F}
stellar population synthesis models (P01), a Salpeter (1955) IMF
(SALP) and the \citet[][CALZ01]{2000ApJ...533..682C} extinction
law. We compared these estimations with those obtained with: (1) the
stellar population models of \citet[][BC03]{2003MNRAS.344.1000B},
\citet[][M05]{2005MNRAS.362..799M} and Charlot \& Bruzual (2009;
CB09); (2) the IMFs of \citet[][KROU]{2001MNRAS.322..231K} and
\citet[][CHAB]{2003PASP..115..763C}; (3) the dust extinction law of
\citet[][CF00]{2000ApJ...539..718C}. These are our main results about
stellar masses:

\begin{itemize}
\item From the comparison of the stellar masses estimated with
  photometric and spectroscopic redshifts we find a 1$\sigma$
  uncertainty of $\sim$0.2\,dex. The distribution of stellar masses as
  a function of redshift for our default modeling assumptions shows
  that 90\% of the galaxies present log(M)$>$10M$_{\odot}$ at z$>$2,
  at the limiting magnitude of our sample ([3.6]$<$23.75).
 
\item We quantified the impact of the choice of different IMFs in the
  estimated stellar masses. For the CB09 models, we found that the use
  of a SALP, KROU, or CHAB IMF introduces constant offsets (with a
  very small scatter) in the estimated stellar masses:
  $\Delta\log$(M)[SALP$-$KROU]$=$0.19\,dex and
  $\Delta\log$(M)[CHAB$-$KROU]$=$$-$0.04\,dex. For the models of P01,
  the difference for a SALP and KROU IMFs depends on the mass, ranging
  from $\Delta\log$(M)$=$0.03\,dex for masses lower than
  $\log$(M)$=$10\,$\mathcal{M}_\sun$ to 0.13\,dex above that threshold.

\item We quantified the impact of using different SPS codes in the
  estimated stellar masses. We found that the new CB09 models predict
  slightly lower masses than the older version, BC03, by
  $\Delta\log$(M)$=$0.04~$\pm^{0.28}_{0.15}$\,dex. Our stellar masses
  estimated with the P01 models are on average larger than those
  obtained with the CB09 models (for a KROU IMF) by
  $\Delta\log$(M)$=$0.15$\pm$0.26\,dex. The estimates with the P01
  library are also larger than those with the M05 SPS by
  $\Delta\log$(M)$=$0.39$\pm$0.34\,dex.  We found slightly lower
  values of this offset for galaxies with
  $\log$(M)$>$10\,$\mathcal{M}_\sun$ ($\sim$0.30~dex). Our default
  modeling assumptions, [P01,SALP,CAL01], predicts comparatively the
  largest stellar masses.  Accounting for all systematic offsets, all
  models are roughly consistent within a factor of 2-3.

\item We quantified the effect of using different treatments of the
  dust extinction by comparing the stellar masses estimated with a
  CAL01 and CF00 extinction laws. The median result is a small
  systematic deviation of 0.03\,dex towards smaller values when using
  CAL01, and a {\it rms} of $\sim$0.20\,dex. This suggest that the
  different treatments of the dust attenuation do not play a major
  role in the estimate of the stellar masses.

\item The comparison of our results with several stellar mass catalogs
  already published in EGS revealed a good agreement despite the
  differences in the modeling technique and in the photometric
  dataset. We found a median offset and scatter of
  $\Delta\log$(M)$=$-0.07$\pm$0.21\,dex and
  $\Delta\log$(M)$=$0.10$\pm$0.25\,dex with respect to the catalogs of
  stellar masses published by \citet{2006ApJ...651..120B} and
  \citet{2007MNRAS.382..109T}, respectively.

\end{itemize}

SFRs were estimated for all galaxies in our sample following a variety
of procedures. First, we calculated the unobscured SFR (the star
formation which is directly observable in the UV/optical) from the
observed luminosity at 280\,nm (SFR$_{UV,obs}$). To get the total SFR
of a galaxy, the former value must be added to the SFR which is not
directly measurable in the UV/optical because of the extinction by
dust. We calculated this SFR from the IR data taken by {\it
  Spitzer}/IRAC and MIPS at 24\mic\, and 70\mic (if available).  The
general procedure consist on fitting the IR photometry at rest-frame
wavelengths $\lambda>$5$\mu$m (usually involving 8, 24 and 70~$\mu$m
data) to the dust emission templates of
\citet[][CE01]{2001ApJ...556..562C} and
\citet[][DH02]{2002ApJ...576..159D} and
\citet[][]{2009ApJ...692..556R}, but we also performed some test by
fitting only MIPS 24~$\mu$m data to the models of CE01\&DH02.

From the best fit to the models, the IR-based SFR for each object was
estimated with 4 different methods: (1) the total infrared luminosity,
L(TIR), integrated from 8 to 1000\mic\, transformed to a SFR with the
factor published by \citet{1998ARA&A..36..189K}. (2) the rest-frame
monochromatic luminosity at 8\,$\mu$m (SFR$_{\mathrm{B08}}$)
transformed to L(TIR) and SFR using the empirical relation described
in \citet{2008A&A...479...83B} and the Kennicutt factor. (3) the
empirical relation given in \citet{2006ApJ...650..835A} between the
rest-frame monochromatic luminosity at 24\,$\mu$m and the SFR
(SFR$_{\mathrm{A-H06}}$). (4) using equation (14) of
\citet{2009ApJ...692..556R}, that relates the SFR
(SFR$_{\mathrm{R09}}$) to the observed flux in the MIPS 24~$\mu$m band
and the redshift. The monochromatic and integrated luminosities were
computed from the average value of the best fit templates. These are
our main results about SFRs:

\begin{itemize}

\item We quantified the differences in the IR-based SFRs obtained with
  the four methods based on the fit to MIPS 24~$\mu$m data only.  The
  SFRs estimates with the models of CE01 and DH02 are compatible
  within a factor of 2, presenting a maximum difference around
  z$\sim$1.5. The estimates of SFR$_{\mathrm{TIR}}$ and
  SFR$_{\mathrm{A-H06}}$ are roughly consistent
  ($\Delta$SFR$\sim$$-$0.18$\pm$0.05\,dex) when the contribution of
  SFR$_{UV,obs}$ is small. SFR$_{\mathrm{B08}}$ gives systematically
  lower values than SFR$_{\mathrm{TIR}}$ for
  SFR$>$20\,M$_{\odot}$yr$^{-1}$ and z$>$1, and higher values for
  lower redshifts and SFRs.  The difference exceeds a factor of 5 for
  SFR$>$1000\,M$_{\odot}$yr$^{-1}$. The overall agreement between
  SFR$_{\mathrm{TIR}}$ and SFR$_{\mathrm{R09}}$ is rather poor, except
  at z$>$1.8 where the differences are lower than 0.05\,dex. The
  reasons for these discrepancies can be found in the differences in
  the relative emission of the cold and warm dust, and in the strength
  of the PAH and silicate absorption. These characteristics can vary
  by up to a factor of $\sim$5 from one set of templates to the other.

\item For each of the methods to estimate the SFR, we studied the
  effect having a better constrained IR SED comparing the SFRs
  computed from IRAC+MIPS, SFR(8,24,70), and just MIPS 24~$\mu$m,
  SFR(24). At low-z, the median values of SFR$_{\mathrm{TIR}}$(24) and
  SFR$_{\mathrm{A-H06}}$(24) tend to underestimate SFR(8,24,70) by
  0.05 and 0.10\,dex, respectively, with an {\it rms} of
  $\sim$0.2-0.3~dex.  At z$\sim$2, the estimates from these two
  methods based on 24~$\mu$m data only are on average $\sim$0.20\,dex
  larger than those obtained with
  SFR(8,24,70). SFR$_{\mathrm{B08}}$(24) presents the opposite trend,
  giving larger values than SFR$_{\mathrm{B08}}$(8,24,70) at z$\leq$1
  (up to 0.18~dex), but remaining mostly unchanged at higher
  redshifts.

\item The relative differences between each of the methods to estimate
  the IR-based SFRs with respect SFR$_{\mathrm{TIR}}$ (described in
  the first item) remains mostly unchanged when using
  SFRs(8,24,70). The values of SFR$_{\mathrm{TIR}}$(8,24,70)
  (best-effort estimate) for a sample of galaxies in common with
  \citet{2008MNRAS.385.1015S} and \citet{2009ApJ...700..183H}, who
  counted with more photometric fluxes in the FIR and (sub-)mm range,
  presented a relatively good agreement within $\sim$0.3\,dex.  At
  z$<$1.2 we find a small deviation of 0.09\,dex in
  SFR$_{\mathrm{TIR}}$(8,24,70) towards underestimating the SFR of
  MIPS-160 detected galaxies. Larger discrepancies, up to 0.5\,dex,
  might arise for individual galaxies due to the use of different
  template sets.

\end{itemize}

In the context of studies of galaxy evolution, our catalog provides a
self consistent sample with a very detailed characterization of the
systematic uncertainties suitable for multiple scientific purposes. It
is also an alternative to other catalogs providing only photometry,
redshifts or stellar parameters alone. Furthermore, our photometric
catalog itself provides a reference point for independent analysis of
the stellar populations.

The multi-band photometric catalog presented in Paper I, jointly with
the photometric redshifts and estimated stellar parameters presented
here are publicly available. We have developed a web-interface, named
{\it Rainbow Navigator}\footnotemark[3], that provides full access to
the imaging data and estimated parameters and allows several other
data handling functionalities.
\footnotetext[3]{http://rainbowx.fis.ucm.es}

\section*{Acknowledgments}

We thank the referee for providing constructive comments and help in
improving the contents of this paper. We thank G. Bruzual and
S. Charlot for allowing us to use their models prior to
publication. We acknowledge support from the Spanish Programa Nacional
de Astronom\'{\i}a y Astrof\'{\i}sica under grant AYA 2006--02358 and
AYA 2006--15698--C02--02. PGP-G acknowledges support from the Ram\'on
y Cajal Program financed by the Spanish Government and the European
Union.  Partially funded by the Spanish MICINN under the
Consolider-Ingenio 2010 Program grant CSD2006-00070: First Science
with the GTC.  Support was also provided by NASA through Contract
no. 1255094 issued by JPL/Caltech. This work is based in part on
observations made with the {\it Spitzer} Space Telescope, which is
operated by the Jet Propulsion Laboratory, Caltech under NASA contract
1407. Observations reported here were obtained at the MMT Observatory,
a joint facility of the Smithsonian Institution and the University of
Arizona. GALEX is a NASA Small Explorer launched in 2003 April. We
gratefully acknowledge NASA's support for construction, operation, and
scientific analysis of the GALEX mission. This research has made use
of the NASA/IPAC Extragalactic Database (NED) which is operated by the
Jet Propulsion Laboratory, California Institute of Technology, under
contract with the National Aeronautics and Space Administration. Based
in part on data collected at Subaru Telescope and obtained from the
SMOKA, which is operated by the Astronomy Data Center, National
Astronomical Observatory of Japan.

\bibliographystyle{aa}
\bibliography{referencias}

\clearpage
\begin{landscape}
\begin{deluxetable}{cccccccccccc}
\setlength{\tabcolsep}{0.010in} 
\tabletypesize{\scriptsize}
\tablewidth{600pt}
\tablecaption{\label{dataredshift} The IRAC-3.6+4.5~\mic\ sample: Photometric redshifts}
\tablehead{ 
\colhead{Object}& \colhead{$\alpha$}& \colhead{$\delta$}& \colhead{zphot-best} &
 \colhead{zphot-EAZY} &
 \colhead{zphot-I06} &
 \colhead{zspec} &
 \colhead{qflag} &
 \colhead{zphot-err} &
 \colhead{Qz} &
 \colhead{N(band)} &
 \colhead{Stellarity}\\ 
(1)&(2)&(3)&(4)&(5)&(6)&(7)&(8)&(9)&(10)&(11)&(12)} 
\startdata

irac003270\_1 &215.43910540 &53.08468920 &1.09 &1.06 &1.04 &0.00000 &2 &0.07 &0.31 &16 &0\\ 
irac003278 &215.42614011 &53.09447161 &0.49 &0.39 &0.58 &0.00000 &0 &0.20 &2.58 &16 &7\\ 
irac003291\_1 &215.44058360 &53.08123980 &0.83 &0.80 &0.84 &0.85700 &4 &0.02 &0.11 &16 &0\\ 
irac003310 &215.42129738 &53.09430607 &0.08 &0.14 &0.15 &0.00000 &0 &0.07 &0.13 &18 &0\\ 
irac003313 &215.43553774 &53.08200958 &1.06 &1.01 &1.09 &0.00000 &0 &0.08 &0.37 &16 &1\\ 
\enddata
\tablecomments{
(1) Object unique identifier in the catalog.\\
(2,3) Right Ascension and Declination (J2000) in degrees.\\
(4) Probability weighted photometric redshift. This is our default value of photometric redshift for SED fitting based estimates.\\
(5) Photometric redshift estimated with the code EAZY (\citealt{2008ApJ...686.1503B}) using the default template
configuration and the $K$-band luminosity prior applied to the [3.6] band. The input photometric 
catalog is the same as for the other redshifts.\\
(6) Photometric redshifts as estimated in \citet{2006A&A...457..841I} from the (5-band) $i'$ selected
catalog of the CFHTLS. This catalog overlaps with the IRAC sample in the central
portion of the mosaic (52.16$^{\circ}<\delta<$53.20$^{\circ}$ \&
214.04$^{\circ}$$<\alpha<$215.74$^{\circ}$).\\
(7) Spectroscopic redshift determination drawn from DEEP2 (\citealt{2007ApJ...660L...1D}; $\sim$8,000 galaxies)
and (\citealt{2003ApJ...592..728S}; LBGs at z$\gtrsim$3).\\
(8) Quality flag of the spectroscopic redshift (4=$>$99.5\%, 3=$>$90\%, 2=uncertain, 1=bad quality). Only
redshifts with qflag$>$2 have been used in the analysis.\\
(9) Uncertainty in zphot-best(4) estimated from the 1~$\sigma$ width of the probability distribution function.\\
(10) Reliability parameter of the photometric redshift estimated with EAZY (see \citealt{2008ApJ...686.1503B}
 for more details); Good quality redshifts are in general Q$_{z}\leq$1.\\
(11) Number of different photometric bands used in to estimate the photometric redshift with {\it Rainbow}, column (4).\\
(12) Sum of all the stellarity criteria satisfied (see \S 5.4 of Paper I). A source is classified as star for Stellarity$>$2.\\
(This table is available in its entirety in a machine-readable in the online version. A portion is shown here for guidance.)}
\end{deluxetable}
\clearpage
\end{landscape}

\LongTables
\begin{landscape}
\begin{deluxetable}{cccccccccccc}
\setlength{\tabcolsep}{0.002in} 
\tabletypesize{\scriptsize}
\tablewidth{0pt}
\tablecaption{\label{datamass} The IRAC-3.6+4.5\mic\ sample:  Stellar Mass estimates}
\tablehead{ 
\colhead{Object}& \colhead{$\alpha$}& \colhead{$\delta$}& 
 \colhead{M(best)} &
 \colhead{M(zspec)} &
 \colhead{z-fit} &
 \colhead{M(P01,KROU)} &
 \colhead{M(BC03,CHAB)} &
 \colhead{M(M05,KROU)} &
 \colhead{M(CB09,CHAB)} & 
 \colhead{M(CB09,SALP)} &
 \colhead{M(P01,CF00)} \\
& & & 
 \colhead{M-err} &
 \colhead{M-err} &
&
 \colhead{M-err} &
 \colhead{M-err} &
 \colhead{M-err} &
 \colhead{M-err} &
 \colhead{M-err} &
 \colhead{M-err} \\
(1)&(2)&(3)&(4)&(5)&(6)&(7)&(8)&(9)&(10)&(11)&(12)\\
& & & 
(13)&(14)& &(15)&(16)&(17)&(18)&(19)&(20)} 
\startdata
irac003270\_1 &215.43892696 &53.08455063 &9.79 &9.79 &1.09 &9.67 &9.64 &9.61 &9.61 &9.84 &9.98\\&&&0.05 &0.05 & &0.04&0.08 &0.07 &0.08 &0.08 &0.07\\ 
irac003278 &215.42614011 &53.09447161 &11.69 &11.69 &0.49 &11.55 &11.66 &11.35 &11.48 &11.74 &11.75\\&&&0.15 &0.15 & &0.14&0.17 &0.16 &0.18 &0.15 &0.17\\ 
irac003291\_1 &215.44043562 &53.08128671 &10.84 &10.84 &0.86 &10.97 &10.64 &10.30 &10.57 &10.63 &10.48\\&&&0.03 &0.04 & &0.04&0.04 &0.03 &0.03 &0.04 &0.04\\ 
irac003310 &215.42129738 &53.09430607 &8.80 &8.80 &0.08 &8.61 &8.92 &8.64 &8.79 &9.04 &8.75\\&&&0.06 &0.06 & &0.06&0.08 &0.08 &0.05 &0.05 &0.06\\ 
irac003313 &215.43553774 &53.08200958 &9.94 &9.94 &1.06 &9.83 &9.64 &9.19 &9.40 &9.63 &9.59\\&&&0.04 &0.04 & &0.05&0.04 &0.05 &0.07 &0.07 &0.04\\ 
\enddata
\tablecomments{
(1) Object unique identifier in the catalog.\\
(2,3) Right Ascension and Declination (J2000) in degrees.\\
(4-13) Stellar mass [log(M$_{\odot}$)] with the associated uncertainty, estimated with our default modeling parameters, [P01,SALP,CAL01], and zphot-best.\\
(5-14) Stellar mass [log(M$_{\odot}$)] with the associated uncertainty, estimated with our default modeling parameters, [P01,SALP,CAL01], forcing the photometric redshift to the spectroscopic value, when available.\\
(6) Redshift used in the fitting procedure of (8,9,10,11 and 12). This redshift refer to zphot-best unless the spectroscopic redshift is available; in that case the redshift is forced to the spectroscopic value.\\
(7-15) Stellar mass [log(M$_{\odot}$)] with the associated uncertainty, estimated with the modeling parameters, [P01,KROU,CAL01], and z-fit.\\
(8-16) Stellar mass [log(M$_{\odot}$)] with the associated uncertainty, estimated with the modeling parameters, [BC03,CHAB,CAL01], and z-fit.\\
(9-17) Stellar mass [log(M$_{\odot}$)] with the associated uncertainty, estimated with the modeling parameters, [M05,KROU,CAL01], and z-fit.\\
(10-18) Stellar mass [log(M$_{\odot}$)] with the associated uncertainty, estimated with the modeling parameters, [CB09,CHAB,CAL01], and z-fit.\\
(11-19) Stellar mass [log(M$_{\odot}$)] with the associated uncertainty, estimated with the modeling parameters, [CB09,SALP,CAL01], and z-fit.\\
(12-20) Stellar mass [log(M$_{\odot}$)] with the associated uncertainty, estimated with the modeling parameters, [P01,SALP,CF00], and z-fit.\\
(This table is available in its entirety in a machine-readable in the online version. A portion is shown here for guidance.)}
\end{deluxetable}
\clearpage
\end{landscape}

\begin{landscape}
\begin{deluxetable}{ccccccccccccc}
\setlength{\tabcolsep}{0.001in} 
\tabletypesize{\scriptsize}
\tablewidth{0pt}
\tablecaption{\label{datasfr} The IRAC-3.6+4.5\mic\ sample: IR-Luminosities and Star Formation Rate estimates}
\tablehead{ 
\colhead{Object}& \colhead{$\alpha$}& \colhead{$\delta$}& 
 \colhead{f(24)} &
 \colhead{f(70)} &
 \colhead{z-fit}&
 \colhead{SFR(0.28)} &
 \colhead{SFR(R09)} &
 \colhead{L(TIR,24)} &
 \colhead{SFR(TIR,24)} &
 \colhead{SFR(TIR,24,CE01)} &
 \colhead{SFR(B08,24)} &
 \colhead{SFR(AH06,24)} \\
 &&& 
 \colhead{err-f}&
 \colhead{err-f}&
 &&& 
 \colhead{L(TIR,best)} &
 \colhead{SFR(TIR,best)} &
 \colhead{SFR(TIR,CE01)} &
 \colhead{SFR(B08,best)} &
 \colhead{SFR(AH06,best)} \\ 
(1)&(2)&(3)&(4)&(5)&(6)&(7)&(8)&(9)&(10)&(11)&(12)&(13)\\
 &&&(14)&(15)&&&&(16)&(17)&(18)&(19)&(20)}
\startdata
irac003270\_1 &215.43892696 &53.08455063 &55 &--- &1.09 &6.0 &21.8 &10.93 &14.6 &15.1 &17.5 &9.4\\&&&13 &--- &&&&10.93 &14.8 &15.1 &18.4 &11.0\\ 
irac003278 &215.42614011 &53.09447161 &--- &--- &0.49 &2.0 &1.7 &-10.34 &-3.8 &-4.7 &-8.1 &-2.2\\&&&--- &--- &&&&--- &--- &--- &--- &---\\ 
irac003291\_1 &215.44043562 &53.08128671 &104 &--- &0.86 &2.5 &17.4 &10.92 &14.4 &14.6 &17.5 &9.3\\&&&12 &--- &&&&10.94 &14.9 &14.6 &18.5 &10.7\\ 
irac003310 &215.42129738 &53.09430607 &61 &--- &0.08 &0.1 &0.0 &8.61 &0.1 &0.1 &0.3 &0.1\\&&&10 &--- &&&&8.51 &0.1 &0.0 &0.2 &0.1\\ 
irac003313 &215.43553774 &53.08200958 &--- &--- &1.06 &3.6 &19.9 &-10.90 &-13.8 &-14.5 &-17.3 &-8.6\\&&&--- &--- &&&&--- &--- &--- &--- &---\\ 
\enddata
\tablecomments{
(1) Object unique identifier in the catalog.\\
(2,3) Right Ascension and Declination (J2000) in degrees.\\
(4,14) Observed flux and uncertainty in MIPS 24~$\mu$m [$\mu$Jy].\\
(5,15) Observed flux and uncertainty in MIPS 70~$\mu$m [$\mu$Jy].\\
(6) Redshift used in the fitting procedure. This redshift refer to zphot-best unless the spectroscopic redshift is available; in that case the redshift is forced to the spectroscopic value.\\
(7) UV based SFR [M$_{\odot}$yr$^{-1}$] estimated from the monochromatic luminosity at 2800\AA\ rest-frame using the calibration of \citet{1998ARA&A..36..189K}.We also refer to this value as SFR$_{UV,obs}$\\
(8) IR based SFRs [M$_{\odot}$yr$^{-1}$] estimated from the observed flux in MIPS 24$\mu$m and the redshift using the formula of \citet{2009ApJ...692..556R}.\\
(9,16) Total IR luminosity [log(L$_{\odot}$)] obtained integrating (from 8-1000$\mu$m) the average of the best fitting templates. (9) is computed from the fit of MIPS 24~$\mu$m data to the models of \citet[][CE01]{2001ApJ...556..562C}, \citet[][DH02]{2002ApJ...576..159D}; (10) is computed from the fit of IRAC-8.0 and MIPS 24 and 70~$\mu$m (best effort) data to the models of CE01, DH02 and \citet[R09]{2009ApJ...692..556R}.\\
(10,17) IR based SFRs [M$_{\odot}$yr$^{-1}$] estimated from L(TIR) using the calibration of \citet{1998ARA&A..36..189K}. (10) and (17) are computed using the same combination of data and models as (9) and (16), respectively.\\
(11,18) Same as 10, but in this case the IR SED is fitted only with the models of CE01. (11) and (18) are computed from the fit of these modeles to MIPS-24\mic\ data only and IRAC-8.0 and MIPS 24 and 70\mic\, respectively.\\
(12,19) IR based SFRs [M$_{\odot}$yr$^{-1}$)] estimated from the monochromatic luminosity at 8$\mu$m rest-frame using the relation of \citet{2008A&A...479...83B}. (12) and (19) are computed using the same combination of data and models as (9) and (16), respectively.\\
(13,20) IR based SFRs [M$_{\odot}$yr$^{-1}$]  estimated from the monochromatic luminosity at 24$\mu$m rest-frame using the realtion of \citet{2006ApJ...650..835A}.  (13) and (20) are computed using the same combination of data and models as (9) and (16), respectively.\\
The uncertainties in the values of L(TIR) and the SFRs can be as high as a factor of 2. The accuracy in (9,16) is limited to two decimal places and one decimal place in (10-13,17-20).
Negative values in the columns (9-13) indicate that the sources are non-detected in MIPS~24$\mu$m. In these cases, the corresponding IR luminosity is estimated from a upper limit of f(24)$=$60$\mu$Jy, and the estimates in (16-20) are not computed.\\
(This table is available in its entirety in a machine-readable in the online version. A portion is shown here for guidance.)}
\end{deluxetable}
\clearpage
\end{landscape}

\label{lastpage}
\end{document}